\documentclass[12pt,preprint]{aastex}
\newcommand\ts{\thinspace}
\usepackage[mathscr]{eucal}
\usepackage{graphics}

\shortauthors{Guyon, Sanders, \& Stockton}
\shorttitle{Near Infrared Adaptive Optics Imaging of QSO Host Galaxies}

\begin{document}

\title{Near-Infrared Adaptive Optics Imaging of QSO Host Galaxies}

\author{O. Guyon\altaffilmark{1,2}, D. B. Sanders\altaffilmark{1}, Alan Stockton\altaffilmark{1}}
\altaffiltext{1}{Institute For Astronomy, University of Hawaii, 2680 Woodlawn Drive, Honolulu, HI 96822.}
\altaffiltext{2}{Subaru Telescope, National Astronomical Observatory of Japan, 650 N. A'ohoku Pl., Hilo, HI 96720.}

\begin{abstract}
We report near-infrared (primarily $H$-band) adaptive optics (AO) imaging  with the 
Gemini-N and Subaru Telescopes, of a representative sample of 32 nearby ($z < 0.3$) 
QSOs selected from the Palomar-Green (PG) Bright Quasar Survey (BQS), in order to 
investigate the properties of the host galaxies.  Two-dimensional modeling and visual 
inspection of the images shows that $\sim$36\% of the hosts are ellipticals, $\sim$39\% 
contain a prominent disk component, and $\sim$25\%  are of undetermined type. 
Thirty percent show obvious signs of disturbance.  
The mean $M_{\rm H}({\rm host}) = -24.82$ (2.1{\ts}$L_{\rm H}^*$), with a range 
 $-$23.5 to $-$26.5 ($\sim$0.63{\ts}$L_{\rm H}^*$ to 10{\ts}$L_{\rm H}^*$).   At 
 $<${\ts}$L_{\rm H}^*$, all hosts have  a dominant disk component, while at 
$>${\ts}2{\ts}$L_{\rm H}^*$ most are ellipticals.   ``Disturbed" hosts are found at all 
$M_{\rm H}({\rm host})$, while ``strongly disturbed" hosts appear to favor the more 
luminous hosts.   Hosts with prominent disks have less luminous QSOs, while the 
most luminous QSOs are almost exclusively in ellipticals or in mergers (which presumably shortly
will be ellipticals).  At $z < 0.13$, where our sample is 
complete at $B$-band, we find no clear correlation between $M_{\rm B}({\rm QSO})$ and 
$M_{\rm H}({\rm host})$.  However, at $z > 0.15$, the more luminous QSOs 
($M_{\rm B} < -24.7$), and 4/5 of the radio-loud QSOs, have the most luminous $H$-band hosts 
($> 7{\ts}L_{\rm H}^*$), most of which are ellipticals.  Finally, we find a strong correlation 
between the ``infrared-excess", $L_{\rm IR} / L_{\rm BB}$, of QSOs with host type and 
degree of disturbance. Disturbed and strongly disturbed hosts and hosts with dominant 
disks have $L_{\rm IR} / L_{\rm BB}$ twice that of non-disturbed and elliptical hosts, 
respectively.  QSOs with ``disturbed" and ``strongly-disturbed" hosts are also found to 
have morphologies and mid/far-infrared colors that are similar to what is found for 
``warm" ultraluminous infrared galaxies, providing further evidence for a possible 
evolutionary connection between both classes of objects.

\end{abstract}

\keywords{instrumentation: adaptive optics --- galaxies:hosts --- galaxies:QSOs}

\section{Introduction}
It is now widely accepted that quasi-stellar objects (QSOs) are powered by accretion onto a 
massive black hole (MBH) within a galaxy. While MBHs are commonly found in galaxies, 
QSOs require the central black hole to be fed by a steady supply of matter. Characterization 
of the host galaxies of QSOs can provide important clues to understand both the formation 
mechanisms and the evolution of QSOs.

Observations of low-z QSOs offer the obvious advantage of increased resolution and 
sensitivity for accurately measuring host properties, and relatively large samples 
are required to both adequately probe the full range of host properties and to explore possible 
relationships between host type and properties of the QSO nucleus.     The pioneering 
optical surveys of low-z QSOs in the mid 1980's, using 4m-class telescopes  \citep{hut84,smi86},
produced several common themes, as well as enigmatic problems that still remain today.
QSOs were found typically to live in luminous ($> L^*$) hosts, with a mixture of host types 
(spirals, barred-spirals, ellipticals), including a significant fraction of disturbed and/or 
strongly-interacting systems.  There was also some hint of a trend in host type with host 
luminosity, with spirals and luminous ellipticals favored at the low and high end, respectively, 
of the range of QSO luminosities.  However, it often proved difficult to distinguish reliably 
between different host types, often due to poor sensitivity or angular resolution or both; in addition, the 
mixture of objects being observed was abnormally weighted toward radio-loud and/or 
X-ray selected objects despite these objects being only a small fraction of the predominant 
radio-quiet population.  Optical QSO surveys with {\it HST}/WFPC2  \citep{bah97,boy99,mclu99,ham02,dun03} 
offered dramatically improved resolution, but for the most part contained similar biases toward 
radio-loud objects, resulting in a preponderance of ellipticals.  
In the end, results from the {\it HST} surveys did not significantly modify
earlier conclusions based on the ground-based surveys. 

More recent surveys of QSO hosts have favored the near-infrared, where the fraction of the 
observed light due to the underlying host galaxy is largest, and ground-based seeing 
can rival that from {\it HST}.   Important large near-infrared QSO surveys include those 
from the ground with 2/4m-class telescopes \citep{dun93,mcl94a,mcl94b,tay96,per01}
and, more recently, those exploiting adaptive optics (AO) \citep{mar01,sur01}, and, finally, those using 
{\it HST}/NICMOS  \citep{mcl01,vei05}.   Of most relevance to our own current work are 
the near-infrared observations by McLeod and co-workers  who have  imaged the host 
galaxies for a large volume-limited sample of optically-selected QSOs from the PGBQS 
similar to the sample and selection criteria we have used.  Other 
near-infrared surveys of QSOs include studies with nearly equal mixtures of radio-quiet and 
radio-loud QSOs \citep[e.g.,][]{dun93,tay96,mar01,per01}, and studies of  ``far-infrared-excess" 
QSOs \citep[e.g.][]{sur01,vei05}.   All of the near-infrared surveys generally agree 
with previous optical surveys in finding that QSOs typically reside in luminous (i.e., $> L^*$) 
hosts, and that spirals represent an increasing fraction of hosts at lower QSO luminosities. 
However, the near-infrared surveys still emphasize the overall diversity of host types, 
and they are still plagued by difficulties with the accurate determination of radial profiles due to a 
combination of problems including bad seeing, unstable PSF, and ultimately the faintness 
of the hosts at large radii.  

Finally, it has now become clear that optically-selected QSOs may be rivaled in their 
numbers by newly discovered populations of near-infrared selected \citep{cut02}, and far-infrared 
selected \citep{low88,san88} QSOs.  It is extremely intriguing that a very high 
fraction ($>70$\%) of the former  show evidence of tidal interactions \citep{hut03}, and 
nearly all of the latter are found to be major mergers of large gas-rich 
galaxies (e.g., \citealt{san96}).  These new findings raise the obvious questions of 
possible evolutionary connections between these different classes of objects, all of which 
seem to be capable of producing luminous QSOs.  It thus became clear to us that 
it might prove extremely useful to revisit the properties of optically-selected QSO hosts
and to search for trends in host type with other properties such as far-infrared 
luminosity and color, which might then prove useful in future studies designed to 
explore possible connections between the familiar optically-selected QSO population and the 
various types of infrared-selected QSOs. 

The availability of AO systems on 8--10 m class telescopes provides  
a potentially important new capability for accurately measuring the distribution of light in 
the host galaxy of QSOs.  In particular, our access to relatively large amounts of observing 
time using the new University of Hawaii near-infrared AO system, Hokupa`a, on 
Mauna Kea, allowed us to carefully design a sensitive and relatively uniform survey, 
primarily at {\it H}-band, of a large representative sample of low-z QSOs randomly selected 
from the optical ({\it B}-band) PGBQS.  The sample was made large enough to include a 
representative sub-sample of the minority population of radio-loud objects, as well as 
to allow for adequate sampling of a wide range of QSO luminosities, but was 
otherwise unbiased.  An additional bonus for our new observations was the existence 
by 2003 of relatively complete and accurate spectral-energy-distributions (SEDs) from radio-to-Xray
wavelengths \citep{san89,elv94}, including most importantly, new far-infrared fluxes for 
most of our targets, which were recently determined from reprocessed {\it IRAS} and {\it ISO} 
data \citep{haa00,haa03}.  

 In this paper we present the first near-infrared imaging study of a 
large number of QSOs (32 objects) made with AO systems on the Gemini 8.2m and Subaru 8.1m 
telescopes on Mauna Kea.   We first describe, in \S \ref{sec:sample} 
how our sample was selected. The observations and data reduction techniques are detailed in 
\S \ref{sec:observations}. The QSO host galaxies are characterized through 2-D fits and visual 
analysis of the images in \S \ref{sec:analysis}.  Section \ref{sec:adopthostclass} presents our 
adopted classification scheme, including descriptions of the individual objects and 
comparisons to the host types derived from previous imaging studies, and summarizes the host fit parameters.  
Section \ref{sec:discussion} discusses how host type depends on other 
properties of the QSO (e.g. degree of disturbance, radio loudness, the near-infrared absolute luminosity 
of the host, and the total far-infrared luminosity), and compares our results with previous studies. 
Our main conclusions are summarized in \S \ref{sec:conclusions}.

As background to our current work, good brief reviews of previous problems and subtleties 
associated with interpreting the host types of 
low-z QSOs can be found in \citet{mcl94a} and \citet{mar01}.  In addition, a more thorough 
review of the wide range of QSO host studies -- their consensus and differences -- including 
theories as to the origin and evolution of QSOs and their possible relation to other types 
of extragalactic objects, can be found in the proceedings of three recent workshops 
specifically devoted to the study of QSO hosts \citep{cle97,marq01,bar06}.   

\section{Sample selection}
\label{sec:sample}
The performance of AO systems is a strong function of the photon flux seen by the wavefront 
sensor (WFS). Although the Hokupa`a (the Institute for Astronomy AO system \citep{gra98} in use on the 
Gemini North Telescope from mid-2000 to mid-2002) and Subaru AO \citep{taka04} systems 
can ``lock'' onto objects as faint as 
$m_{\rm V} \approx 18$ under good seeing, only very partial wavefront correction can be achieved for 
such faint objects. To ensure reasonably good image quality (FWHM$<0\farcs2$), our sample was limited to 
bright QSOs (approximately $m_{\rm V}<16$). This observational constraint would result in a strong 
bias towards intrinsically highly luminous QSOs if the sample were not redshift-limited. Moreover, 
observation of higher redshift QSOs is made more difficult for the purpose of this work by the 
decrease in spatial resolution, and especially surface brightness, with increasing distance. Our 
sample is hence limited to $z \approx 0.3$.  

The ideal sample for our current observations proved to be the Palomar-Green Bright Quasar Sample
\citep[BQS: ][]{sch83}.  Although there are more recent and larger QSO surveys, the BQS still contains 
the majority of the nearest and brightest optically-selected, UV-excess QSOs for study from the 
northern sky, and its magnitude limit was close to our desired brightness limit for good AO performance.  There 
are also extensive multi-wavelength observations in the literature for the BQS sample, 
which potentially allows us to further explore the relationship of the QSO host properties to other 
properties of the QSO.  

The BQS contains 92 QSOs with $M_{\rm B} {\rm(Vega)} < -23$ 
\citep[in the $H_o = 50, q_o = 0$ cosmology used in][]{sch83}, in the 10,714 deg$^2$ surveyed region 
of the sky at $\vert b \vert > 30^\circ$ and $\delta >  -14^\circ$.  Forty-seven of these QSOs met our 
final redshift and brightness selection criteria ($z \leq 0.3$, $B_{\rm Vega}  \leq 16.3$),  of which we were
eventually able to observe 58\% (28/47) during our two year observing campaign on Mauna Kea. 
These 28 QSOs from the BQS were essentially selected at random from the complete sub-sample of 47,
depending only of the RA range observable during each observing run.  

Four additional objects were eventually added to the list of 28, bringing the total number of 
observed targets to 32.  Two objects from the BQS with $M_{\rm B} > -23$
(PG1119+120 and PG 1126-041) were successfully observed during one of the earliest observing 
runs, while we were still defining the scope of the overall AO campaign.  Rather than discard 
the data, we decided to include these objects here for two reasons. First they allow 
us to further explore the low luminosity end of the QSO sample, and second, we noted early on 
that the uncertainties in the $B$-band magnitudes listed in the original BQS are large enough such that 
these two objects might have properly been in our sample.  Two additional objects were added to 
our observing list in the latter half of our overall campaign, when it was realized that we had only two 
``bona-fide'' radio-loud QSOs (RLQs) in our list of observed objects.  One of these RLQs, 4C31.63
($z = 0.295$), was at the time it was added, the highest redshift object observed, but still within our 
planned redshift limit for the survey, and its $B$-band magnitude was above the average overall threshold  for 
the BQS.  The second RLQ, PG2251+113 ($z = 0.326$), was above our stated redshift cutoff, but satisfied  
our $B$-band limit and was the only RLQ from the BQS that could be observed at the RA range available 
for observations during 2001, September.   The addition of these two RLQs allowed us to bring 
the total fraction of RLQs in our observed sample ($5/32 \approx 16\%$) up to the ratios more 
typically found (i.e., $\sim$15\%) in large QSO surveys. 

\placetable{tab:sourcelist}

The complete list of our 32 observed targets is given in Table 1 (Vega magnitudes are used in
this table and throughout this paper).  The listed $B$-band magnitudes have been 
updated where possible with newer more accurate values than originally given in \citet{sch83}.  The 
absolute magnitudes are also updated to an $H_{\rm o} = 75, \Omega_{\rm M} = 0.3, \Omega_\Lambda = 0.7$, 
cosmology:  the $M_{\rm B} < -23$ QSO criteria used by \citet{sch83} now corresponds to $M_{\rm B} < -22.1$.
We note that the updated $B$-band magnitudes resulted in three additional objects, PG0838+770, PG0844+349 and PG1229+204, falling below the absolute magnitude cutoff, bringing the total number of lower luminosity objects to five.

\section{Observations and data reduction}
\label{sec:observations}
\subsection{Instruments}
Both the Hokupa`a AO system, built by the University of Hawaii and used on the Gemini North 
telescope, and the Subaru Telescope AO system, were used for this project. For both systems, 
a 36-element curvature sensor is used with a 36-element bimorph mirror. Curvature AO systems 
are traditionally very efficient on faint targets, partly thanks to the use of photon-counting devices 
(Avalanche Photo Diodes) with no readout noise. This allows these systems to guide on 
$m_{\rm V}=18$ sources and to routinely provide good correction (better than 0\farcs2) on $m_{\rm V}=16$ 
point sources, which makes them well suited for QSO host galaxy imaging.
The AO-corrected beam was fed to QUIRC for Gemini and to IRCS for Subaru. Both cameras 
use $1024 \times 1024$ detectors for near-IR imaging.

\subsection{Choice of filter and target}
The contrast between the QSO and a typical host galaxy is better (lower) at about 1 $\mu$m 
(in the rest frame) than either in the visible or the longer infrared wavelengths. This is largely 
due to a local minimum near 1 $\mu$m in the rest-frame radio-to-Xray spectral energy 
distributions (SEDs) of nearly all QSOs, both RLQs and RQQs, except for the strongly 
beamed Blazars---e.g., Fig.~2 of \citet{san89}, and also Fig.~1 of \citet{mcl95}.   Moreover, the 
performance of AO systems is optimal in the near infrared: at shorter wavelengths, the number 
of actuators is too small to sample the wavefront properly. At longer wavelengths, the
high thermal background precludes the observation of low-surface-brightness sources, and the 
diffraction limit of the telescope restricts the achievable angular resolution. Observations in the 
near-IR are therefore well-suited for QSO host studies from the ground.

The seeing was constantly estimated during the observations by image quality monitoring and 
monitoring of the control signals of the AO system. Most of the observations were 
made using the $H$ filter, which provides the optimal resolution under good seeing conditions. 
When the seeing was bad (corrected FWHM in $H$ band larger than 0\farcs2), the $K^\prime$ filter was 
preferred to the $H$ filter, and usually allowed high resolution imaging (FWHM$<0\farcs2$) even 
under worse than average seeing, but at the expense of a stronger thermal background. A few 
observations were done in the $J$ band on exceptionally good nights. Fainter QSOs were 
preferentially observed under good seeing conditions, to avoid running out of bright targets 
when the seeing was average or worse than average.  These precautions resulted in a 
surprisingly homogeneous set of images (angular resolution about 0\farcs15) considering 
the wide range of seeing conditions and QSO magnitudes of the sample.

\placetable{tab: observations_log}

Table 2 gives the complete summary of our observing log for all of the data that are
presented in this paper, including filter, exposure times, seeing (FWHM), and telescope used.
Twenty-six objects were eventually observed at $H$, and 8 of these 26 were also observed 
at either $J$ or $K^\prime$, or both.   Four objects were observed at $K^\prime$ only, one 
at $J$ only, and one at both $J$ and $K^\prime$.  Twenty-eight objects were observed at Gemini, 
and four objects were observed at the end of our campaign (Feb, 2003) using Subaru. 

\subsection{Reference PSF observations}
For each target, a list of about six PSF candidates, all within a few arcminutes of the quasar, was 
prepared prior to observations. Each of those was briefly observed (without closing the AO loop) 
to test if it could, with the help of the WFS neutral density filter set, reproduce the WFS photon 
counts generated by the quasar. This allowed us to match the photon counts of the quasar to 
within 10\% for most of the objects in our sample. The AO loop parameters (membrane stroke 
and loop gain) were not modified between observations of the target and the PSF.

For most of the targets, the PSF total integration time was much smaller than the QSO total 
integration time. Because of the variability of the PSF presented in \S\ref{sec:PSFstab}, the 
residuals of the 
PSF subtraction are limited by effects that do not average well with time, and there is little gain 
in observing the PSF star for longer than a few minutes. However, long exposure times are 
required to record the faint structure of the host galaxy that are outside the central part of 
the PSF, where PSF subtraction is not essential (where the PSF wings are fainter than the host 
galaxy). When possible, the PSF star was observed with the same frame integration time as 
the QSO to use the PSF frames for sky extraction.

\subsection{PSF stability---lessons learned}
\label{sec:PSFstab}
In this section, we characterize the PSFs delivered by Gemini+Hokupa`a and the Subaru AO 
system and investigate their variability. This preliminary work is essential for attempting the 
characterization of the QSO hosts in our data.

The PSF delivered by an AO system is a function of :
\begin{itemize}
\item{{\bf 1. Guide star V magnitude and color of the object imaged.}} The brightness of the 
star, as seen by the WFS, strongly affects the performance of the AO system. Due to the 
wavelength-dependence of the seeing (and the optics of the camera), PSFs are also expected 
to be color-dependant. 
\item{{\bf 2. Atmospheric seeing.}} Under poor seeing conditions, the FWHM of the corrected 
PSF is larger. In addition, the brightness of the large halo surrounding the PSF core increases 
with the turbulence. 
\item{{\bf 3. Instrumental variables.}} Wind-powered telescope shake, change in the optical 
path, and the primary and secondary mirror wavefront all modify the PSF.
\item{{\bf 4. AO loop parameters.}} Loop gain and membrane stroke (for curvature systems) 
set the speed of the AO loop and the optical gain of the WFS respectively, and therefore affect 
the PSF.
\end{itemize}

\begin{figure}
\epsscale{1}
\plotone{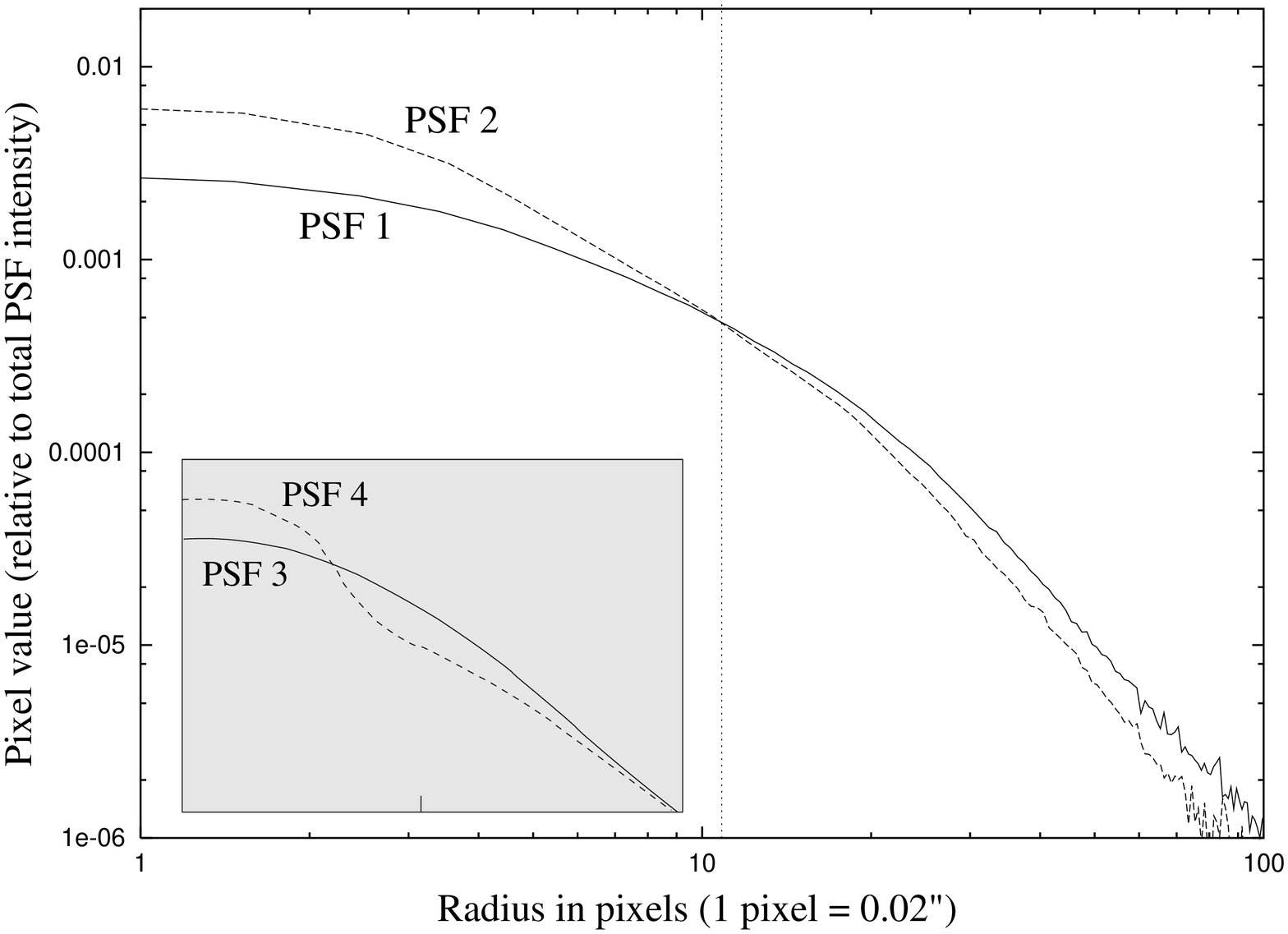}
\caption{\label{fig:psfcmpprof} PSF radial profile variations. The 2 radial profiles plotted in this figure 
(PSF 1 and PSF 2) are derived from 30 s exposures of the same star taken 5 min apart and demonstrate 
the changes in the PSF radial profile due to seeing variations. In the small inset, the log-log radial profile 
of 2 PSFs of different brightness (PSF 4 is brighter than PSF 3) but under the same seeing conditions 
illustrate the effect of the PSF brightness on the radial profile.}
\end{figure}

Despite efforts to match the brightness (as seen by the WFS) of the reference PSF and the 
QSO, small discrepancies in the two flux levels are unavoidable. More importantly, we found 
that the variations in seeing always limit the accuracy of the PSF subtraction when the flux of 
the PSF and the QSO are matched to better than 10\%. Figure \ref{fig:psfcmpprof} shows 
how the radial profile of the PSF can change rapidly (timescale of about a minute) due to 
seeing variations. The far wings of the PSF (1\arcsec\ to 2\arcsec\ in Fig.~\ref{fig:psfcmpprof}) 
show how much 
light is scattered by small-scale turbulence (high spatial frequencies of the seeing), which is 
not corrected by the AO system. In this figure, it can be seen that these wings are much higher 
for PSF1 than for PSF2. Consequently, because the turbulence was stronger during the 
acquisition of PSF1, the AO correction is poorer and, inward of about 0\farcs2 (dotted vertical 
line), the surface brightness of the PSF is lower. The ``crossing point,'' for which the loss of 
surface brightness due to the poor AO correction exactly equals the gain of surface brightness 
due to the increased scattering by small-scale turbulence, is set by the number of actuators of 
the AO system. The surface brightness at this crossing point is insensitive to seeing 
variations. Figure \ref{fig:psfcmpprof} also shows how the brightness of the PSF affects its 
radial profile: for brighter PSFs, the AO correction is better and flux is taken from an 
``intermediate'' annulus between 0\farcs1 and 0\farcs3 and put into the central diffraction 
core of the PSF. The far wings of the PSF are left unchanged because of the limited number 
of actuators of the AO system. During real observation, a combination of PSF variations due 
to changing seeing and different source brightness will affect the radial profile.

\placefigure{fig:psfcmpprof}

\begin{figure}
\epsscale{1}
\plotone{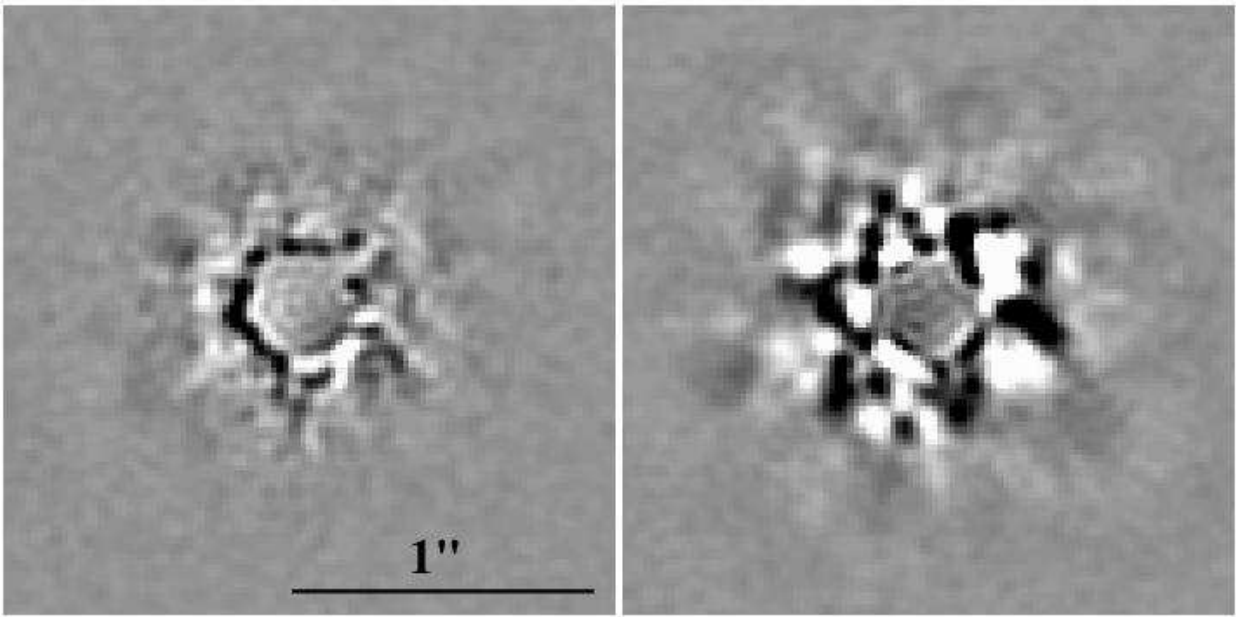}
\caption{\label{fig:rot}Effect of the Cassegrain rotator angle variations on the PSF stability. The left image 
is the difference between two consecutive 10 s images of a star without any Cassegrain rotator rotation. 
The right image is the difference between two consecutive 10 s images between which the Cassegrain 
rotator was rotated by 15$^\circ$. The intensity scale is identical in both images.}
\end{figure}

Although the observer has no control over the seeing, it is possible to reduce PSF variations 
by avoiding any change in the instrument during observations. For example, the AO loop 
parameters should be kept constant. During this observing program, it was discovered that 
the PSF stability could be significantly improved by turning off the Cassegrain rotator of the 
Telescope. At the Gemini and Subaru telescopes, which have alt-azimuth mounts, the 
Cassegrain units carrying the AO systems are usually rotated to preserve the field orientation 
on the detector. The primary and secondary mirror create static wavefront aberrations. If 
these static aberrations are rotating in front of the AO system, which has preferential directions 
due to the geometry of the actuators and also introduces some higher order aberrations, the 
resulting rotational shear between the incoming wavefront and the aberrations of the AO system 
changes the structure of the PSF. This very complex PSF variation is difficult to predict since 
the individual sources of aberrations are poorly known. Figure \ref{fig:rot} illustrates this effect 
through the comparison of two differenced PSFs, one for which the Cassegrain rotator angle 
was kept unchanged between the two exposures, and the other for which a 15$^\circ$ rotation 
occurred between the two exposures. Because some of the aberrations come from the primary 
and secondary mirror and some come from the AO system and the camera, a PSF obtained 
after rotation of the Cassegrain rotator cannot be reproduced by a rotation of a PSF obtained 
at a different Cassegrain rotator angle.

\placefigure{fig:rot}

\subsection{Decoupling of the PSF and the host galaxy}
We have seen above that turning off the Cassegrain rotator makes the PSF more stable and 
therefore, improves the accuracy of the PSF subtraction. There is a second advantage to this 
technique: the PSF and the image of the host galaxy can be decoupled through differential 
rotation.

When attempting to detect faint structures very close to the QSO, we are usually very sensitive 
to small variations in the PSF structure, such as ``flares'' of the PSF. These ``flares'' are more 
frequent when observing faint targets such as QSOs, because the limited number of incoming 
photons doesn't allow fast recovery from events such as a wind-induced telescope shake. It is 
possible to encounter such a flare on the object and not on the reference PSF, leading to the 
false detection of a structure in the host galaxy. Careful visual examination of individual frames 
can help detecting these artifacts but is not fully reliable. When observing with the Cassegrain 
rotator off, the PSF has a constant orientation on the detector while the object rotates, which 
unambiguously decouples the object from the PSF variations. For a ``flare'' to be misidentified 
as a structure in the host galaxy, it would need to rotate in the PSF at a speed reproducing the 
field rotation. Because we observed most objects continuously for more than an hour, the total 
rotation during the observation is usually sufficient to efficiently decouple the object from the PSF.

Because of the improved PSF stability when observing without the field derotator, most of our 
observations were done with the Cassegrain rotator turned off. The geographical coordinates 
of the telescope, the UT time of the exposure, the coordinates of the target and the angle of 
the Cassegrain rotator are required, for each exposure, to recover the field orientation.

This observing mode requires the exposure time to be short enough to avoid rotational smearing 
of the image. The limit we set on the rotational smearing corresponds to half a full width at half 
maximum (FWHM) at the edge of the detector when the target is at the center. This limit allows 
exposure times of 20 s to 30 s on most of the sky. 

For some objects, detector saturation was reached in a few seconds. Because of the relatively long 
readout time of the camera (10 s) on Gemini/Hokupa`a, we chose to obtain a series of 
short unsaturated exposures and a series of longer exposures which were essential to reach 
a good flux sensitivity. We were careful to obtain the longer exposures when the field rotation 
was slow.

\subsection{Host galaxy characterization with AO images}
AO PSFs are time-variable and have relatively strong and extended wings compared with
{\it Hubble Space Telescope} ({\it HST})
PSFs, but the image of the underlying host is made sharp, which allows accurate retrieval of 
the radial profile in regions where the wings of the central PSF are not significantly brighter 
than the host galaxy. Depending on the objects of our sample, we find that it is possible to 
accurately recover the host radial profile outward of about 0\farcs2 (for hosts which account 
for a large fraction of the total flux) to 2\arcsec\ (for hosts which are much fainter than the central 
point source)  with classical PSF subtraction techniques. This corresponds to about 0.5 kpc to 
2 kpc for most of the objects in the sample. Thanks to the large collecting area of 8 m class 
telescopes, the host radial profile can usually be recovered to up to 5\arcsec\ to 10\arcsec. Beyond 
this angular separation, the error due to imperfect sky subtraction becomes larger than the 
host: due to the small field of view of the images, the sky level cannot be accurately decoupled 
from the faint outer part of the host galaxy.

The good angular resolution (0\farcs1 to 0\farcs2, which is comparable to or better than
that of {\it HST} at the same wavelength) and depth of QSO images obtained on a 8 m 
class telescope equipped with an AO system makes it possible to image small and faint 
structures in the host. This is very important for this project, which relies on the ability to 
characterize the morphology of the host beyond the disk vs. elliptical galaxy classification. 

\subsection{Image coaddition and PSF synthesis}
For each target, the frames were processed and coadded by a data reduction pipeline written 
in C. Flat field and bad pixel maps were created from a set of dome flat exposures with 
increasing exposure time. The sky frame was obtained from the dithered QSO and nearby 
PSF exposures.

The persistence effect of the detector was corrected for by ignoring the pixels that have been 
recently strongly exposed to light. An empirical formula was used to select such pixels from 
the pixel value and the time of the exposure of the 10 preceding frames.

\begin{figure}
\epsscale{1}
\plotone{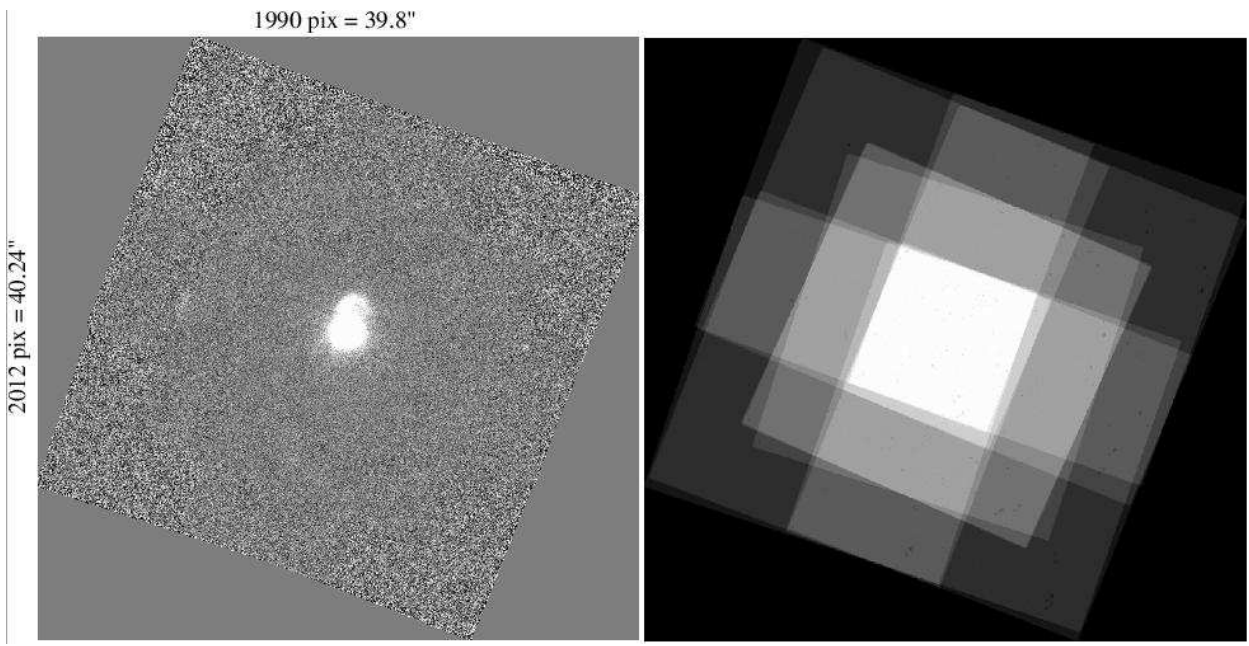}
\caption{\label{fig:exposuremap}Result of the coaddition of the 33 frames (30 s exposure per frame) of 
PG1700+518 (left). On the right, the total exposure for each pixel of the image is represented and 
ranges from 990 s (central part) to 0 s (corners).}
\end{figure}

\placefigure{fig:exposuremap}

The first steps described above (flat fielding, bad pixel identification and sky subtraction) 
were applied to both the PSF and the QSO frames. For each QSO, a coadded image was 
generated by derotating each QSO frame before coaddition. As shown in 
Fig.~\ref{fig:exposuremap}, because of the combination of dithers, rotations and bad pixels 
masking (or pixels ignored due to the persistence of the detector), the total exposure time 
varies across the final image. 

A reference PSF frame is obtained through coaddition of individual PSF frames, but without 
field derotation. To avoid increasing the noise of the PSF-subtracted image, the far wings of 
the PSF were replaced by synthetic PSF wings generated from a fit of the wings radial profile. 
The PSF image matching the coadded QSO image was obtained by coadding multiple copies 
of the reference PSF frame with the proper rotation angles and relative exposure times.

\section{Analysis and Results}
\label{sec:analysis}
\subsection{Images}
Images are shown in figure \ref{fig:qsoimages1} for each object of our 
sample. The flux in the raw images (left) is dominated by the bright central nucleus, although 
the host galaxy is usually visible without PSF subtraction, thanks to the good contrast and high 
angular resolution delivered by AO. The PSF-subtracted images (middle) reveal more host galaxy 
details, especially in the central part of the image.

\placefigure{fig:qsoimages1}

The quality of the PSF subtraction is highly variable through the sample, and depends upon 
many factors (see \S\ref{sec:PSFstab}). In some cases, the reference PSF used for subtraction 
is less peaked (and, therefore, has a comparatively brighter ``halo'' component) than the actual 
PSF during observation. If the host galaxy vs central nucleus luminosity ratio is low, the residual 
of the PSF subtraction is then negative in an annulus surrounding the positive residual at the 
center of the image (see \S\ref{sec:2Dana} for a description of the 2-D analysis algorithm used 
to subtract the PSF): this effect can be seen on PG0026+129, PG0838+770, PG1411+442. 
PSF subtraction errors due to poor stability of the AO control also occur when the source is faint, 
the atmospheric turbulence is strong and/or clouds vary the source brightness during the 
observation. These effects are responsible for the poor PSF subtraction of PG0050+124, 
PG1116+215 and PG1302$-$102. During the observation of PG1617+175, an avalanche photodiode 
of the AO wavefront sensor burned out, suddenly creating a bright extension in the PSF 
delivered by the system, as can be seen in the PSF subtracted image.

Many of the images also show artefacts due to sky subtraction residuals. These features are usually faint and beyond 5\arcsec\ from the QSO. They are most visible in the images of PG0804+761 (curved feature 4\arcsec\ North-West of the QSO and compact artefacts 10\arcsec\ due East, North, West and South of the QSO), PG0838+770 (about 5\arcsec\ to 7\arcsec\ South of the QSO), PG0844+349 (straight edge and extended feature 6\arcsec\ East of the QSO), PG0923+201 (straight feature running North-South about 5\arcsec\ East of the QSO), PG1001+054 (faint extended fringes 5\arcsec\ to 10\arcsec\ South of the QSO), PG1119+120 (faint extended fringes mostly in 10\arcsec\ South-West of the QSO), PG1307+085 (4 faint features in a 10\arcsec\ square pattern centered on the QSO), PG1626+554 (wide 5\arcsec-radius arc centered in the QSO, mostly visible in the northern half of the image) and PG2214+139 (faint broad features in a 8\arcsec\ square pattern centered on the QSO).

\subsection{Photometry}
The flux within a 5\farcs5 diameter circular aperture centered on the QSO was measured for each object, with an
accuracy of about $\pm$5\% at the 1 $\sigma$ confidence level. This aperture size allows easy comparison
with previous measurements by \citet{neu79} and \citet{neu87}. For the objects observed in non-photometric
conditions, only a lower limit for this flux could be determined. 

\placetable{tab:photom}

Table \ref{tab:photom} compares our photometry with other measurements from the literature 
and from the 2MASS archive. No correction was applied to account for differences between 
the 2MASS $K_s$ and Mauna Kea $K^\prime$ photometric systems, as these two filters are very similar \citep{toku02}.

The values shown in Table \ref{tab:photom} are consistent with previous findings that show 
that many of the QSOs observed in this sample appear to show some variability in the near-infrared. 
The amplitude of variability shown in this table is in good agreement with \citet{neu89}: 
both sets of data identify PG1001+054, PG1426+015, 
PG1613+658, PG1617+175 and PG2130+099 as the most variable QSOs of our sample. Since the observations
made under non-photometric conditions need to be calibrated with other photometric measurements of the same object,
the variability of QSOs can cause errors in the estimate of the host galaxy luminosity. In order to minimize
such errors, the time difference between our observations and the photometric measurement should be kept small.
2MASS measurements, rather than older measurements from \citet{neu79} or \citet{neu87}, were therefore used for 
photometric calibration of objects for which no reliable photometry could be derived with our observations. In one
case (PG1116+215), the lower limit obtained with our data is higher than the 2MASS measurement: this lower limit 
was then adopted for photometric calibration. We note that PG0050+124 was observed under very poor conditions: poor seeing and large absorption by clouds resulting in highly variable adaptive optics performance. We did not attempt to perform radial profile or $\chi^2$ analysis for this object, which explains why no photometry was adopted in Table \ref{tab:photom}.

\subsection{Two-dimensional $\chi^2$ analysis of the hosts}
\label{sec:2Dana}
The 2-D analysis relies on a least-square fitting of the image according the the following equation:
\begin{equation}
\label{equ:chi2}
\chi^2 = \sum_{\rm pixels} \frac{(Image-(PointSource+Host)\otimes PSF)^2}{\sigma^2}.
\end{equation}
Pixels too close to (distance less than 1\arcsec\ to 2\arcsec\ depending on the QSOs), or too far 
from (more than 8\arcsec\ to 10\arcsec) the center were excluded from the fit because PSF 
calibration errors dominate the host's signal in the inner region of the image, and sky calibration 
is unreliable in the outer parts of the image (due to the small field of view of the detector). 
Rejecting the central part of the image in the fitting algorithm could allow solutions where the 
total flux of the image is not preserved: the PSF contribution could be increased to a point where 
its total flux exceeds the total flux in the raw image to fit the host galaxy beyond the inner edge 
of the fitting region. To avoid this effect, the total flux of the PSF component was not allowed to 
exceed the total flux in the raw image.

Bright companions, when present, were also excluded from the fit. The value of $\sigma$ was 
directly measured on the image. This 2-D fit can have up to 9 free parameters: 1 for the intensity of 
the central source, 4 for the bulge component (modeled in this work as a single de-Vaucouleurs 
$r^{1/4}$ profile) and 4 for the disk component (modeled as an exponential profile). A simulated annealing 
minimization routine was used to find the absolute minimum. 

\placetable{tab:2Dfitresult1}

The $\chi^2$ minimization routine was first run with the full 9 free parameters to yield 
$\chi_{\rm exp+r^{1/4}}^2$, the minimum value of $\chi^2$ for a exponential + $r^{1/4}$ 
de Vaucouleurs profile host model. Then, $\chi_{\rm r^{1/4}}^2$ ($r^{1/4}$ host only) and 
$\chi_{\rm exp}^2$ (exp host only) were computed, with 5 free parameters 
in each case.  The resulting reduced $\chi^2$ values for each of these three cases are 
listed in Table 4 for all of our targets.  It is worth noting here that the noise component 
$\sigma$ in equation \ref{equ:chi2} is often poorly characterized, 
due to the difficulty of estimating the temporal variations of the PSF wings. Thus the reduced $\chi^2$ 
values listed in Table 4 cannot be reliably compared between targets, but different fits obtained 
from the same image can still be compared to classify the host galaxies. 

For each object, the 1-D radial profiles extracted from the 2-D image and the corresponding 2-D 
fit models (point nucleus + host model convolved by the PSF) are shown in 
Fig.~\ref{fig:qsofits1} to \ref{fig:qsofits11}. The 1-D profile of the residual error, which is also 
shown in the same figures for the 3 fits, tends to become large at small angular distance from 
the center because of errors in the PSF subtraction. The domain within which the fit was 
performed is shown by 2 vertical red lines. 

Due to the limited field of view of the instrument (20\arcsec),
estimations of sky background level and the PSF are prone to errors beyond $\approx$8\arcsec\ radius. This 
explains the peculiar behavior of some profiles beyond $\approx$8\arcsec\ (for example, the abrupt drop at 10\arcsec\ in the PSF profile for PG1440+356), and justifies our choice to 
exclude the outer part of the profile for host galaxy characterization.

\placefigure{fig:qsofits1}
\placefigure{fig:qsofits2}
\placefigure{fig:qsofits3}
\placefigure{fig:qsofits4}
\placefigure{fig:qsofits5}
\placefigure{fig:qsofits6}
\placefigure{fig:qsofits7}
\placefigure{fig:qsofits8}
\placefigure{fig:qsofits9}
\placefigure{fig:qsofits10}
\placefigure{fig:qsofits11}

\subsection{Adopted host classifications}
\label{sec:adopthostclass}
\subsubsection{Classification scheme} 
\label{sec:classscheme}
In deciding which of the three $\chi^2$-model fits listed in Table 4 best characterizes 
the host galaxy, the following criteria were used.
Unless the 2-D fit strongly favors a 2-components bulge+disk host
(i.e.  if $\chi_{\rm r^{1/4}+exp}^2 < min(\chi_{\rm exp}^2,\chi_{\rm r^{1/4}}^2)-0.1$ and no
more than 70\% of the total host flux is allocated to either the
bulge or disk in this bulge+disk fit), the smallest of the
$\chi_{\rm r^{1/4}}^2$ and $\chi_{\rm exp}^2$ is adopted as the result of the
2-D fit.   The adopted fits for each object and each filter used are listed in columns 2--4 
of table \ref{tab:hostclass}. 

\placetable{tab:hostclass}

{\it However}, in a number of cases, the difference in the $\chi^2$ values 
for the exponential vs. the  $r^{1/4}$ de Vaucouleurs profile were small 
enough to render the choice of a particular host model somewhat meaningless.
In the cases where $0.02 < |\chi_{\rm r^{1/4}}^2-\chi_{\rm exp}^2| \leq 0.05$, we consider the
result of the fit to to favor one model, but with very low confidence (noted in parentheses in table
\ref{tab:hostclass}).  And in those cases where $|\chi_{\rm r^{1/4}}^2-\chi_{\rm exp}^2| \leq 0.02$, or in the case of 
bad PSF subtraction, we consider the fit to be inconclusive (a question mark in table \ref{tab:hostclass}).

Since the overall QSO host classification scheme used in this work takes into account 
{\it both} the results of the 2-D $\chi^2$-model analysis and morphological information 
visually extracted from the images, table \ref{tab:hostclass} also lists information obtained from 
visual inspection of the host images, including the presence or absence of ``Bars,'' 
(spiral) "Arms,'' and the overall degree of ``Dist(urbance),'' respectively.  Additional visual information 
is given in the Notes.  

Each host was finally classified using a combination of information which included 
the best-fit $\chi^2$-model ``type" and the visual appearance.  The final ``Adopted Type" listed in 
col. 8 of  table \ref{tab:hostclass} was either ``D$_{\rm p}$" = Disk present, ``E" = Elliptical, or 
``B+D" = Bulge+Disk, except for those objects with  insufficient and/or ambiguous  information from the $\chi^2$-model 
fits, or objects with ``strongly disturbed" disks, where a ``?"  was listed.   
We note here that since the 2-D fitting analysis does not include the central
part of the image (due to PSF variability), the type ``Disk present'' does not exclude the presence of a bulge (which
would be hidden in the central part of the image), and the ``Disk present''/'``Bulge+Disk'' distinction is therefore
unclear. ``Elliptical'' and ``Bulge'' (in the Bulge+Disk type) refer to a $r^{1/4}$ for all 2-D fits performed in this work. 
Strongly disturbed hosts were not classified according to host type, since such a classification would be difficult and unreliable.

While the degree of disturbance is estimated purely from visual inspection of the images, the 
rules adopted for the host ``adopted type'' classification were as follows:
\begin{itemize}
\item{If a bar or spiral arms are obvious in the image, the host is classified as a ``Disk present'', or a 
``Bulge+Disk'' (the decision between ``Disk present'' and ``Bulge+Disk'' is made using the 2-D fitting 
analysis results).}
\item{If no spiral structure or bar is visible in the image, the host classification is given by the 
2-D $\chi^2$-model fitting analysis. If the host has been imaged in several spectral bands, and the corresponding 
2-D $\chi^2$-model  fitting analysis do not agree, the fitting result obtained for the spectral band with the best 
PSF subtraction is adopted.}
\end{itemize}

It is worth restating here that we have chosen to give priority to the features 
visible in the images rather than the results of the 2-D $\chi^2$-model fits, primarily 
because the 2-D $\chi^2$-model fits are prone to errors.  It can 
be seen from the 1-D profiles in Fig.~\ref{fig:qsofits1} to \ref{fig:qsofits11} that for many 
objects, the 2-D fit does not produce a reliable results - we found that changing the fit domain 
often changes the result of the fit.   Thus, it was reasoned that, for example, in several objects 
where a bar or spiral arms were clearly visible in the host images, the host should not be 
classified as an elliptical, even if the 2-D fit suggests so.   It also should be noted that not all 
objects gave ambiguous results.   Some objects are quite easy to classify, as the $\chi^2$-model fit results 
are clear and agree with the images.  
But overall, the images show that the $\chi^2$-model fits  often give misleading 
results, and in $\sim$19\% of the objects (6/32) the hosts are so disturbed that a 
$\chi^2$-model  fit is of little relevance.

\subsubsection{Notes on individual objects}
\label{sec:notes}

Before presenting the final fit parameters corresponding to the host ``adopted type"
listed in column 8 of Table \ref{tab:hostclass}, it is worth comparing our $\chi^2$-model fit results
and observed visual features with results from previous published studies of these objects. 
Each object is discussed separately in the paragraphs below, including comparisons with 
previous analyses by different authors and their published ``host types,'' and including 
remarks about the various difficulties encountered in our own analysis, which often led to 
ambiguities in the final choice of host type. 

\begin{itemize}
\item{{\bf \objectname[]{PG 0026+129}}\\}
The image shows an obvious host, slightly elongated east-west, with no apparent structure. 
The radial profile analysis fails to identify clearly the galaxy type, but gives host magnitudes 
ranging from $m_{\rm H} = 13.43$ (bulge+disk fit) to $m_{\rm H} = 13.69$ (disk fit), and nucleus 
magnitudes ranging from $m_{\rm H}=13.02$ to $m_{\rm H}=13.05$. Although our image does not 
allow classification of this host, the {\it HST}/NICMOS observations of \citet{mcl01} suggest that it
has an $r^{1/4}$-law profile. A non-stellar apparent companion is visible 4.8\arcsec\ from 
the central nucleus ($m_{\rm H}=18.5$, $PA=22\arcdeg$).

\item{{\bf \objectname[]{PG 0050+124}}\\}
This object is also known as I Zw 1 and is often classified as a Seyfert galaxy, particularly
since the host shows clear spiral arms \citep{bot84,sur98,zhe99,can01}. In our observations, 
the spiral arms are 
visible in the host galaxy across the field of the detector (20\arcsec). Although 
the observing conditions for this object were poor (making it impossible to perform an efficient 
PSF subtraction or photometry), the morphology is unambiguously that of a spiral host. 
\citet{can01} have found spectroscopic evidence for continuing star formation in the disk as
well as morphological evidence for weak interaction with the small companion to the west.
\citet{sco03} find a mass in molecular gas of $\sim10^{10}$ M$_{\odot}$, and recent 
observations with the BIMA interferometer \citep{sta04} resolve individual giant molecular 
clouds in a ring structure from about 1.2 to 3.5 kpc from the nucleus.

\item{{\bf \objectname[]{PG 0157+001}}\\}
The host galaxy of PG 0157+001 (= MRK\,1014) is large and bright, and 
it shows 2 prominent
spiral-like features that extend from $\lesssim1$ kpc to at least 28 
kpc (limited by the edge of the
detector, 10\arcsec\ from the nucleus), as well as a bright bulge. 
This galaxy has been the
subject of numerous ground-based and {\it HST} imaging studies (e.g., 
\citealt{mac84,sto87,hut90,
sur98,mcl99,can00}). The spiral features are clearly not conventional 
spiral arms (see Fig.~2 of
\citealt{can00}); instead, as many of the previous studies have 
emphasized, they are almost
certainly tidal tails generated in a past major merger.  These tails 
provide strong evidence
that both of the merging galaxies themselves had cold stellar disks.
The host galaxy is highly asymmetric,
with the northern tail brighter than the southern tail. Several 
bright knots can be seen in the
northern tail, which are also clearly seen in {\it HST} images (e.g., 
\citealt{mcl99}). A small,
bright companion is visible to the south-west at a projected distance 
of about 25 kpc.
The host is well fitted as a $m_{\rm H} = 13.30$ bulge+disk galaxy 
with a
$m_{\rm H}=13.41$ central nucleus, which is
compatible with the structure seen in the image. \citet{can00} 
confirm the presence of significant
populations of young stars, particularly along the leading edge of 
the northern tail.

\item{{\bf \objectname[]{PG 0804+761}}\\}
This host is best described as a ``bar'' about 5 kpc in radius. An extended (kpc size) faint 
companion is visible to the East of the QSO, at a projected distance of 4 kpc from the QSO. 
The curved structure visible in the image about 8\arcsec\ to the North-West of the QSO is a 
artifact from the camera. No reliable 2-D fit was found for this galaxy in $K^\prime$, and the fit in $H$ 
suggests an elliptical galaxy. However, the bar indicates that this galaxy contains a prominent disk. 
The only previous published imaging study of this object is
that of \citet{mcl94a}, who found a profile consistent with a disk galaxy from their ground-based
$H$-band imaging. \citet{sco03} detect molecular gas mass of about $3\times10^9$ M$_{\odot}$
in this galaxy.

\item{{\bf \objectname[]{PG 0838+770}}\\}
The central part of the host of PG0838+770 is a 9 kpc radius bar. A noticeable twist of this bar is 
seen in our image, and a ~20 kpc long arm is starting from the East end of the bar. This arm 
seems to connect both ends of the bar, thus forming the northern half of a ring immediately 
surrounding the bar. Visible images \citep{sur01} show both the bar and two spiral arms outside the ring. An 
extended (about 2.5 kpc diameter) apparent companion is also visible to the northwest at 
a projected distance of 12 kpc. Although the host is best fitted by a $r^{1/4}$ law, the bar 
and spiral arms show that a substantial disk is present.

\item{{\bf \objectname[]{PG 0844+349}}\\}
Two symmetric spiral arms are clearly visible, extending to about 9 kpc radius in our image. 
\citet{hut90} describe the host as a ``barred spiral, with some complex outer structure.'' 
Our 2-D fit, which suggests a dominant bulge, is uncertain because of poor PSF subtraction. 

\item{{\bf \objectname[]{PG 0923+201}}\\}
Although some previous studies have listed this object as interacting \citep{mcl94b} or
possibly interacting \citep{hut89}, apparently because of the bright galaxy projected $\sim10\arcsec$
to the south, the host is unambiguously a very symmetric elliptical. It has an apparent extended 
companion at a projected distance of 7 kpc (2\farcs2) almost due north. This host was also 
identified as an elliptical from {\it HST}/WFPC2 observations \citep{mcl99}. The 2-D fit shows that 
the host galaxy ($m_{\rm J}=15.34$, $m_{\rm H}=14.73$) has a low ellipticity (about 0.1), and the central 
nucleus is about 1 magnitude brighter than the galaxy.

\item{{\bf \objectname[]{PG 0953+414}}\\}
This host was identified as a spiral from ground-based imaging by \citet{hut89}; however, all
subsequent imaging studies, including {\it HST}/WFPC2 observations of \citet{bah97} and
\citet{mcl99}, have found the host too faint to classify reliably. This is also true for our data. 
The brightest part of the host is visible as a faint extension to the south of the QSO in our image.
\citet{mcl99} find marginal evidence favoring an elliptical profile, but they also claim a possible
tidal arm to the southwest, which would indicate the presence of a disk.

\item{{\bf \objectname[]{PG 1001+054}}\\}
Our image shows a highly elongated host (long axis radius of 4 kpc) which is probably a bar 
but could also be a small inclined disk. The 2-D fit suggests an elliptical galaxy, but we choose 
to classify the host as a ``disk present'' from visual inspection of the image. \citet{sur01} see only a small, 
uniform host from ground-based optical and near-IR imaging.

\item{{\bf \objectname[]{PG 1004+130}}\\}
The quasar, also known as 4C\,13.41 and PKS\,1004+13, is a powerful steep-spectrum radio
source. The host galaxy is large, bright and smooth, and is shown to be an elliptical by the 2-D fit. \citet{bah97}
also classify the host as a bright elliptical from {\it HST}/WFPC2 imaging. 

\item{{\bf \objectname[]{PG 1116+215}}\\}
The PSF subtraction is poor for this object. Nevertheless, the host
appears to be asymmetric, with several companions: a compact companion at a projected 
distance of about 4 kpc to the west, a faint extended (2 kpc diameter) companion at 5.5 kpc 
projected distance to the south, a compact companion (star?) at 12.7 kpc to the south-southwest, 
and an elongated bright galaxy at 33 kpc to the West (outside the field of view of the image shown 
in Fig.~5). The 2-D fit to the host suggests that it is a disk, although there is considerable uncertainty
in the determination. \citet{bah97} list the host as ``probably an elliptical.'' 

\item{{\bf \objectname[]{PG 1119+120}}\\}
The host is bright and smooth, with a north-south elongation. An elongated companion is visible 
2.6 kpc (2\farcs7) to the north, projected on the host galaxy. Optical images \citep{sur01} show 2 
faint extended rings on the east and west sides of the brightest part of the host. These features 
are too faint to be detected in our Near-IR images. Our 2-D fitting gives a bulge+disk decomposition
and shows that the host galaxy is almost a magnitude brighter than the central nucleus. 

\item{{\bf \objectname[]{PG 1126$-$041}}\\}
Our image shows a bright and large host with spiral arms extending beyond 10 kpc. The 2-D fit 
suggests the presence of a bulge of luminosity comparable to the disk component. A bright apparent 
companion, at a projected distance of 6.6 kpc (5\farcs7), is clearly resolved and shows a small elongation.

\item{{\bf \objectname[]{PG 1211+143}}\\}
This host is too faint to be classified in this work. \citet{tay96} have concluded from their ground-based
$K$-band imaging that the host is dominated by a disk.

\item{{\bf \objectname[]{PG 1229+204}}\\}
This host was identified as a spiral from ground-based visible observations \citep{hut92} and 
{\it HST}/WFPC2 observations \citep{ham02}. Our observations show a prominent bar (radius is 5 kpc), 
with faint spiral arms starting from their ends. The result of the 2-D fit is a bulge+disk model about 
one magnitude brighter than the central nucleus. We classify this host as a ``disk present'' because of the 
structures visible in our image.

\item{{\bf \objectname[]{PG 1302$-$102}}\\}
The quasar is a strong flat-spectrum radio source. 
Two compact, but resolved, bright companions are within 10 kpc of the QSO. Although decoupling 
the closer companion and the host in the 2-D fitting analysis is difficult, visual inspection of the image 
suggests that the host of PG 1302-102 may be a somewhat asymmetrical elliptical galaxy. The 2-D fit favors 
a $r^{1/4}$ model about 1.5 magnitudes fainter than the central nucleus, which we adopt for the 
classification of this host. \citet{bah97} and \citet{ham02} (working from the same {\it HST}/WFPC2
image) both suggest an elliptical host.

\item{{\bf \objectname[]{PG 1307+085}}\\}
Our image shows a smooth symmetrical host with no apparent structure, and the 2-D fit shows that it 
is an elliptical galaxy with no detectable elongation. Analysis of {\it HST}/WFPC2 images gives a
similar result \citep{bah97,ham02}. The four small ojects at the corners of a perfect 10\arcsec\ by 10\arcsec\ square centered on the QSO are sky subtraction artefacts.

\item{{\bf \objectname[]{PG 1309+355}}\\}
This host is smooth, symmetric and moderately elongated in the North-South direction. Although it has 
been described as an early-type spiral from {\it HST}/WFPC2 observations \citep{bah97,ham02}, 
our 2-D fit favors an elliptical host. {\it HST} observations also give a closely $r^{1/4}$-law profile, 
but pick up weak but definite spiral features, which are not seen in our image probably 
because of the longer wavelength.  Classification for this host is therefore somewhat uncertain, 
although we adopt the ``Elliptical'' classification 
for consistency with the criteria adopted for the rest of our sample.

\item{{\bf \objectname[]{PG 1351+640}}\\}
Our image shows a large smooth host, slightly elongated in the north-south axis. Ground-based 
observations in the visible show a smooth host, best fitted as an elliptical \citep{hut92}. Our 2-D analysis 
does not allow classification of this galaxy. Deep optical imaging \citep{sur01} shows an faint extension 
of the host to the west. This feature is marginally detected in our image, and appears as a subtle light 
excess about 5\arcsec\ west of the QSO.

\item{{\bf \objectname[]{PG 1411+442}}\\}
Highly disturbed host, with a bright elongated companion at a projected distance of 3.6 kpc 
(2\farcs1). Three long tidal arms or loops are visible in our image. This host had previously 
been classified as a spiral \citep{hut92}. Deep optical images \citep{sur01} show that the tidal 
features extend out to at least 30 kpc. Our 2-D fit does not allow classification of this host, and gives 
different answers in different bands. We therefore classify this object as highly disturbed.

\item{{\bf \objectname[]{PG 1426+015}}\\}
Strongly disturbed host, with a bright companion at a projected distance of 4.1 kpc (2\farcs5). The 
host galaxy is asymmetrical: its south-west side is brighter and shows tidal features. Our 2-D fit favors 
an elliptical host, but the apparent tidal features lead us to classify this object as strongly disturbed.

\item{{\bf \objectname[]{PG 1435$-$067}}\\}
Our image shows a smooth host galaxy. The northern half of the host is slightly brighter than the 
southern half at a radius of about 2 kpc. A companion is resolved (about 2 kpc diameter) at a projected 
distance of 10.8 kpc (4\farcs8) to the west. Our 2-D fit suggests an elliptical, which we adopt for the 
host classification.

\item{{\bf \objectname[]{PG 1440+356}}\\}
Two symmetric spiral arms are visible in this host, without any obvious evidence of disturbance. The 
2-D fit for this object is of poor quality, and its result (high surface brightness disk galaxy of small size) 
is probably driven by a mismatch in the PSF or the background sky level. \citet{sur98} classify the
host as a spiral with a bar.

\item{{\bf \objectname[]{PG 1613+658}}\\}
This bright host galaxy shows a very complex structure, with prominent tidal arms and a bright 
resolved apparent companion projected 5.1 kpc (2\farcs2) west of the QSO. This structure was
first reported by \citet{hut92}. Similar structure is 
shown by an {\it HST}/NICMOS image \citep{mcl01}. Deep optical images \citep{sur01} show 
tidal arms extending out to 35\arcsec. 

\item{{\bf \objectname[]{PG 1617+175}}\\}
The PSF subtraction is poor because of an APD failure in the AO system. Our image shows a 
smooth host and the apparent asymmetry in the image is most likely due to PSF subtraction 
residuals. An elliptical host is favored by the 2-D analysis, and adopted for host classification. 
Due to the poor quality of the PSF subtraction, this classification might not be reliable.

\item{{\bf \objectname[]{PG 1626+554}}\\}
Smooth host. The 2-D analysis suggests an elliptical host, but the reference PSF for this object 
is not very good, and the host classification might not be reliable. The rings visible in the 
residual wide-field image of Fig.~\ref{fig:qsoimages1}g might be due to sky subtraction residuals. 

\item{{\bf \objectname[]{PG 1700+518}}\\}
Highly disturbed host, with an interacting companion 7 kpc (1\farcs6) north of the QSO. A tidal arc, 
especially bright in the north and west, is associated with the companion. Several bright knots 
are visible within this tidal feature. Fainter extended features to the south-west and south-east 
of the QSO are visible in our image. Despite the high quality of our data, no reliable classification 
is obtained by our 2-D fit for this highly disturbed object. \citet{sto98} obtained imaging with an 
earlier version of the Hokupa`a AO
system at the Canada-France-Hawaii telescope, detecting the companion and its arc. They also
found spectroscopic evidence for a young (85 Myr) stellar population in the companion. 
\citet{hin99}, from their {\it HST}/NICMOS imaging of this object, suggest that the companion is
a ring galaxy, produced by a small-impact-parameter collision.

\item{{\bf \objectname[]{PG 2130+099}}\\}
This spiral host shows two symmetric, high-contrast spiral arms extending out to at least 6\farcs5 
from the QSO. A non-stellar apparent companion is visible 9\arcsec\ to the east-southeast of the 
QSO. These features are also seen in ground-based and {\it HST}/WFPC2 optical
imaging \citep{hut92,sur98}, which also show evidence for a central bulge. Strangely, \citet{tay96} 
classify this object as an elliptical, although with
marginal confidence. Despite the poor PSF subtraction in our image of this object, the 2-D fit 
correctly identifies this host contains a prominent disk. There is a good match between the fit parameters 
obtained in the $H$ and $K^\prime$ images, both indicating that the host is almost 2 magnitudes fainter 
($\delta m_{\rm H} = 1.56$, $\delta m_{\rm K^\prime} = 1.96$) than the central nucleus.

\item{{\bf \objectname[]{4C 31.63}}\\}
As the name indicates, this quasar is a strong radio source.  The host is smooth and elongated, 
with 3 small apparent companions (projected distances are 11 kpc 
(2\farcs5), 12 kpc (2\farcs8) and 20 kpc (4\farcs7)) within 5\arcsec\ of the QSO. The companion 
12 kpc to the South is of stellar appearance. Our image suggests that, in the central part of the 
host (at about 2\arcsec\ from the QSO), the southeast side of the galaxy is brighter than the 
northwest side. However, the effect is subtle and may be due to a PSF artifact; we therefore
have chosen to classify this host as undisturbed.  The 
2-D analysis suggests that the host is an elliptical. The classification by \citet{tay96} is ambiguous, 
whereas \citet{ham02} classify the host as an elliptical from an {\it HST}/WFPC2 image. Spectroscopy of 
the host \citep{mil02} also support classification as an elliptical galaxy.

\item{{\bf \objectname[]{PG 2214+139}}\\}
This host has a very peculiar radial profile, although the 2-D analysis suggests that it is an elliptical. 
Sharp concentric rings surround the the central QSO at radii ranging from 2.5 kpc to 5 kpc, and 
are especially obvious in the north-west part of the host. These shells are probably tidal debris 
from a recent merger.

\item{{\bf \objectname[]{PG 2251+113}}\\}
The quasar is a 4C and Parkes radio source.
The smooth host is moderately elongated in the east-west direction, and it has a non-stellar 
apparent companion 4\farcs3 to the North. The 2-D analysis shows a slight preference for a disk, 
but the classification is to be considered somewhat unreliable due to poor PSF subtraction. It would be quite
unusual for a radio source of this power to have a host dominated by a disk, and \citet{hut92}
report an $r^{1/4}$ profile for it.

\item{{\bf \objectname[]{PG 2349$-$014}}\\}
This highly disturbed host shows 3 tidal arms and a close companion at 5 kpc (1\farcs7) 
projected distance. The small companion is composed of a bright compact nucleus and two fainter 
knots about 0\farcs3 on either side of it, to the East and West. Spectroscopy \citep{mil02} has 
confirmed that the companion and QSO redshifts match. This host was identified as a 
elliptical undergoing strong interaction from {\it HST}/WFPC2 observations \citep{bah97}, which is in 
agreement with our 2-D fit and our image. Due to the features visible in our image, we chose to 
classify this host as strongly disturbed. The quasar itself is a strong Parkes radio source.

\end{itemize}

\subsubsection{Fit parameters for adopted host type}
\label{sec:hostfitparameters}

Table \ref{tab:2Dfitresult} lists the value of 
the $\chi^2$-model fit parameters (for the ``adopted host" type listed in col. 8 of table \ref{tab:hostclass}) according for the following host decomposition :
\begin{equation}
\label{equ:hostfit}
I(r,\theta) = I_e \times 10^{-3.3307 \times \left(\frac{f(r,\theta,e_e,PA_e)}{r_e}\right)^{\frac{1}{4}}} + I_0 \times e^{-\frac{f(r,\theta,e_0,PA_0)}{r_0}}
\end{equation}
where
\begin{equation}
f(r,\theta,e,PA) = r \times \sqrt{1.0+sin^2(\theta-PA) \times \left(\frac{1}{(1-e)^2}-1\right)}.
\end{equation}

\placetable{tab:2Dfitresult}

For all objects, we verified that the minimization algorithm gives results independent of the 
starting point, and that $\chi_{\rm exp+r^{1/4}}^2$ is smaller than $\chi_{\rm r^{1/4}}^2$ and $\chi_{\rm exp}^2$. 
For ``E"  and ``D$_{\rm p}$" types in col. 8 of table \ref{tab:hostclass} only the first or second term, respectively, 
on the right-hand side of equation \ref{equ:hostfit} is relevant, while for type ``B+D" all 9 parameters are 
used in the fit.  When the host classification is uncertain (``?" in col. 8 of table \ref{tab:hostclass}), the 
9 parameters of the bulge+disk fit decompositions are listed in Table \ref{tab:2Dfitresult}.

\section{Discussion}
\label{sec:discussion}
The results presented in the previous section illustrate both the overall success of our survey in 
being able to detect the great majority of the host galaxies of nearby PGQSOs, as well as the 
difficulties encountered when attempting to fit model radial surface brightness profiles to individual 
hosts.  The two year timeframe of the survey provided sufficient opportunity to revisit objects 
where the initial data had been compromised by poor seeing and/or problems with the 
stability of the PSF.  In the end we were able to reliably detect the host galaxies in all but 
four of our targets (28/32), with sufficient signal-to-noise to allow fitting of the radial surface 
brightness profiles out to radii typically $>$10 kpc -- a tribute in large part to the light 
gathering power of 8-m telescopes.  However, despite extreme attention to detail, the 
stability of the PSF was still a limiting factor in preventing accurate subtraction of the 
central point source, thus compromising our ability to determine the host light profile a 
radii $<${\ts}1-3{\ts}kpc, thus making the fit result rather sensitive to the choice of  inner 
radius limit.   Our model fits also proved to be somewhat sensitive to the exact choice of 
outer radius limit, thus in the end our refusal to use  $\chi^2$-fits alone, and our attempt 
to give at least equal weight to visual inspection of the images, in the final determination 
of  ``host type".  Despite these caveats, there are still several important statistical results 
and trends that clearly emerge from the data.

\subsection{Classification statistics---host types and absolute magnitudes}

Twenty-eight of our 32 objects had detected hosts for which we attempted make a 
classification, and we discuss the  distribution of  ``host type" and host absolute magnitude 
for these 28 objects below.  The remaining 4 objects were not classified due either to 
having hosts which were too faint or having data which suffered from a bad PSF (see 
footnote to Table  \ref{tab:hostclassstatnir}).  Table \ref{tab:hostclassstatnir} summarizes 
the number of objects and their mean H-band absolute magnitudes for each ``host type" -- 
``E", ``D$_{\rm p}$", ``B+D" -- as well as for those objects where a ``host type" could not 
be reliably determined (i.e. ``Unknown"), and also by ``degree of disturbance" -- 
``Non-disturbed", ``Disturbed",  ``Strongly disturbed".  These classifications follow 
the category listings as used  in column ``Dist.'' and column ``Adopted type'', respectively of Table \ref{tab:hostclass}.  

\placetable{tab:hostclassstatnir}

The percentage distributions of ``host type" and the relationship between ``degree of disturbance" 
and the ``host type" can be seen from the graph in Fig.~\ref{fig:hosttype}.  It is immediately 
apparent that the hosts of the sample of PGQSOs in our survey are not characterized by a 
single dominant type.  For the subsample of 19 objects with classified host type 
(``E", ``D$_{\rm p}$", ``B+D" in table \ref{tab:hostclass}), there appears to be an almost
equal split between objects with no detected disk component (``E")  and objects with 
detected disks (``D$_{\rm p}$" and ``B+D").   It is also apparent that a significant fraction, 
(9/28 = 30\%), of the hosts of PGQSOs show clear signs of disturbance and/or are strongly 
disturbed.   If we then make the reasonable assumption that the ``disturbed", and even 
more likely the ``strongly disturbed", hosts also have a disk component, then the percentage 
of objects with a detectable disk component could be as large as 57\% (16/28) to 68\% (19/28). 

\placefigure{fig:hosttype}

Table  \ref{tab:hostclassstatnir} also lists the median and average $M_{\rm H}$ values, and 
the 1$\sigma$ uncertainty in the average for both the host  and nucleus,  for the total 
sample (28 objects) and for each ``host type" subsample.   These values have been 
computed directly from the adopted fits for each object as listed previously in Table 
\ref{tab:2Dfitresult}.  The median and mean host magnitudes typically differ by less 
than 0.3{\ts}mag, with no obvious systematic offset.  In what follows we will use the mean 
values to characterize the properties of the total sample and the ``host type" subsamples. 

For our entire sample (28), the mean host absolute $H$-band magnitude is 
$M_{\rm H} = -24.82$, which is equivalent to $\sim 2.1{\ts}L_{\rm H}^*$ [assuming $M_{\rm H}^* = 24.00$ \footnote[3]{Derived by adopting  the mean observed value of  $H\!-\!K = 0.22$ for the hosts and an observed value of $M_{\rm K}^* = -24.22$  from \citet{mob93}, corrected for our assumed value of $H_0 = 75${\ts}km{\ts}s$^{-1}$Mpc$^{-1}$.}].  Our value of $\sim 2.1{\ts} L_{\rm H}^*$ for 
the mean host luminosity is almost identical to the values of $\sim 2{\ts} L_{\rm H}^*$ 
and 2.3{\ts}$L_{\rm H}^*$ found respectively by \citet{mcl94b}, and \citet{sur01}  from 
their earlier ground-based observations of samples of QSOs selected 
from the same BQS sample of PGQSOs, and using the same $z \lesssim 0.3$ redshift cutoff as 
our current sample.  The latter study also finds a range of 0.5--7.5{\ts}$L_{\rm H}^*$ for 
the host galaxies, very similar to our measured range of $\sim$0.63--10.0{\ts}$L_{\rm H}^*$. 

Mean $H$-band host magnitudes and luminosities for samples of PGQSOs at $z < 0.3$, 
and for quasars in general, are not that abundant in the literature, but the few reported 
measurements all seem to fall generally in the range $\sim${\ts}1--3{\ts}$L_{\rm H}^*$, 
with the lower value being found for low-luminosity QSOs and the higher value being 
found for more luminous QSOs (e.g., \citealt{mcl94a,mcl94b,mcl01}).  
It is interesting to note that this range of host $M_{\rm H}$  is {\it nearly identical} to the range in the 
average $M_{\rm H}$ found for our different ``host type" {\it sub-samples}.  
Table \ref{tab:hostclassstatnir} shows that the lowest mean value---$M_{\rm H} = -23.78$ 
($\sim 0.8 L_{\rm H}^*$)---is found for hosts of type D$_{\rm p}$, i.e., hosts which have 
a dominant disk component, while the mean value for the ellipticals (type E) in our 
sample is $M_{\rm H} = -25.24$ ($\sim 3.1 L_{\rm H}^*$).   The similarity in range and mean  
{\it H}-band luminosities for different host type with previous mean values versus QSO luminosity 
could be explained, for example, if there were a clear correlation of host type with QSO 
luminosity, and indeed we do see such a correlation, as discussed below. 

\subsection{Host properties vs. QSO luminosity}

A more complete representation of the range of properties of our sample of PGQSOs is given by 
the data shown in both Fig.~\ref{fig:MHhMB_class} and Fig.~\ref{fig:MBvsMH_class}.  The color 
coding and shading of the symbols used to represent the individual objects is the same in 
each of the figure panels, and provides an aid to easily see the mean and range of 
host and nuclear properties for objects with different host type and degree of disturbance. 

\placefigure{fig:MHhMB_class}

Figure \ref{fig:MHhMB_class} (top panel) clearly shows a distinct change in host type with 
increasing $M_{\rm H}$ of the host.  Hosts with dominant disk components  
(i.e. ``D$_{\rm p}$'') are preferentially clustered at the low end 
(i.e., $\sim$sub-$L_{\rm H}^*$ to $L_{\rm H}^*$)
of the observed range, whereas E-type hosts dominate at above-$L_{\rm H}^*$.  Objects 
which have both a relatively large bulge component and a detectable disk (i.e ``B+D") are 
distributed more like E-type hosts, although the relatively small number (3) of ``B+D" hosts limits our ability to say much more, other than that the they cluster near $\sim L_{\rm H}^*$.
 In general, Fig.~\ref{fig:MHhMB_class} 
(top panel), shows that disturbed and strongly disturbed hosts are found at all values of 
host $M_{\rm H}$, although there seems to be a tendency for the strongly disturbed hosts 
to be found at the upper end of our observed range of $M_{\rm H}$.  Finally, we comment on the 
apparent bimodal distribution of host $M_{\rm H}$ in figure \ref{fig:MHhMB_class}, simply by 
noting that with just a relatively small shift in the computed host $M_{\rm H}$ for a few objects, 
this could just as easily be a single distribution with a high luminosity tail.  
 
Figure \ref{fig:MHhMB_class} (bottom panel) also shows a change in the distribution of 
host type with increasing $M_{\rm B}$ of the QSO, similar to what is seen with increasing 
host $M_{\rm H}$ (top panel)---E-type hosts dominate at $M_{\rm B} < -23$ while 
at $M_{\rm B} > -23$ hosts with prominent disks (``D$_{\rm p}$" and ``B+D") appear to 
become an increasing dominant fraction of all objects with decreasing B-band luminosity 
of the QSO. 

Several earlier imaging studies of low-redshift QSOs (i.e., $z < 0.3$) have also commented 
on the fact that lower luminosity QSOs appear to have a  larger fraction of hosts with a 
distinguishable disk component, while high luminosity QSOs seem to reside primarily 
in elliptical hosts.  The collective series of papers by McLeod and coworkers 
\citep{mcl94a,mcl94b,mcl95,mcl01} used both ground-based and {\it HST} near-infrared imaging data of a PG sample 
that is most similar to ours to conclude that  high luminosity QSOs seem to predominantly 
live in elliptical hosts while the host types of low-luminosity QSOs are more likely to show 
evidence for disks.   Independently, Dunlop and coworkers \citep{tay96,mclu99,dun03}, 
using a combination of ground and space-based, near-infrared and optical imaging data 
for a mostly different sample of QSOs, reached generally similar conclusions when they 
emphasized the almost exclusive presence of elliptical hosts for the higher luminosity 
QSO in their sample, while also pointing out the increasing number of disks found in the 
hosts of the lower-luminosity, radio-quiet QSO in their sample.  And recently, \citet{ham02} 
provided a comprehensive summary of their {\it HST} archival study of 71 QSO host galaxies 
with WFPC2 images in F606W or redder filters, which concluded that ``elliptical hosts are 
typically twice as luminous as spiral hosts", although they also noted that the luminosity distributions 
of elliptical and spiral hosts for the radio-quiet QSOs were ``generally compatible".   

\placefigure{fig:MBvsMH_class}

A more comprehensive presentation of our data is given in Fig.~\ref{fig:MBvsMH_class}, 
which shows the relationship of the computed 2-D fit value of $M_{\rm H}$  (column 5 of Table 
\ref{tab:2Dfitresult})  
and the observed $M_{\rm B}$ (column  5 of Table \ref{tab:sourcelist}) for each QSO, along with 
the symbol for host type,  the measured redshift and an indicator for radio loudness (``RLQ"). 
In addition to showing the trends of host type versus  $M_{\rm H}$ and  $M_{\rm B}$ discussed above, 
Fig.~\ref{fig:MBvsMH_class} also shows how the ratio of these two quantities (which was summarized 
by host type in cols. 9-11 of  Table \ref{tab:hostclassstatnir}) varies over the sample. 

One interesting question is whether there is any evidence for a correlation between 
QSO luminosity and the luminosity of the host galaxy.  The results shown in 
Fig.~\ref{fig:MBvsMH_class} suggest not, at least for hosts with 
$-23.5 > M_{\rm H}({\rm host)} > -25.5$ (i.e. $\sim${\ts}0.6 - 4 $L_{\rm H}^*$(host));   
if anything, the mean ratio $M_{\rm B} / M_{\rm H}({\rm host})$ versus $M_{\rm H}$(host), 
in this range of $M_{\rm H}$(host), seems to {\it decrease} with increasing $M_{\rm H}$(host), 
but this result is not significant because of small number statistics.   However, we also note 
that that all of our objects with 
$M_{\rm H}{\rm (host)} > -25.5$ are at $z < 0.15$, and the stated $B$-band limit ($B \sim 16.16$) of the 
BQS \citep{sch83} implies that we should still be complete out to this redshift for QSOs with 
$M_{\rm B} < -21.3$.   For hosts with $M_{\rm H}{\rm (host)} < -25.5$,  Fig.~\ref{fig:MBvsMH_class} 
does seem to imply that they produce more luminous QSOs. However, this result is clearly 
compromised by incompleteness due to the $B$-band limit of the BQS; 
all of the sources with $M_{\rm H}{\rm (host)} < -25.5$ lie 
at redshift $z > 0.13$, which excludes detection of QSOs with $M_{\rm B} > -23$. 

As was pointed out by the referee, what does seem to be significant from 
Fig.~\ref{fig:MBvsMH_class} is that even the most luminous galaxies with disks do not 
seem to harbor really luminous QSOs. It would seem that it may be necessary to
have a merger violent enough to completely destroy the disks of the progenitor galaxies 
to produce the most luminous QSOs.   

\subsection{Radio-Loud vs. Radio Quiet hosts}

The number of radio-loud quasars (RLQs) in our sample is relatively small (5/28 = 18\%), but 
it is representative of the fraction found in previous studies of large samples of QSOs, 
which show that RLQs are only 
$\sim$15\% of the total population of optically-selected QSOs. 
Although we did not specifically distinguish RLQs and RQQs when computing 
mean host $H$-band luminosities for different ``host types" in Table \ref{tab:hostclassstatnir},  
the RLQ hosts are explicitly identified in Fig.~ \ref{fig:hosttype}--\ref{fig:MBvsMH_class}.   
It is immediately obvious that RLQs live in the most luminous QSOs  (i.e., $M_{\rm B} < -24$) with 
the most luminous hosts (i.e., $M_{\rm H}({\rm host}) \lesssim -26$ corresponding to 
$\gtrsim 6${\ts}$L_{\rm H}^*$).  These facts, coupled with our previously discussed trends of host type with 
QSO $B$-band and host $H$-band luminosity, are consistent with previous published results which 
claim that RLQs are typically 1-2 mags more luminous than RQQs in the optical, and that RLQs are 
exclusively found in elliptical hosts, which in turn are usually 
much more luminous---typically 1-2 magnitudes, depending on whether the observations 
are in the optical or near-infrared, respectively---than the mean values found 
for the hosts of RQQs (e.g., \citealt{hut84,smi86,mclu99,ham02,dun03}).  

Although the origin of the observed dichotomy in the mean properties of RLQ and RQQ hosts 
is somewhat beyond the scope of our current study, it is nevertheless worth restating the 
obvious -- that  QSO surveys biased toward radio-selected objects will necessarily find 
more luminous,  elliptical hosts when compared with pure optically selected samples of QSOs.

\subsection{Degree of disturbance of host galaxy}

One of the expected strengths of AO observations of QSOs using 8m telescopes was the ability to 
better characterize the ``host type", and in addition to detect faint extended structure in the host 
which may have escaped detection in previous imaging observations with smaller telescopes.  
Although the majority of our host classifications are in general agreement with previous 
results (see \S\ref{sec:notes}), our images were more likely to allow us to distinguish 
asymmetric features (e.g. putative tidal debris, central bars, etc.) from, for example, more symmetric 
spiral structure, and to better separate out compact, circumnuclear structures from the bright 
QSO nucleus. We note that since we rely on morphological features such as tidal arms to classify hosts as ``not disturbed'', ``disturbed'' or ``strongly disturbed'', this classification is relatively insensitive to usually very symmetrical PSF subtraction residuals. For example, the host type of PG 1626+554 and PG 2251+113 are unknown due to poor PSF subtraction, but the smooth host images obtained lead us to classify them as ``not disturbed''.

According to our analysis, 30\% (9/30) of the classified QSO hosts were categorized as being ``disturbed" or ``strongly disturbed" in Table  \ref{tab:hostclass}.
The 6 hosts classified as ``strongly disturbed" show asymmetric features reminiscent of tidal structure, as might be due, for example, to the advanced merger of two spiral disks. Putative ``tidal arms" are clearly seen in PG0157+001, PG 1411+442, PG 1613+658, PG 1700+518 and PG 2349$-$014, while PG 1426+215 has a clearly asymmetric disk.  The three additional objects listed as ``disturbed" generally show fainter asymmetric features, (plausibly faint tidal debris) whose effect on the 1-D radial profile is small enough to still allow us to distinguish between different host types.  The 4 ``disturbed" hosts include two ``D$_{\rm p}$'' and one ``E''.

In addition to the 9 hosts classified as either ``disturbed" or ``strongly disturbed", there are 
3 hosts that were found to have a prominent well-defined bar, whose size in each case was 
$\sim${\ts}5{\ts}kpc radius. In two of these cases the bar actually dominated the integrated luminosity 
of the host.  The hosts of two of these barred galaxies (PG0804+761, PG1001+054), were 
classified as having a dominant disk profile  (``D$_{\rm p}$"), while the third (PG1229+204), was 
classified as ``B+D". It is unclear whether these bars have any relation to the processes 
that produced the putative tidal features in the ``disturbed" and ``strongly disturbed" hosts,  
but if they do, then the percentage of such hosts rises to 40\% (12/30). 

Our finding of 30\% for the fraction of   ``strongly disturbed" + ``disturbed" hosts (40\% if ``well-defined bars" 
are included) is similar to the range of values, typically 25\%-50\%, quoted for ``morphologically peculiar", and/or 
``interacting" hosts in previous surveys of 
low-z QSOs (e.g., \citealt{hut84,smi86,mcl94b,mar01,sur01}).  Although it is not easy to determine if 
the definitions used to define degree of disturbance in previous surveys are the same as used here, 
a comparison of those objects from our current survey which have been included in previous surveys 
suggests that we are in fact using similar criteria.  

The breakdown by ``host type" of ``disturbed", ``strongly disturbed" and ``well defined bar" systems 
is given in Fig.~\ref{fig:hosttype}.  Only one such object (PG 2214+139) with obvious ``concentric 
shells" is classified as an elliptical.  The remainder of the ``disturbed" and ``well defined bar" systems 
have obvious disks, as presumably do the ``strongly disturbed" objects (see above).  Taken together, 
11 of these 17 systems (65\%) have features which are interpreted as being indicative of obviously disturbed,  
non-axisymmetric structure.   Such structure has in the past been seen as evidence for interaction/mergers
as a possible trigger for producing the QSO.  

The breakdown by $M_{\rm H}$(host) and $M_{\rm B}$(total) of ``disturbed", ``strongly disturbed" and 
``well defined bar" systems is given in Fig.~\ref{fig:MHhMB_class}.  In sum, there appears to be no 
clear evidence for a correlation with either $M_{\rm H}$(host) or $M_{\rm B}$(total) when these
three classes are considered together.  However, if we consider only the ``strongly disturbed" 
systems, there {\it is} a clear preference for  these to be found in the most luminous 
(i.e., $\gtrsim 2 L_{\rm H}^*$) hosts.

\subsubsection{Close compact companions or super-star-cluster knots}

In addition to asymmetric structures, we also found a small, bright knot in the inner disks of 
9 hosts  (see notes in  Table \ref{tab:hostclass}).  These knots were typically 
within a few kpc (projected distance) of the QSO.  They are being discussed here since we 
believe that the evidence points to the knots being associated directly with the host galaxy,
as opposed, for example, to being projected foreground or background objects at much 
larger distances from the QSO.   

The terms ``companion" or  ``close companion"  have been widely used in the literature to 
describe a range of distance scales around QSOs where {\it galaxies} at the same redshift as 
the QSO are found.  For example comparative studies of the environment around QSOs 
typically find that, at low redshifts, QSOs live in regions of moderate galaxy overdensity, 
typical of compact or small groups (e.g., \citealt{yee84,hec84}).  Closer in, typically 
within a projected nuclear separation  of 
$\lesssim 30-50${\ts}kpc  between the companion and QSO nuclei, are numerous reports of close companion
galaxies, clearly distinguishable as a separate galaxy, but often with reported evidence of 
a tidal interaction with the QSO host (e.g., \citealt{sto82,hec84,hut84,bah97,boy99,mar01}).  
At the closest 
distances (typically nuclear separations $\lesssim 10${\ts}kpc), where the putative 
close companion galaxy strongly overlaps the QSO host, as is the case in most of the 
strongly disturbed systems, including our own, the close companion is 
then plausibly the second object involved in an advanced merger with the QSO host.  This 
last case is similar to what is meant here by the term ``close companion", except in 
our case we also clearly distinguish a relatively bright ``compact knot" which could either 
be the nucleus of the companion galaxy, {\it or} possibly a super star-cluster, perhaps 
triggered by the merger.  

We have referred to these objects (the ones within 10kpc projected separation) as ``close compact companions" even though 
redshift is available for only one of the 9 ``knots", since it is likely that 
these ``knots" are indeed at the same redshift as the QSO rather than simply being 
unrelated foreground or background objects.  
Five of the 9  QSO hosts with ``close compact companions"  are the same 5 hosts classified as 
being ``strongly disturbed", and one of the remaining four is classified as being ``disturbed".  
Several of the companions  are also observed to be strongly elongated (e.g., in PG 1119+120, 
PG 1411+441, PG 1426+215, PG 2349$-$014).  These facts strongly suggest that the 
knots are directly related to a merger event. 

However we remain somewhat agnostic 
whether these objects represent the second nucleus of a nearly merged galaxy pair, 
or, instead, a super star cluster perhaps triggered by the merger.  
Although, the ``knots"  appear to be too bright to be ``normal" star-formation knots, and are 
therefore good candidates to be distinct galaxy nuclei, they are in fact similar 
in brightness to the putative ``super star cluster" knots that have often been found 
in the circumnuclear regions of ``infrared-excess QSOs" (e.g., \citealt{sur01})
and that appear to be nearly ubiquitous in the inner disks of ``warm ultraluminous 
infrared galaxies" (e.g. \citealt{sur99}).  All but one of the ``knots" seen in our current QSO sample 
have measured $M_{\rm H}$ in the range $-20.1$ to $-22.3$, and in each case this translates into 
an absolute $H$-band luminosity in the range 3.5--5.5{\ts}mag fainter than the 
host galaxy, and the most luminous knot ($M_{\rm H} = -23.18$), is still 3.3{\ts}mag fainter 
than its host  (PG1302$-$102), and is still within the upper range of luminosities observed 
for the star forming knots in ``warm ULIGs".

\subsection{Mid/Far-infrared properties of QSO hosts}

In addition to trying to better categorize the host types for our sample of QSOs, we were also 
interested in investigating their mid- and far-infrared emission properties, and in particular 
to see if a relationship exists between the infrared luminosity of the QSO and morphology of the host.

Interest in the infrared properties of QSOs has been fueled by suggestions of an evolutionary 
connection between QSOs and ultraluminous infrared galaxies (ULIRGs), where both types of objects 
represent different phases in the end stage of the merger of two relatively equal mass gas-rich 
spirals (see review by \citealt{san96}).  Key to such a connection has been the suggestion  
that those QSOs with the strongest mid/far-``infrared excess" emission seem to have ``disturbed" 
hosts (e.g., \citealt{san96,boy96,can00,cle00,haa00,sur01}), whose properties are somewhat similar 
to the hosts of  ``warm" ULIRGs.   For the QSOs,  ``infrared-excess"  has been defined 
using the mean radio-to-Xray spectral energy distributions (SEDs) of 
QSOs \citep{san89,elv94}.  The data have been interpreted as showing that the 
SED is dominated by two thermal emission ``bumps".  In addition to the well-known ``blue bump" (BB)
of optical/UV emission at wavelengths $\sim{\ts}0.01 - 1.0${\ts}$\mu$m, 
thought to be associated with thermal emission from an accretion disk, there also exists an 
 ``infrared/submillimeter bump" at wavelengths $\sim{\ts}1.0 - 500${\ts}$\mu$m, which is 
 interpreted as being primarily due to dust re-radiation of emission from either the accretion disk 
 or from embedded circumnuclear star formation.  The ``infrared excess" is then defined to be 
 the ratio of luminosities in the two ``bumps",  $L_{\rm IR} / L_{\rm BB}$.  

\placetable{tab:rad_prop}

The recent availability of reprocessed ISO observations of QSOs \citep{haa00,pol00,haa03}, plus the release of 
the {\it IRAS} Faint Source Catalog \citep{mos92}, have now made it possible 
to more accurately estimate the strength as well as the color temperature of the 
``infrared/submillimeter bump'' for the BQS QSOs. In particular, we have 
been able for the first time to compile fluxes in four wavebands (12$\mu$m, 25$\mu$m, 60$\mu$m and 
100$\mu$m)  for the majority (18/32) of our BQS sources, and in 2-3 of these wavebands 
for an additional 10 sources.  Three sources have only one flux measurement, but with sensitive  
upper limits in the remaining 3 wavebands.  Only one source has just upper limits in 
all four wavebands.  These measurements are listed here in Table \ref{tab:rad_prop} where they 
have been used to compute an infrared luminosity, $L_{\rm IR} \equiv L (8-1000)\mu$m, and 
the luminosity ratio, $L_{\rm IR} / L_{BB}$, for each source following the definitions 
and prescriptions found in Table 1 of \citet{san96}.  

\placefigure{fig:MBvsMH_class}

\placetable{tab:hostclassstat}

Using the data in Table \ref{tab:rad_prop}, we compare the infrared/optical ratio, 
$L_{\rm IR}/L_{BB}$, of the QSO hosts versus host type in Fig.~\ref{fig:LirLbb_class}, and 
tabulate the mean and median values of $L_{\rm IR}$ and median $L_{\rm IR}/L_{BB}$, 
for different host type and degree of disturbance in Table \ref{tab:hostclassstat}.  
Figure \ref{fig:LirLbb_class} suggests a fairly strong correlation 
between $L_{\rm IR} / L_{\rm BB}$ and host type,  with nearly all ellipticals (``E") 
having the lowest observed values of $L_{\rm IR}/L_{BB}$.  Table \ref{tab:hostclassstat} 
shows that Es have median values of $L_{\rm IR}/L_{BB}$ slightly less than half that of 
hosts with detectable disks (``D$_{\rm p}$" + ``B+D").  The previously observed correlation between 
non-disturbed hosts and E host type then implies (as shown in Table \ref{tab:hostclassstat}) 
that the median $L_{\rm IR}/L_{BB}$  ratio for ``disturbed" and ``strongly disturbed" hosts is 
also approximately twice the median ratio for ``non-disturbed" hosts.

A strong correlation between IR luminosity and host morphology had previously been suggested
by \citet{cle00}, although a close comparison between his ``quiet/disturbed" classification 
and the images presented here shows that the images he used lacked the angular 
resolution and/or depth to reliably classify the host galaxies: for example, PG1411+442 and 
PG1700+518, two of the most disturbed objects in our sample, were classified as ``quiet'', while 
PG0923+201 and PG0844+349 were classified as disturbed.

\placefigure{fig:IRcolcol}

Another method of characterizing the mid/far-infrared emission of QSOs has been to use the 
mid/far-infrared color temperature (e.g., \citealt{lip94,can01}).  The color-color plot shown in 
Fig.~\ref{fig:IRcolcol}  adopts the same definitions used in \citet{can01}, who first pointed out 
that nearly all (8/9) of the ``transition QSO" they studied---i.e., objects with an ``intermediate position" 
in the far-infrared color-color diagram, in-between the empirical location of optically-selected QSOs 
and far-infrared-selected  ULIRGs---were undergoing strong tidal interactions indicative of major 
mergers.    Our new results are similar, in that all 8 of our ``disturbed" and ``strongly disturbed" QSOs 
for which we could reliably compute far-infrared colors, fall in the ``transition region" between the two 
limits marked by the lines denoting ``power-law" and ``black-body" emission.   Additionally, we find only 
objects we classified as ``non-disturbed" above the ``power-law" continuum line.  We interpret this as 
additional evidence for a positive correlation between fading evidence of tidal structure (i.e., increasing 
age of the merger), and decreasing $L_{\rm IR} / L_{\rm BB}$, since the latter is correlated with 
increasing far-infrared color temperature. 

\section{Conclusions}
\label{sec:conclusions}

We have used AO imaging on the 8.1m Gemini-North and 8.2m Subaru 
telescopes to acquire near-infrared (mostly $H$-band) images of an unbiased 
sample of 32 nearby ($z < 0.3$) optically selected QSOs from the PGBQS in order 
to investigate the luminosity and morphology of the host galaxies, and to compare the 
properties of the host with properties of the QSO.   The  $B$-band 
absolute magnitudes of our QSO targets span a factor of $\sim${\ts}23, 
from $\sim{\ts}M_{\rm B} = -22.12$ (the minimum threshold for ``bona-fide QSOs" according to the definition 
given by \citealt{sch83}) to $M_{\rm B} = -25.53$.  Three sources have absolute luminosities 
slightly below the minimum threshold and are traditionally classified as ``Seyfert{\ts}1 nuclei".  
Given that this threshold is somewhat arbitrary, we have included these three 
objects in the current study.  Following the definitions of \citet{kel94}, twenty-seven 
of our sources were radio-quiet and 5 were radio-loud.

Our imaging campaign in general was quite successful, given that we were able to 
detect and characterize the host type for 28 of our sources.  Our basic results can be 
summarized as follows:

\begin{enumerate}
\item  The mean $H$-band magnitude (and corresponding $L_{\rm H}^*$) 
of the host galaxies for our sample of QSOs was 
$M_{\rm H}({\rm host}) = -24.82$ ($\sim{\ts}2.1{\ts}L_{\rm H}^*$), 
with a range of $-23.5$ to $-26.5$ ($\sim{\ts}0.63 - 10.0{\ts}L_{\rm H}^*$).

\item  For the sample of 28 classified hosts, the distribution of ``best-fit" host types was 10 (36\%) ellipticals (``E"), 3 (11\%) bulge+disk (``B+D"), 8 (29\%) disk dominated (``D$_{\rm p}$"), and 7 (25\%) of indeterminate type.

\item There appears to be a strong correlation between host type and the $H$-band absolute 
magnitude of the host.  Sub-$L_{\rm H}^*$ hosts all have a dominant (``D$_{\rm p}$") 
or strong disk component (``B+D");  $1-2{\ts}L_{\rm H}^*$ hosts appear to be equally divided 
between ellipticals (``E") and disks (``D$_{\rm p}$" or ``B+D"); $>{\ts}2{\ts}L_{\rm H}^*$ hosts 
are mostly ellipticals (``E").   A similar, but somewhat weaker trend is found for the $B$-band
absolute magnitude of the QSO.

\item Thirty percent (9/30) of our detected hosts were classified as either being ``disturbed" (3) 
or ``strongly disturbed" (6).  An additional 3 objects were found to have strong well defined bars, bringing 
the total fraction of objects with obvious disturbed and/or non-axisymmetric structure up 
to 40\% (12/30).  In total, these objects are relatively equally distributed over all values of 
$M_{\rm H}({\rm host})$, and $M_{\rm B}$(total), although in detail, the  ``strongly disturbed" 
objects are much more likely to be found among the most luminous hosts (i.e. $\gtrsim 2 L_{\rm H}^*$), 
with ``disturbed" and ``well defined bars" more likely to be associated with less luminous hosts 
(i.e. $\lesssim 2 L_{\rm H}^*$). 

\item For the subsample of 18 QSOs with redshifts $z < 0.13$, which is the redshift completeness 
limit corresponding to the $m_{\rm B_{\rm lim}} \sim 16.2$ limit of the PGBQS \citep{sch83}, we find 
{\it no} obvious correlation between the absolute magnitude of the QSO,  $M_{\rm B}$(total), 
and the absolute magnitude of the host galaxy, $M_{\rm H}({\rm host})$ .   
However at $z > 0.15$, all 8 sources in our sample have the largest host luminosities, 
$M_{\rm H}({\rm host}) < -25.5$ ( $\gtrsim{\ts}4{\ts}L_{\rm H}^*$),  {\it and} the largest QSO luminosities, 
$M_{\rm B}({\rm total}) < -23.8$, suggesting that  the most luminous QSOs are found among the most 
luminous hosts.   

\item  The mean properties of the hosts of radio-loud quasars (RLQs) are clearly different from 
the mean properties of the hosts of radio-quiet QSOs (RQQs).  All 5 of the RLQs are among the 
10 sources in our sample at $z > 0.17$.   All 5 RLQs are among the 10 most luminous QSOs 
(i.e. $M_{\rm B} < -24$), and for the 4 RLQs where we were able to reliable measure the host, all 
4 are among the 5 most luminous hosts  (i.e. $\gtrsim{\ts}6{\ts}L_{\rm H}^*$).  
Three of the 5 RLQ hosts are classified as ellipticals, one host 
suffers from a bad PSF but otherwise shows no visible sign of disk structure, and one is strongly disturbed 
and cannot be classified.  These results are in keeping with the paradigm which suggest that RLQs 
are preferentially found in giant elliptical hosts with the most luminous QSOs.

\item We find a strong correlation between the infrared excess, $L_{\rm IR} / L_{\rm BB}$, of the 
QSO and its host type and host degree of disturbance.  This ratio is twice as large, on average, 
in hosts with strong 
disk components as in elliptical hosts, and likewise is twice as large, on average, in ``disturbed"
+ ``strongly disturbed" hosts as in ``non-disturbed" hosts.  We also find that ``disturbed" and 
``strongly disturbed" hosts have mid/far-infrared colors that place them in a region of the 
far-infrared color-color 
plane in between the power-law that is characteristic of the majority of QSOs,  and the black-body that is 
characteristic of the majority of ULIRGs.  

\end{enumerate}
 
Our results concerning the mean and range of host luminosities, the general breakdown of host types,  
the fraction of objects with disturbed features, 
and the differences in the mean properties of the hosts of 
RLQs versus RQQs, are in general agreement with previous studies of the host properties of low-z QSOs. 
Our mean and range of $H$-band host luminosities are almost identical to the values reported earlier 
by \citet{mcl94b} and \citet{sur00}, respectively, in their studies of a similar sample of PGQSOs selected 
from the BQS \citep{sch83}.  One advantage of our AO observations with 8m-class telescopes is that we 
have perhaps been able to give more emphasis to trying to accurately characterize the host properties 
of individual sources, and to search for systematic trends in host type versus other properties of the QSO. 

Perhaps the most interesting new results from our survey are those concerning the detailed distribution of 
disturbed hosts versus the luminosity of the host and the luminosity of the QSO, and the apparent 
correlation of infrared excess with host type.  Our results suggest that the percentage of disturbed 
hosts is 
significant (15-30{\ts}\%) across the range of observed host luminosities ($\sim 0.6-10{\ts}L_{\rm H}^*$), 
even though the main host type clearly changes from hosts with prominent disks to hosts 
which appear to be luminous ellipticals.  Whether disturbed hosts represent a particular phase is the 
lifetime of virtually all QSOs or whether they represent an inherently separate class of QSOs is 
somewhat unclear. The answer to this question depends on the answers to two subsidiary questions:
(1) What is the time scale for QSO activity relative to that for the persistence of recognizable
signs of disturbance due to a strong interaction or merger? and (2) What is the true space density
of QSOs with bolometric luminosities above some specific value and the fraction of these that
show some level of disturbance indicating a recent merger?  
Our study has some bearing on this second question.  However, it now seems likely that UV-selected
optical QSO samples, such as the one studied here, may represent considerably less than half of the 
total QSO population at a comparable bolometric luminosity. 
In addition, recent work by \citet{jes05} comparing QSOs found in the Sloan Digital Sky 
Survey with the PG sample
shows that the PG sample is seriously incomplete even in terms of its own specified selection criteria.
This incompleteness suggests two
imperatives for future work. First, it seems essential to develop a sample that is substantially
complete to a specific bolometric QSO luminosity over some large volume. This effort will necessarily
require complementary techniques covering wavelength regimes at least from X-rays to the far IR.
Second, it will be necessary to carry out imaging studies similar to those we have done for the
PGBQS on representative subsamples of QSOs identified by these other techniques, for it is quite likely
that, for example, the fraction showing signs of mergers may vary for samples selected by different
criteria. A recent step in this direction has been taken by \citet{hut03} for QSOs with $z<0.3$ 
found in the 2MASS survey, for which they find an extraordinarily high fraction ($>70$\%) show
disturbances indicating tidal interactions.

 \acknowledgments
We thank the referee for a careful reading of the paper and useful comments to help improve both
its content and its presentation.  We also thank Gabriela Canalizo for a helpful conversation.
The Hokupa`a AO observations were supported by members of the University of Hawaii AO group: 
P. Baudoz, O. Guyon and D. Potter. Support for Hokupa`a came from the National Science Foundation
under grant no.~AST 96-18852.
Our results were based in part on observations obtained at the Gemini Observatory, which is operated by the
Association of Universities for Research in Astronomy, Inc., under a cooperative agreement
with the NSF on behalf of the Gemini partnership: the National Science Foundation (United
States), the Particle Physics and Astronomy Research Council (United Kingdom), the
National Research Council (Canada), CONICYT (Chile), the Australian Research Council
(Australia), CNPq (Brazil) and CONICET (Argentina).  The results were also based in part on data 
collected at Subaru Telescope, which is operated by the National Astronomical Observatory of Japan.

\clearpage

\clearpage

\clearpage

\clearpage

\begin{figure}
\epsscale{1.0}
\plotone{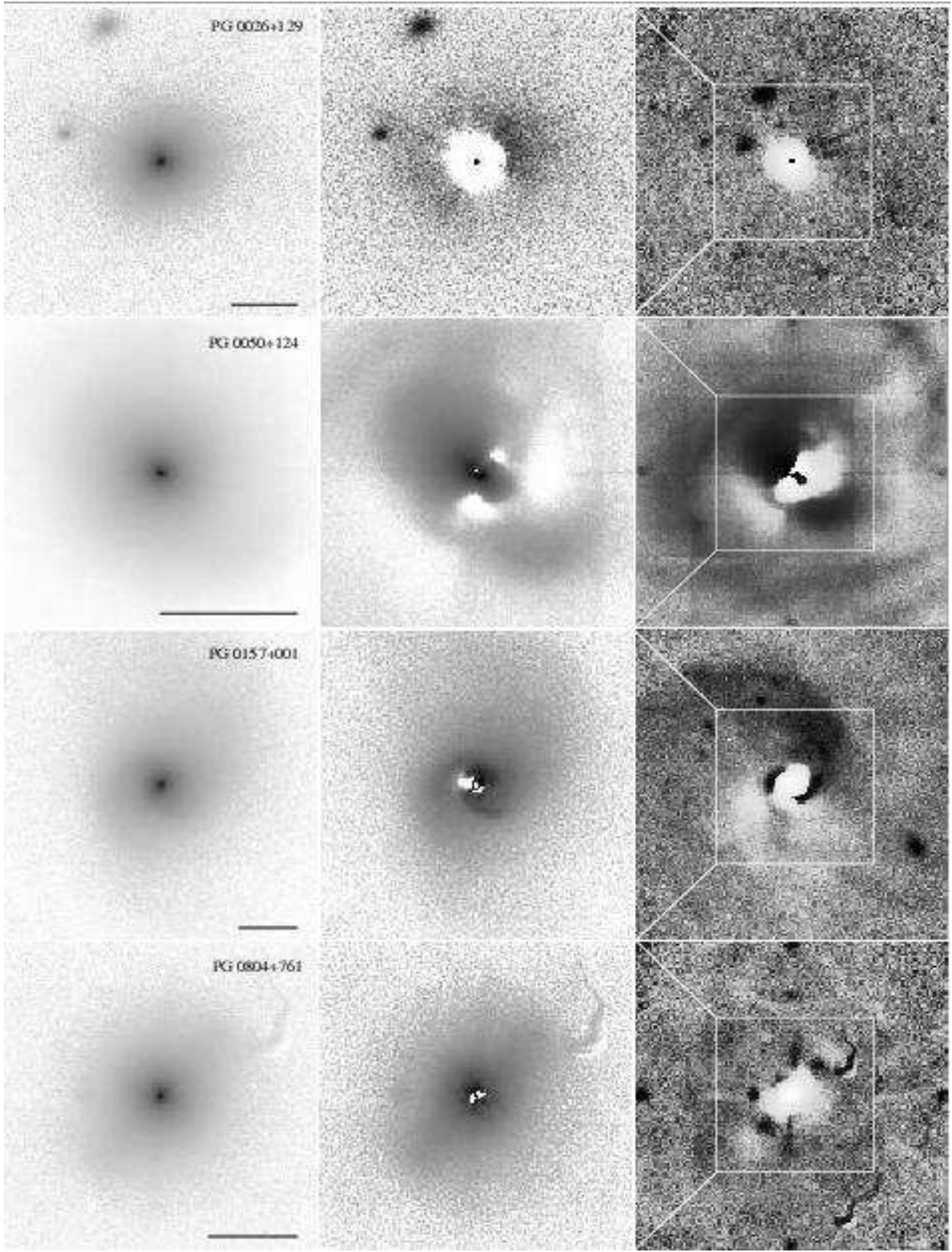}
\caption{\label{fig:qsoimages} Raw (left), PSF-subtracted (center) and PSF+host model subtracted (right) 
images of the QSOs. The images are $10\arcsec\times10\arcsec$ (left and center) and 
$20\arcsec\times 20\arcsec$ (right). The black bar in the left images is 5 kpc long. North is up, East is left on all images.}
\end{figure}
\clearpage

\begin{figure}
\epsscale{1.0}
\plotone{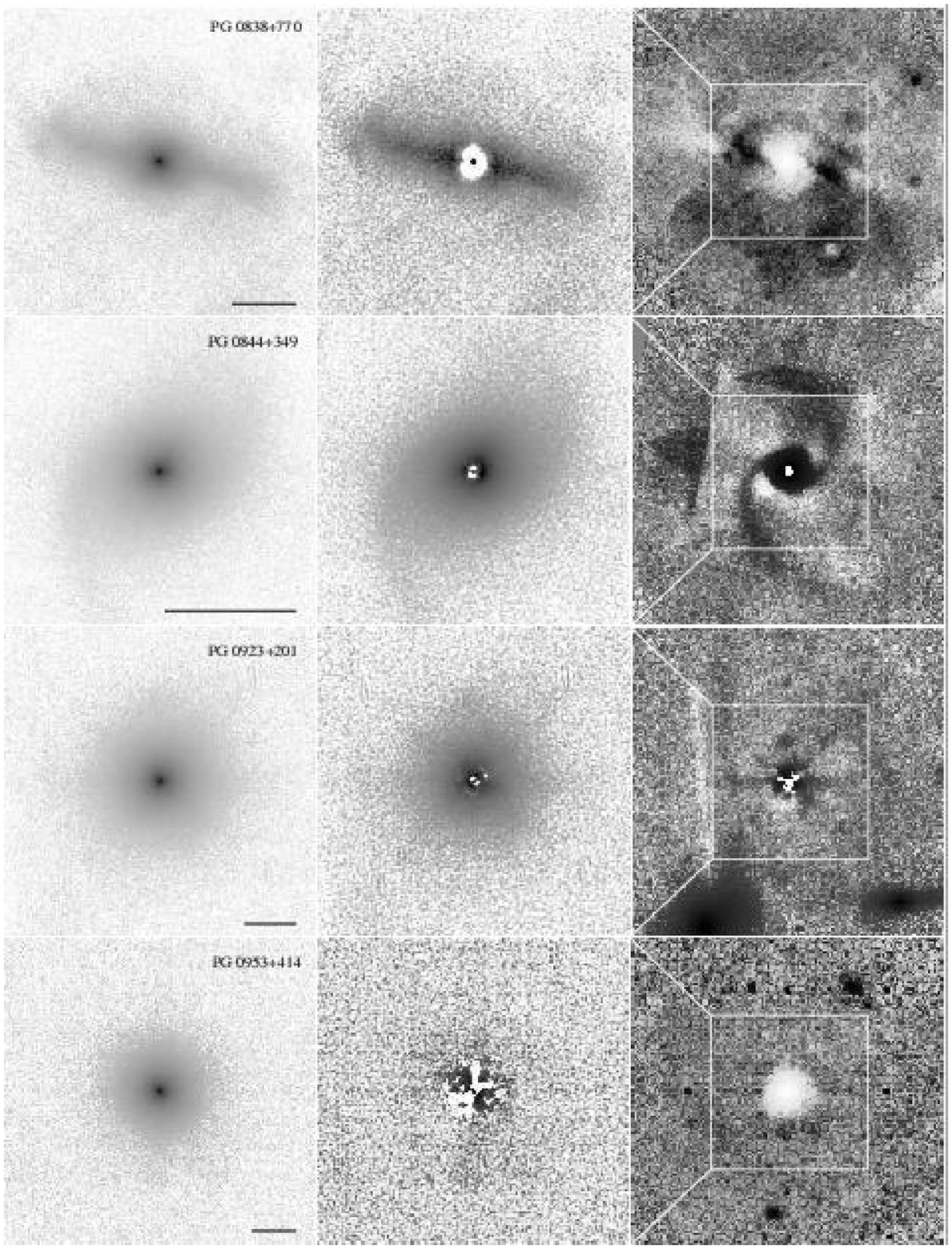}
\centerline{Fig. 4. --- Continued.}
\end{figure}
\clearpage

\begin{figure}
\epsscale{1.0}
\plotone{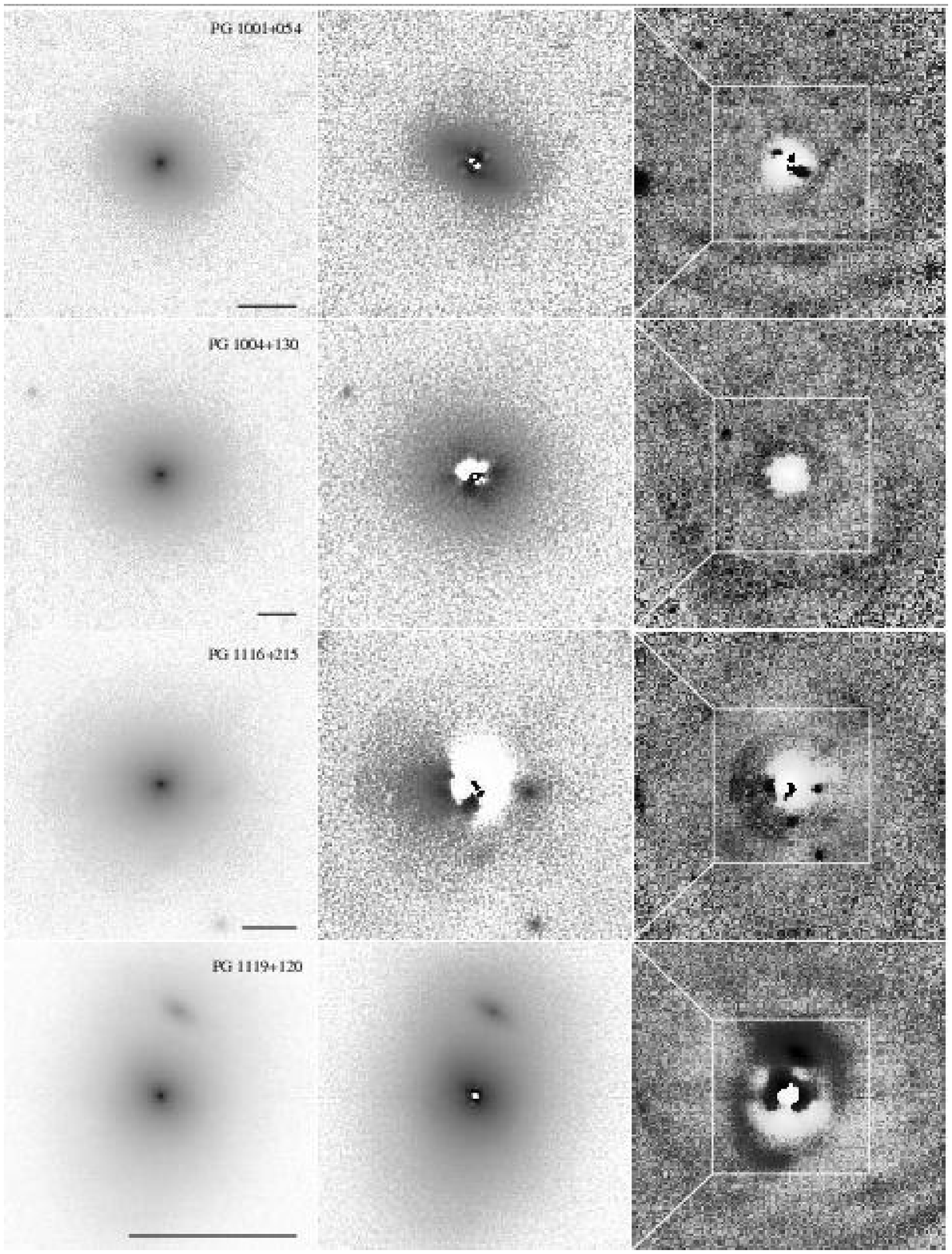}
\centerline{Fig. 4. --- Continued.}
\end{figure}
\clearpage

\begin{figure}
\epsscale{1.0}
\plotone{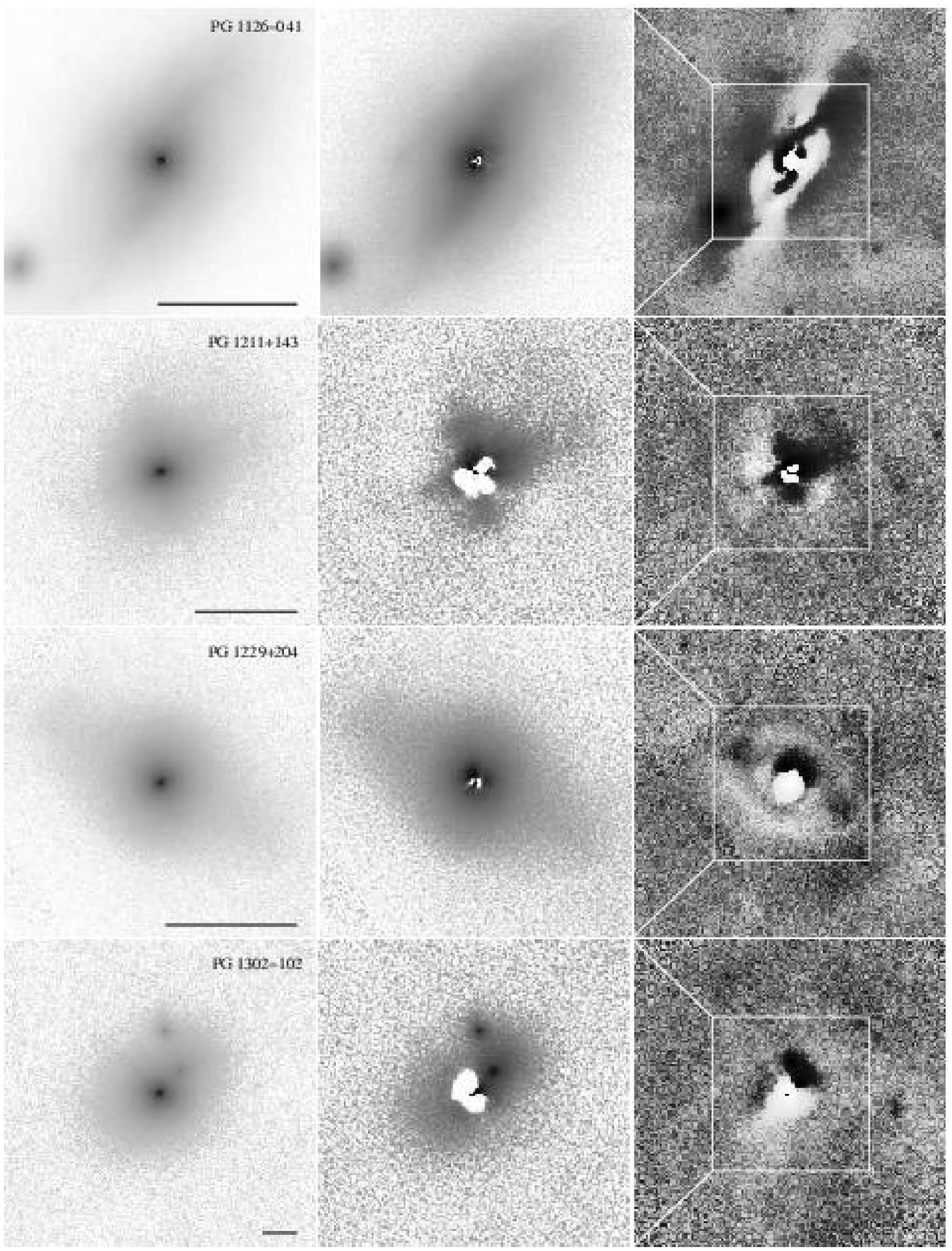}
\centerline{Fig. 4. --- Continued.}
\end{figure}
\clearpage

\begin{figure}
\epsscale{1.0}
\plotone{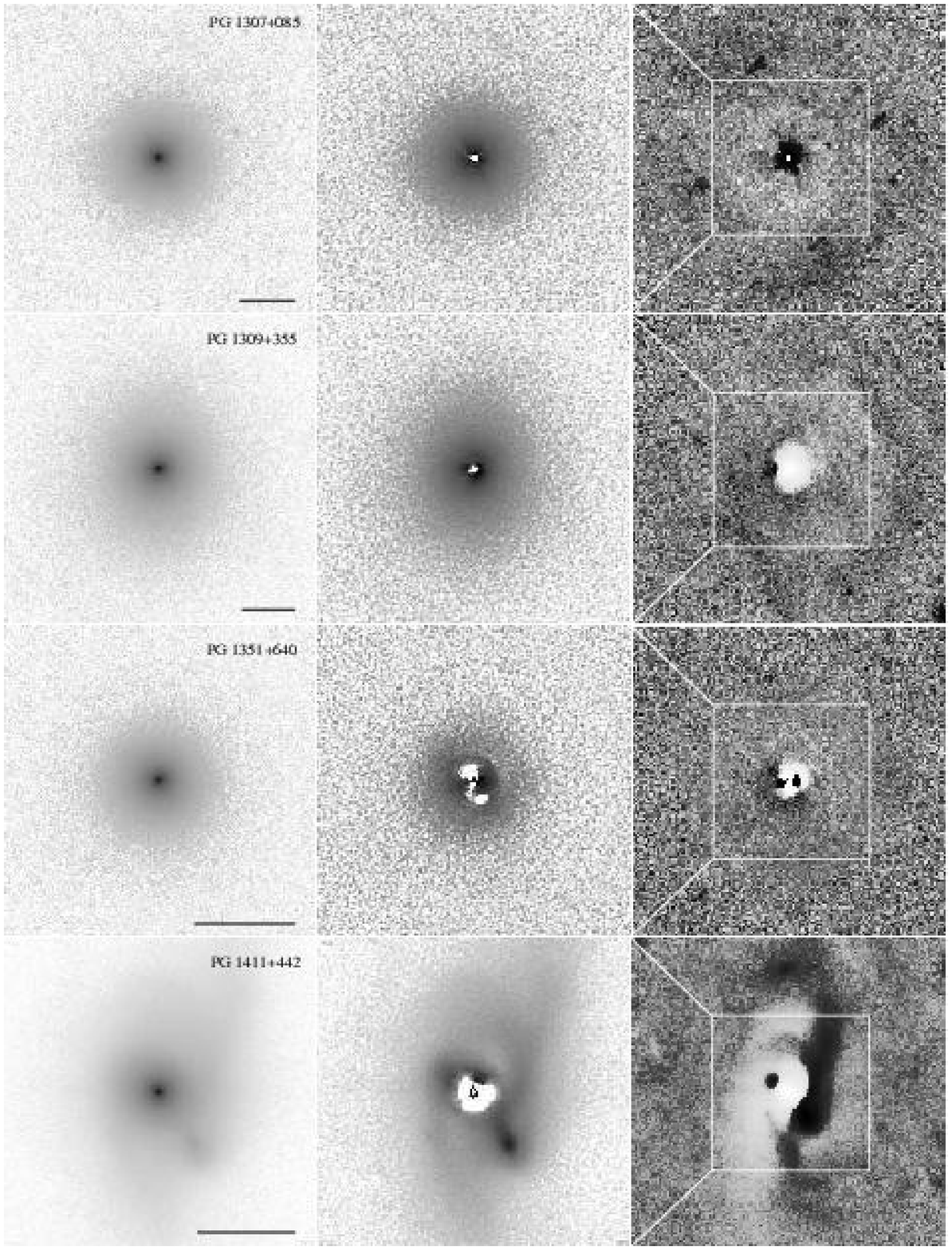}
\centerline{Fig. 4. --- Continued.}
\end{figure}
\clearpage

\begin{figure}
\epsscale{1.0}
\plotone{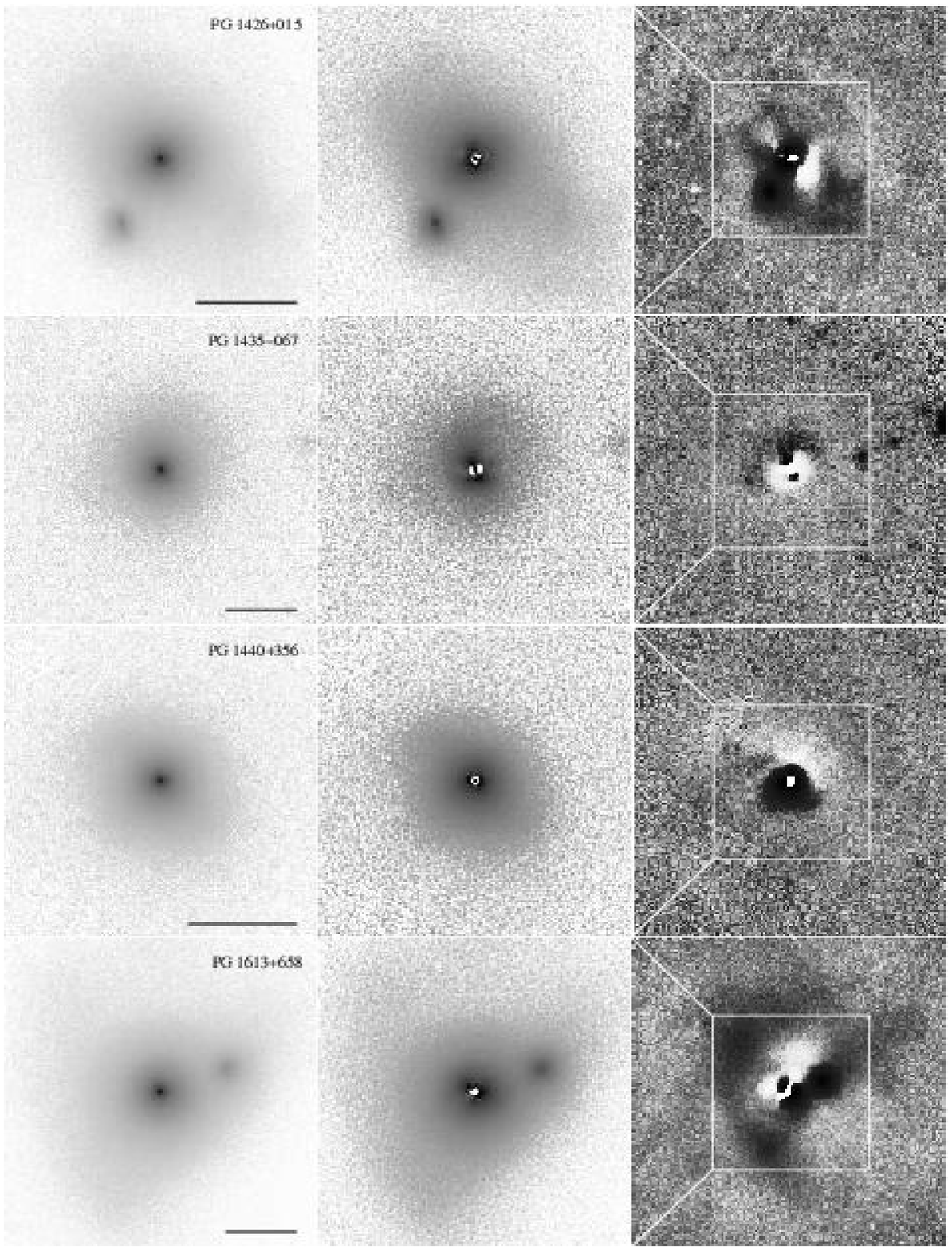}
\centerline{Fig. 4. --- Continued.}
\end{figure}
\clearpage

\begin{figure}
\epsscale{1.0}
\plotone{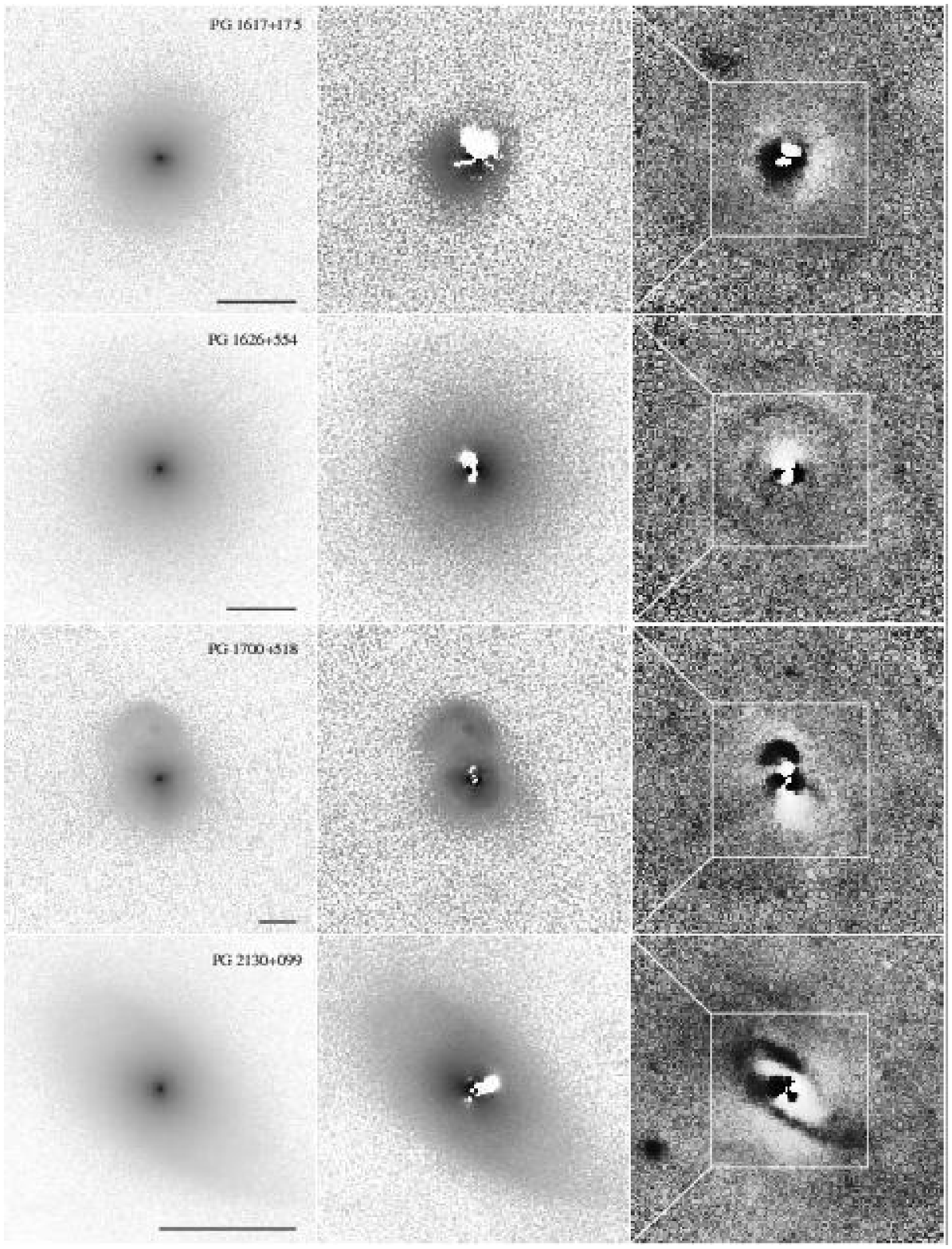}
\centerline{Fig. 4. --- Continued.}
\end{figure}
\clearpage

\begin{figure}
\epsscale{1.0}
\plotone{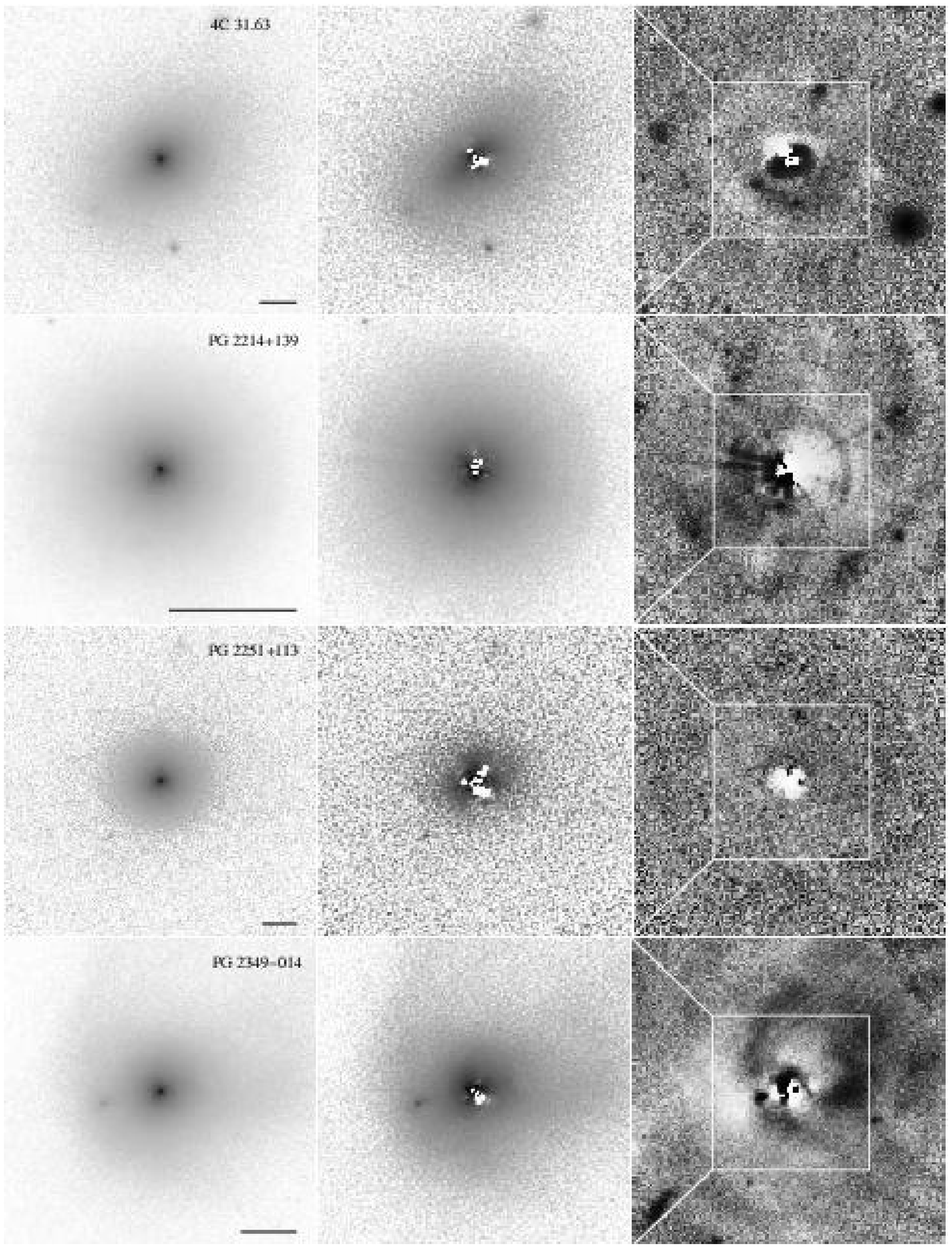}
\centerline{Fig. 4. --- Continued.}
\end{figure}
\clearpage

\begin{figure}
\epsscale{1.0}
\plotone{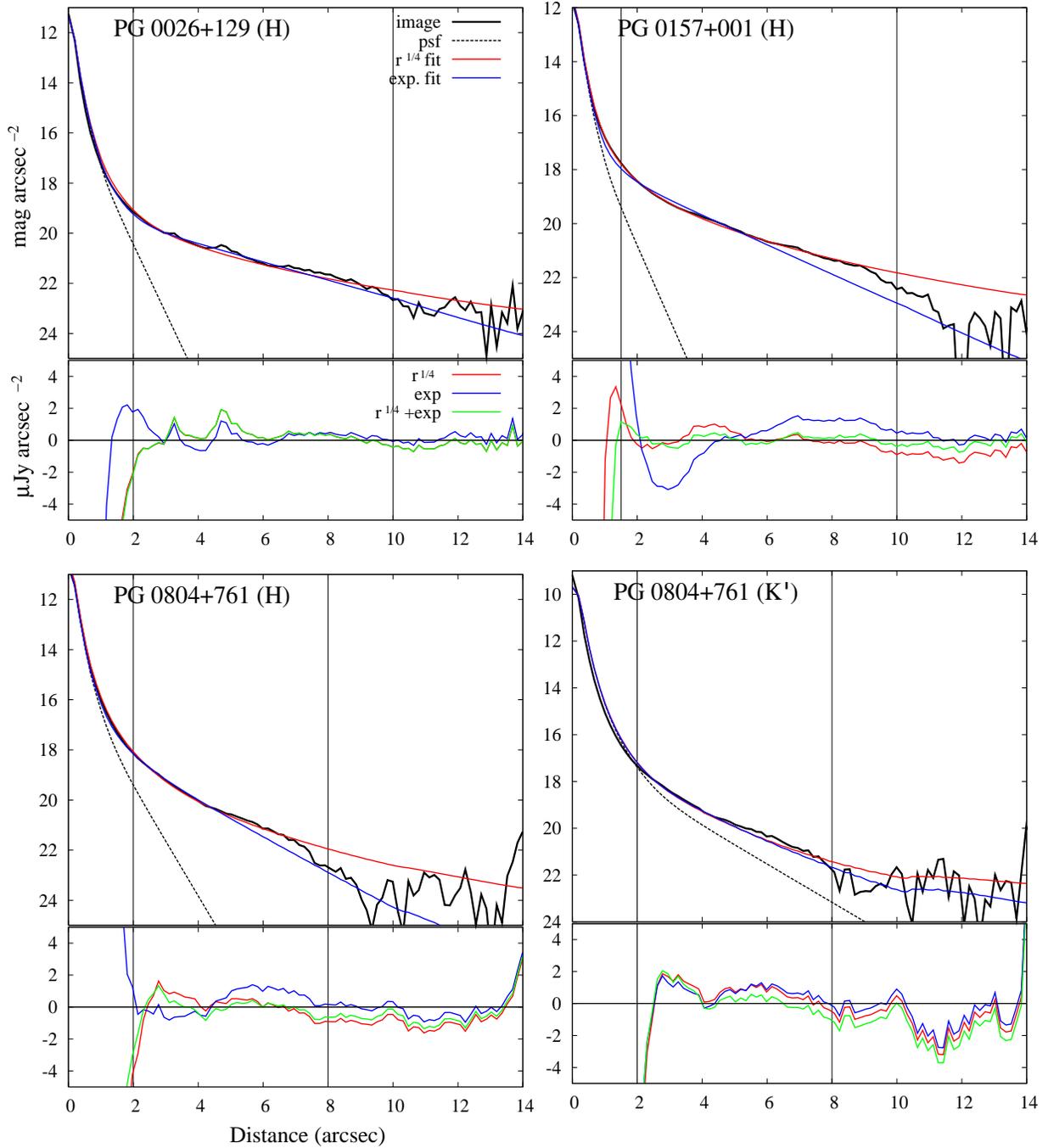}
\caption{\label{fig:qsofits1} 1-D radial profiles of the 2-D raw images and reference PSF (top), and residual errors 
of the $r^{1/4}$ and exponential 2-D fits.}
\end{figure}
\clearpage

\begin{figure}
\epsscale{1.0}
\plotone{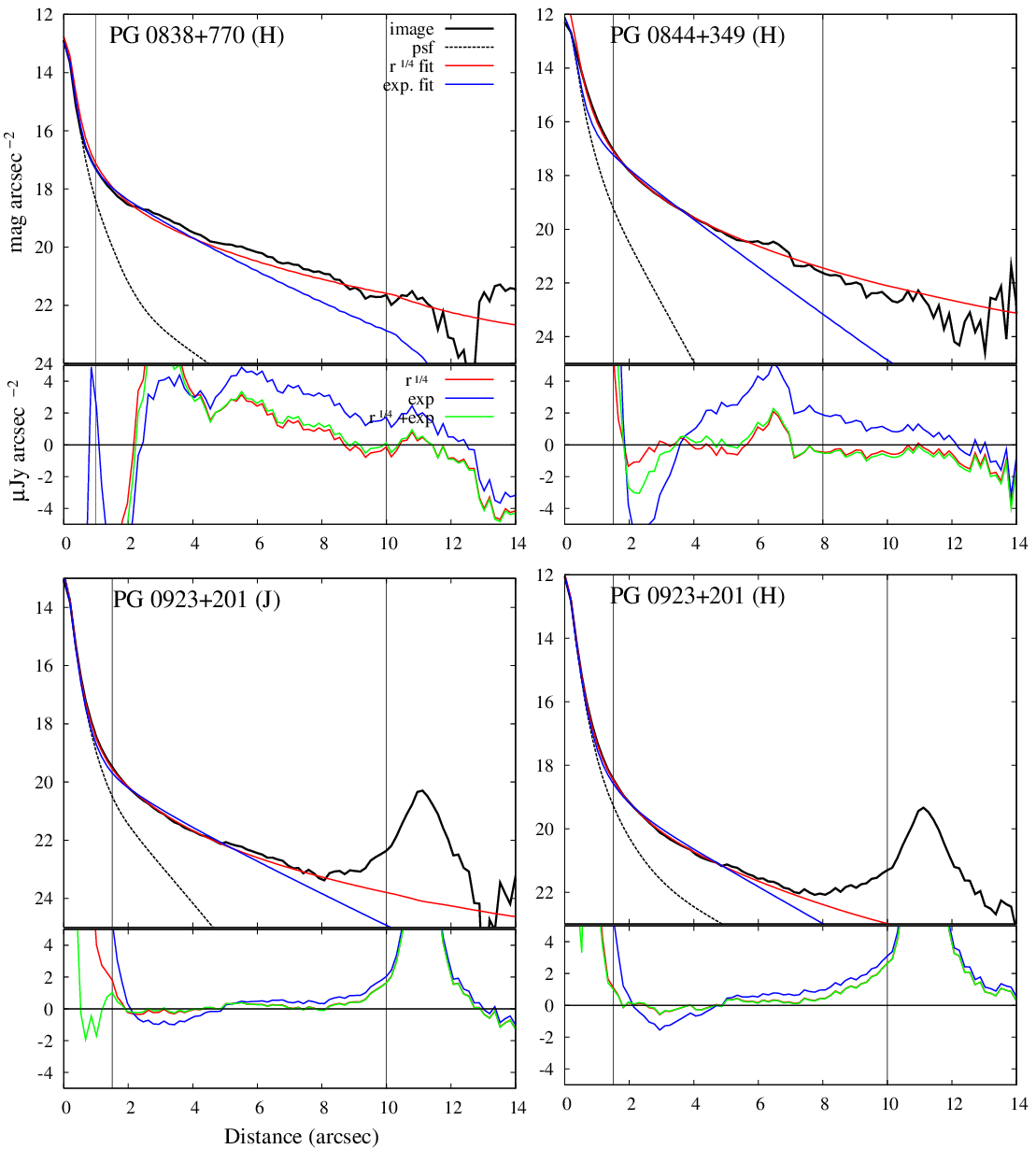}
\centerline{Fig. 5. --- Continued.}
\end{figure}
\clearpage

\begin{figure}
\epsscale{1.0}
\plotone{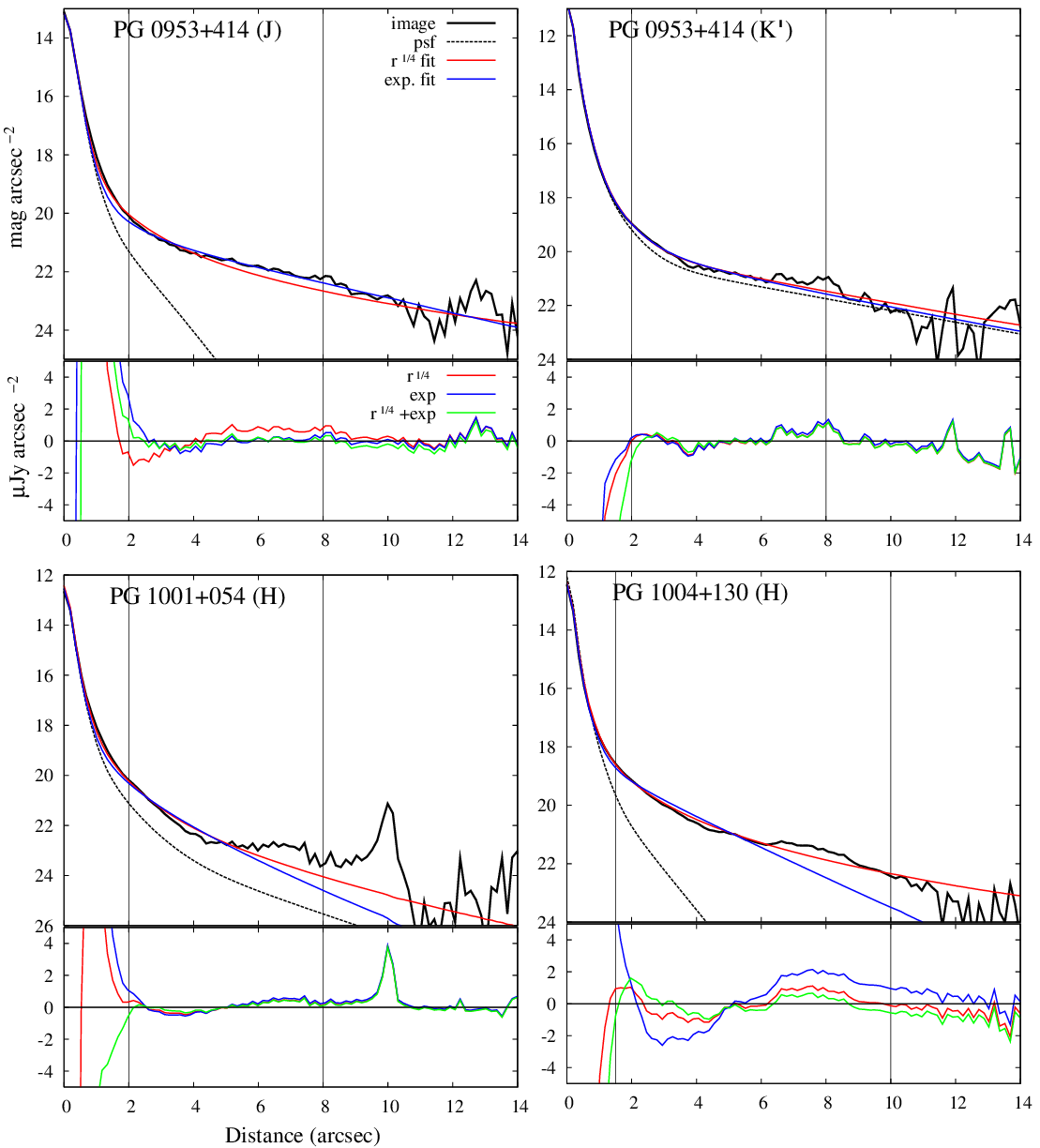}
\centerline{Fig. 5. --- Continued.}
\end{figure}
\clearpage

\begin{figure}
\epsscale{1.0}
\plotone{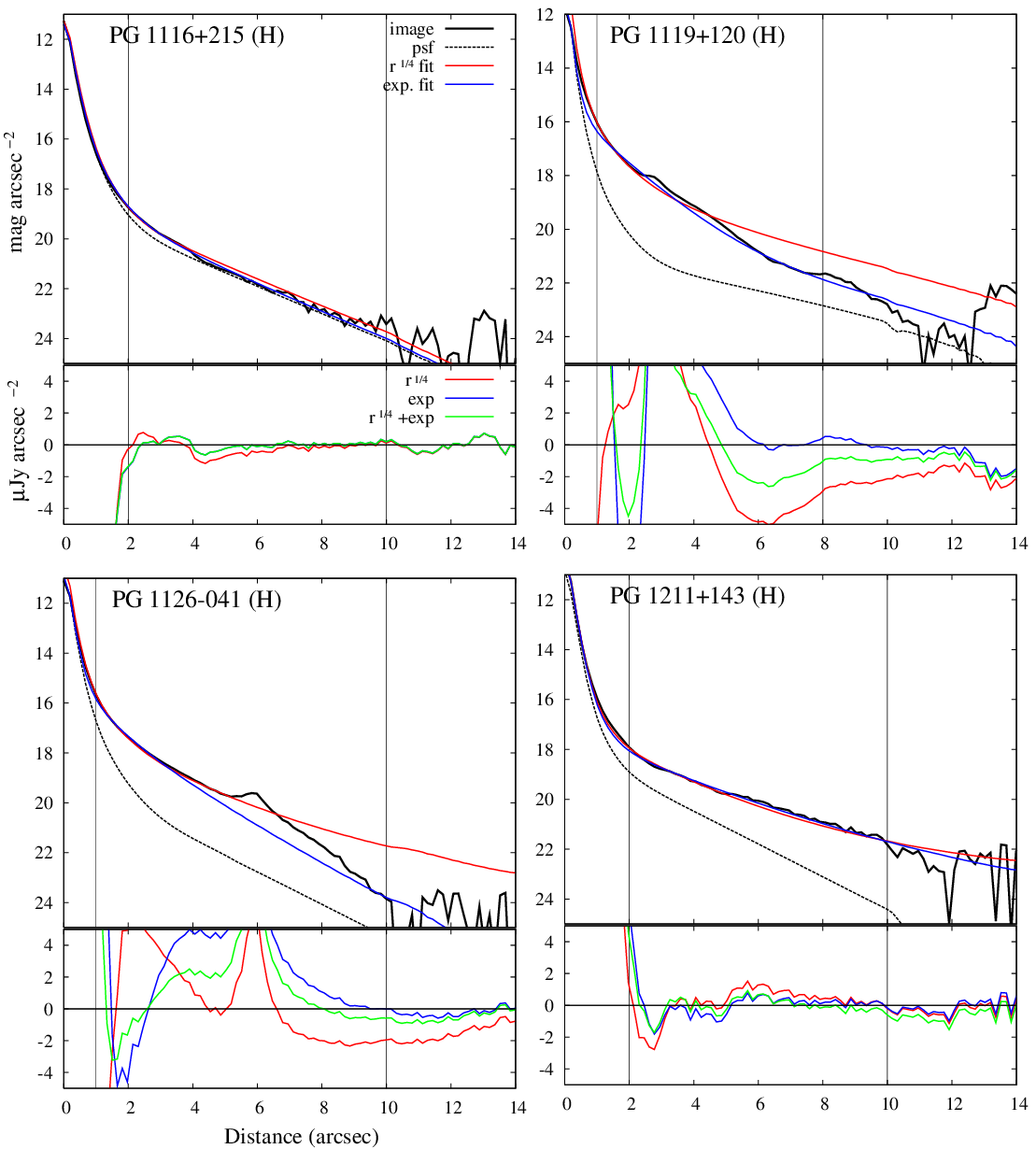}
\centerline{Fig. 5. --- Continued.}
\end{figure}
\clearpage

\begin{figure}
\epsscale{1.0}
\plotone{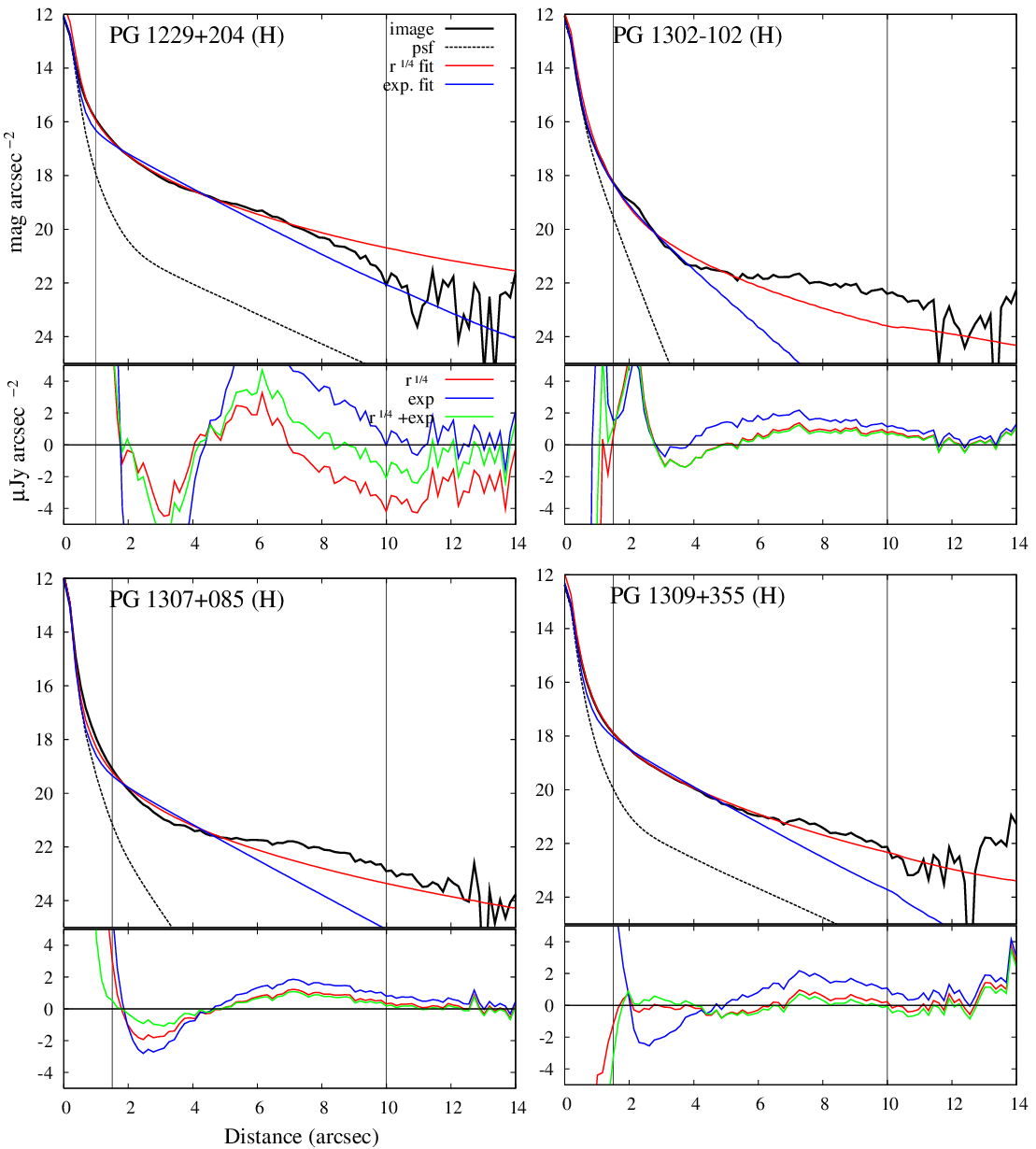}
\centerline{Fig. 5. --- Continued.}
\end{figure}
\clearpage

\begin{figure}
\epsscale{1.0}
\plotone{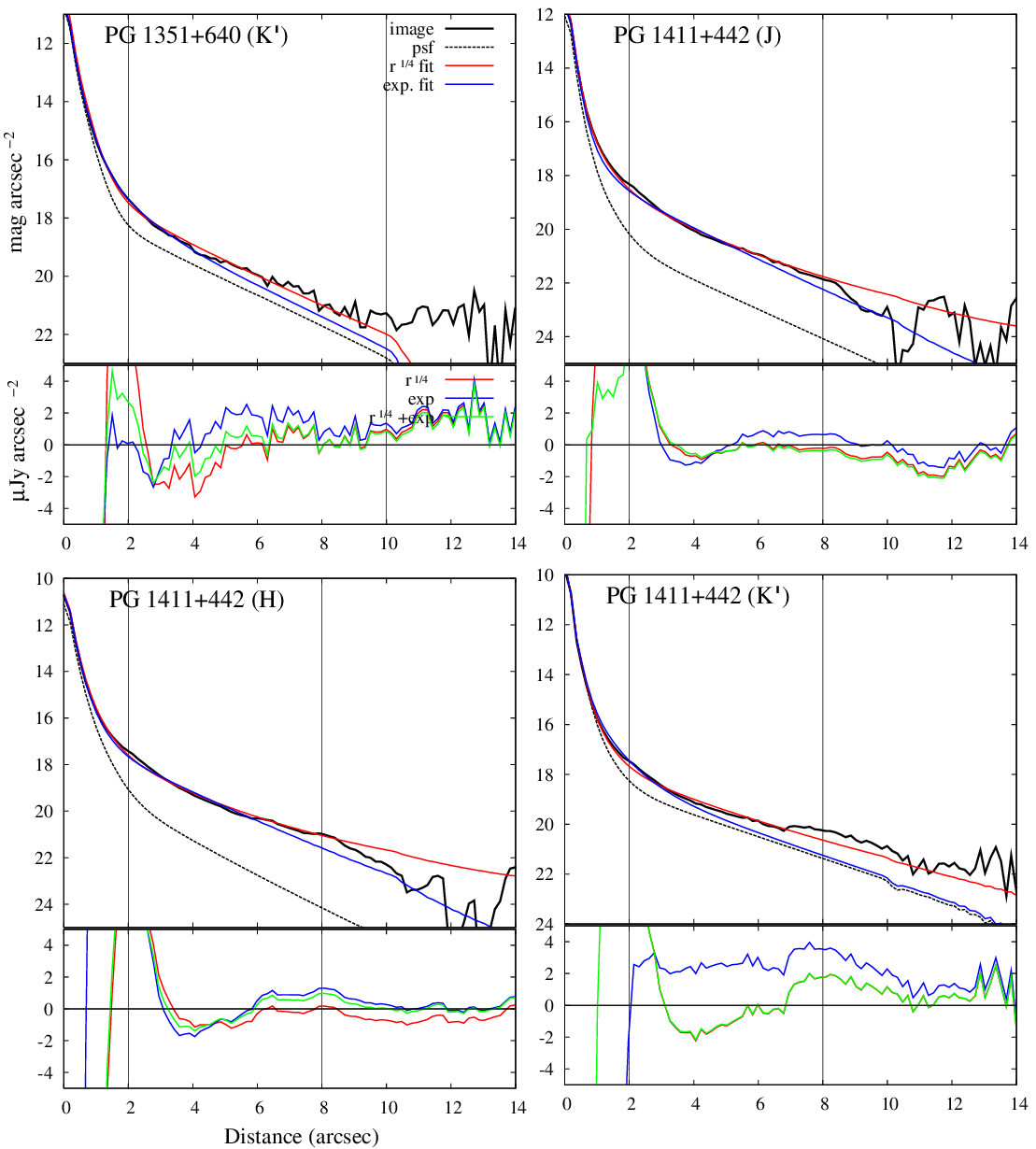}
\centerline{Fig. 5. --- Continued.}
\end{figure}
\clearpage

\begin{figure}
\epsscale{1.0}
\plotone{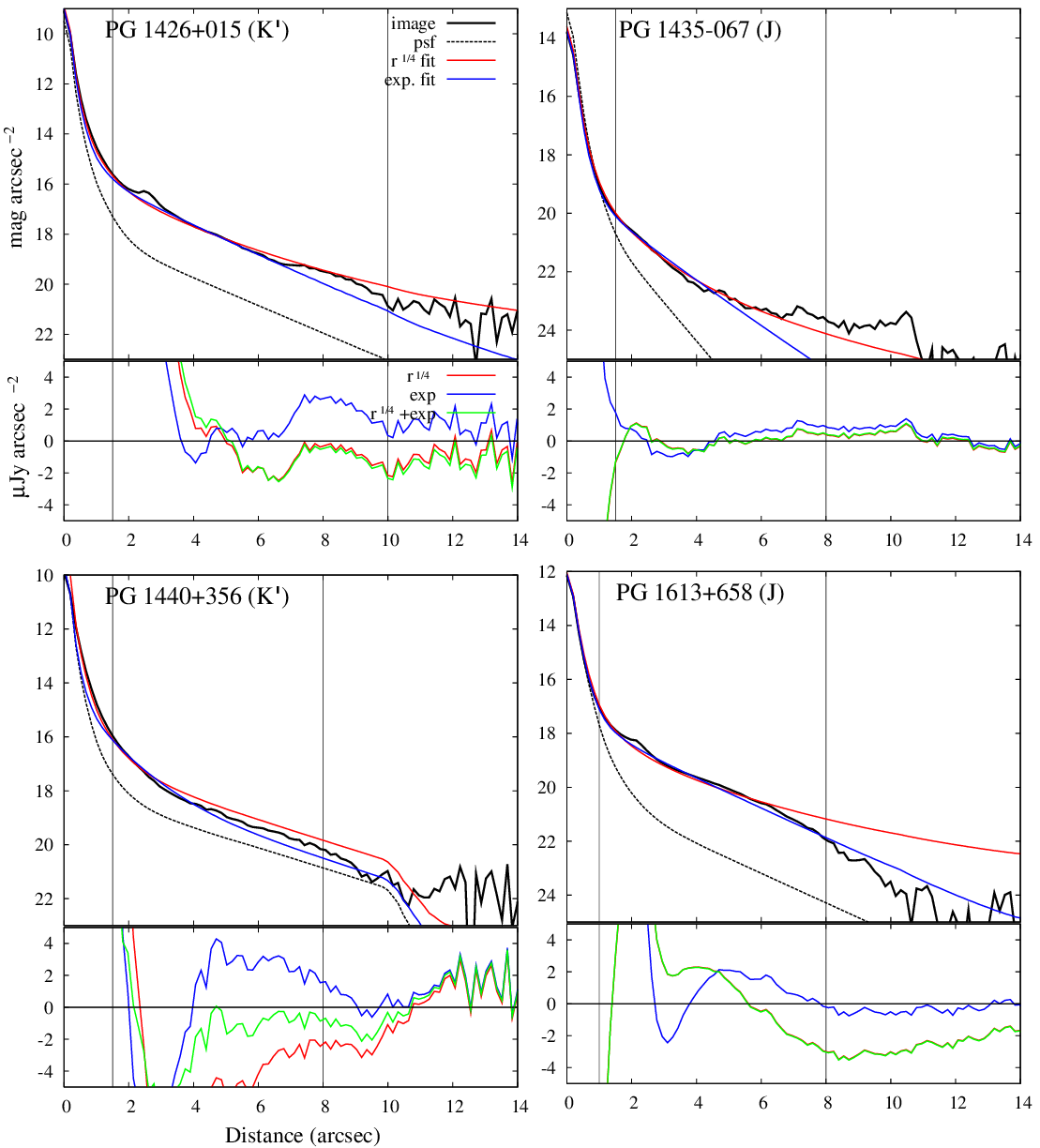}
\centerline{Fig. 5. --- Continued.}
\end{figure}
\clearpage

\begin{figure}
\epsscale{1.0}
\plotone{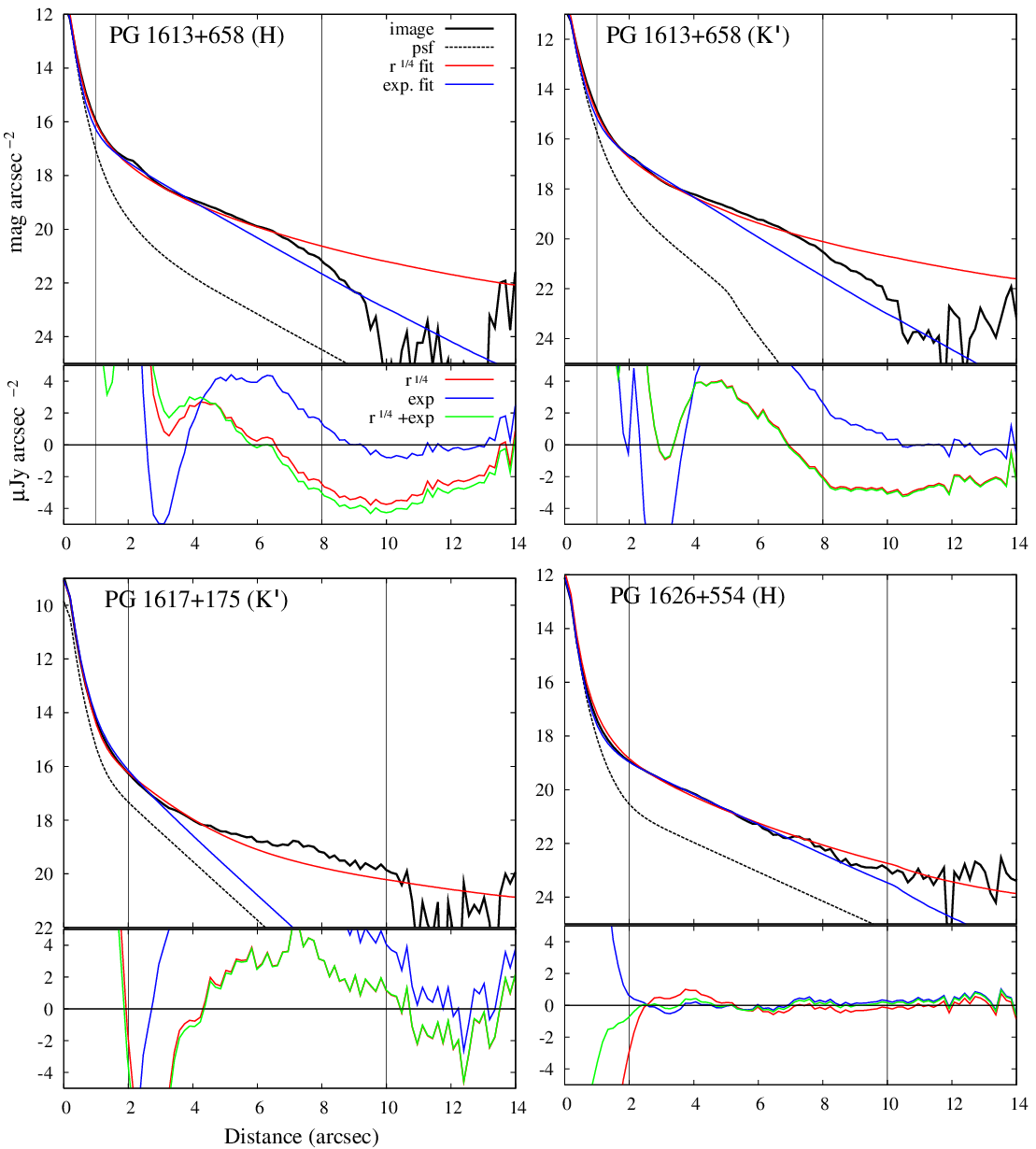}
\centerline{Fig. 5. --- Continued.}
\end{figure}
\clearpage

\begin{figure}
\epsscale{1.0}
\plotone{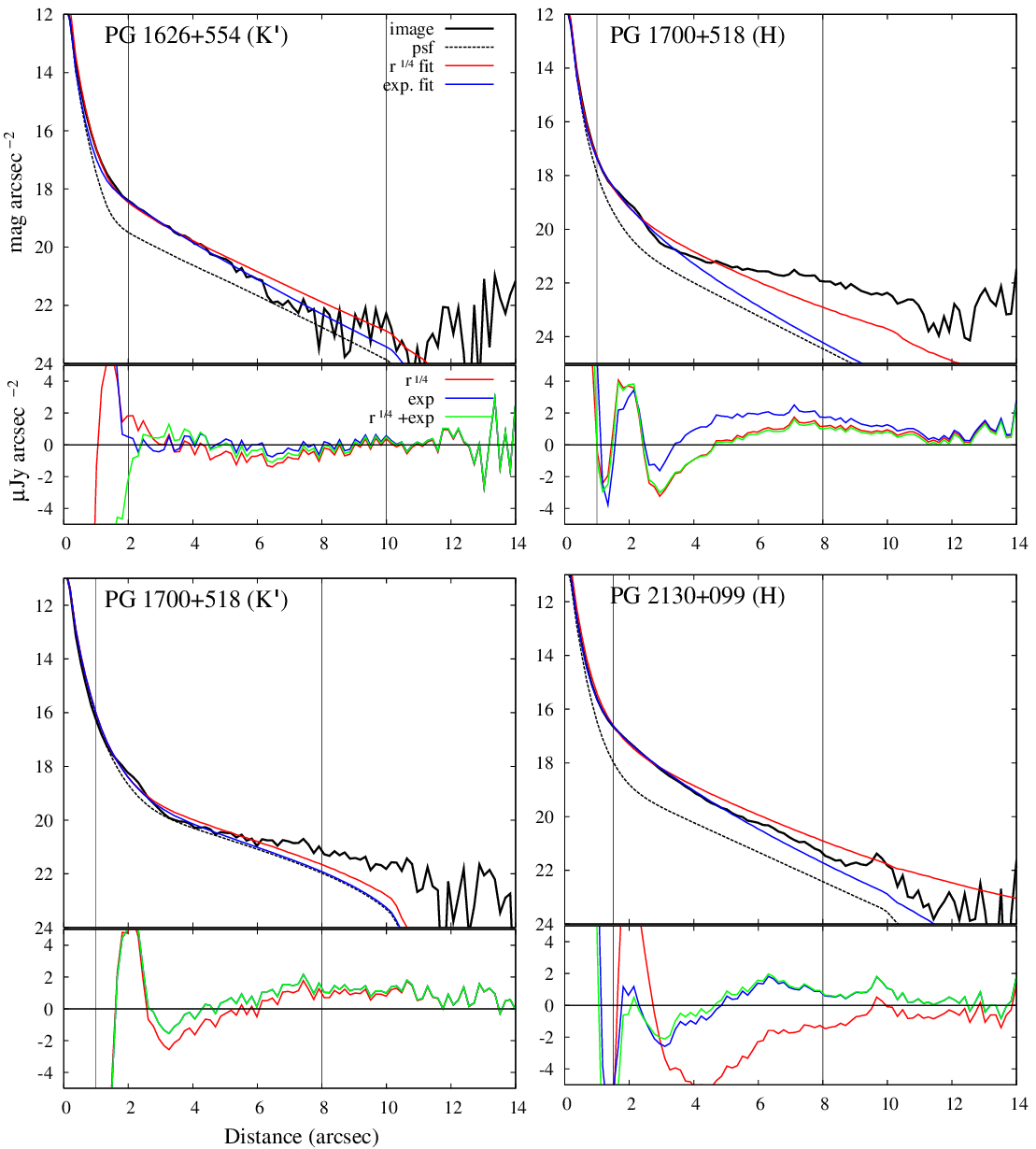}
\centerline{Fig. 5. --- Continued.}
\end{figure}
\clearpage

\begin{figure}
\epsscale{1.0}
\plotone{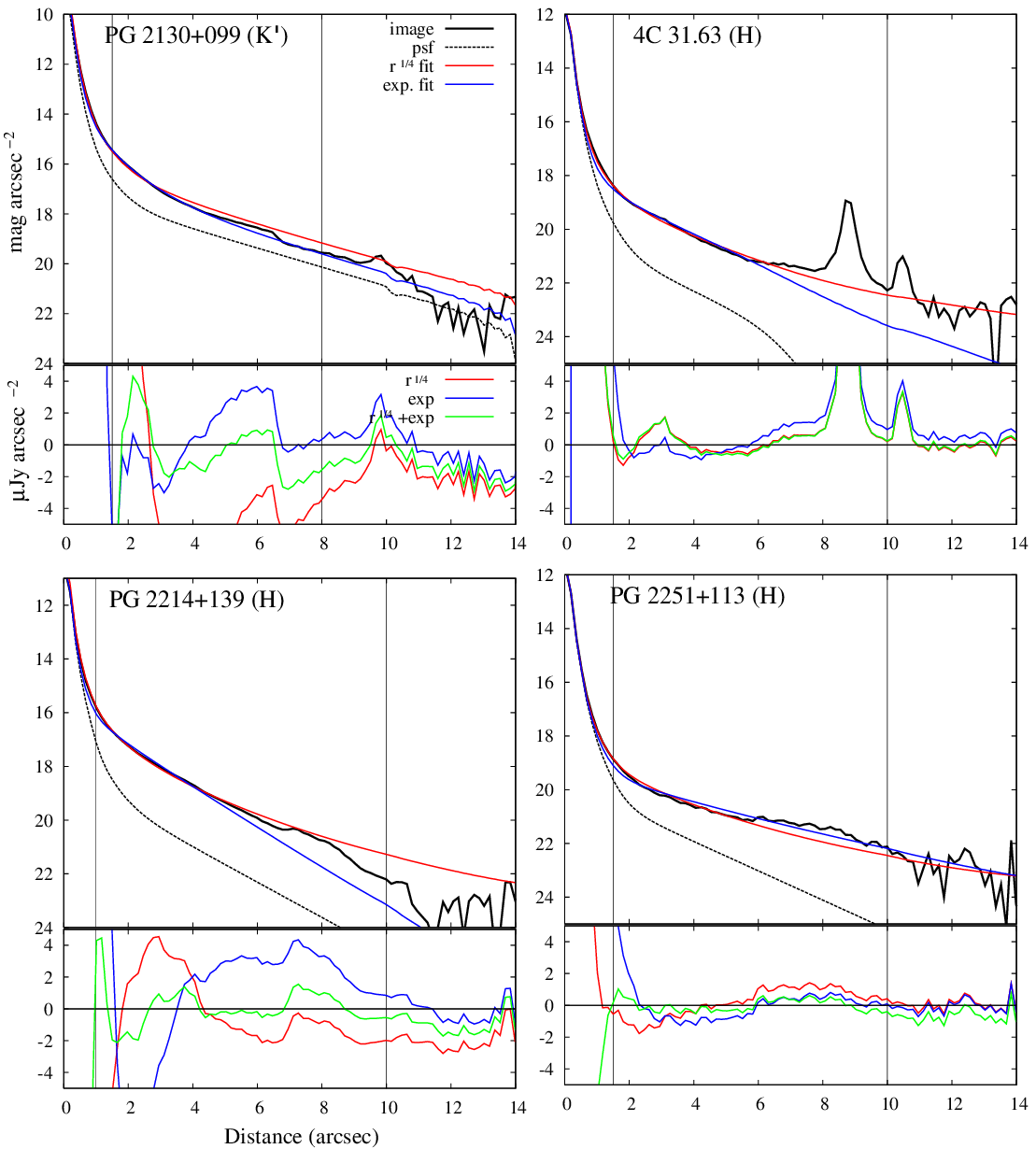}
\centerline{Fig. 5. --- Continued.}
\end{figure}
\clearpage

\begin{figure}
\epsscale{0.5}
\plotone{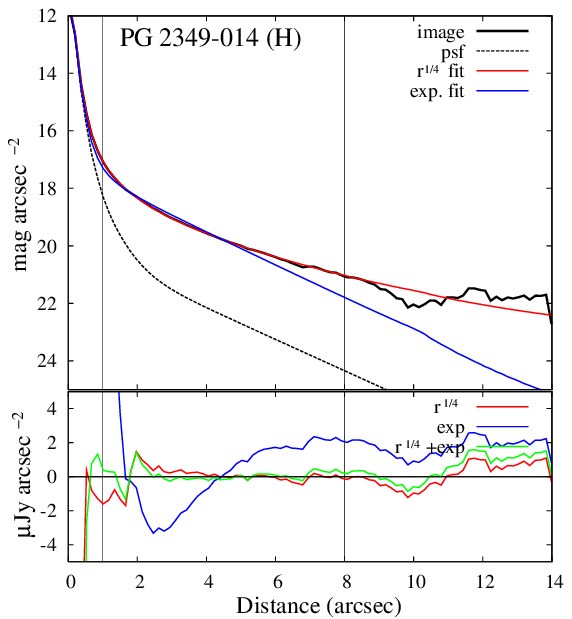}
\centerline{Fig. 5. --- Continued.}
\end{figure}
\clearpage

\begin{figure}
\epsscale{1.0}
\plotone{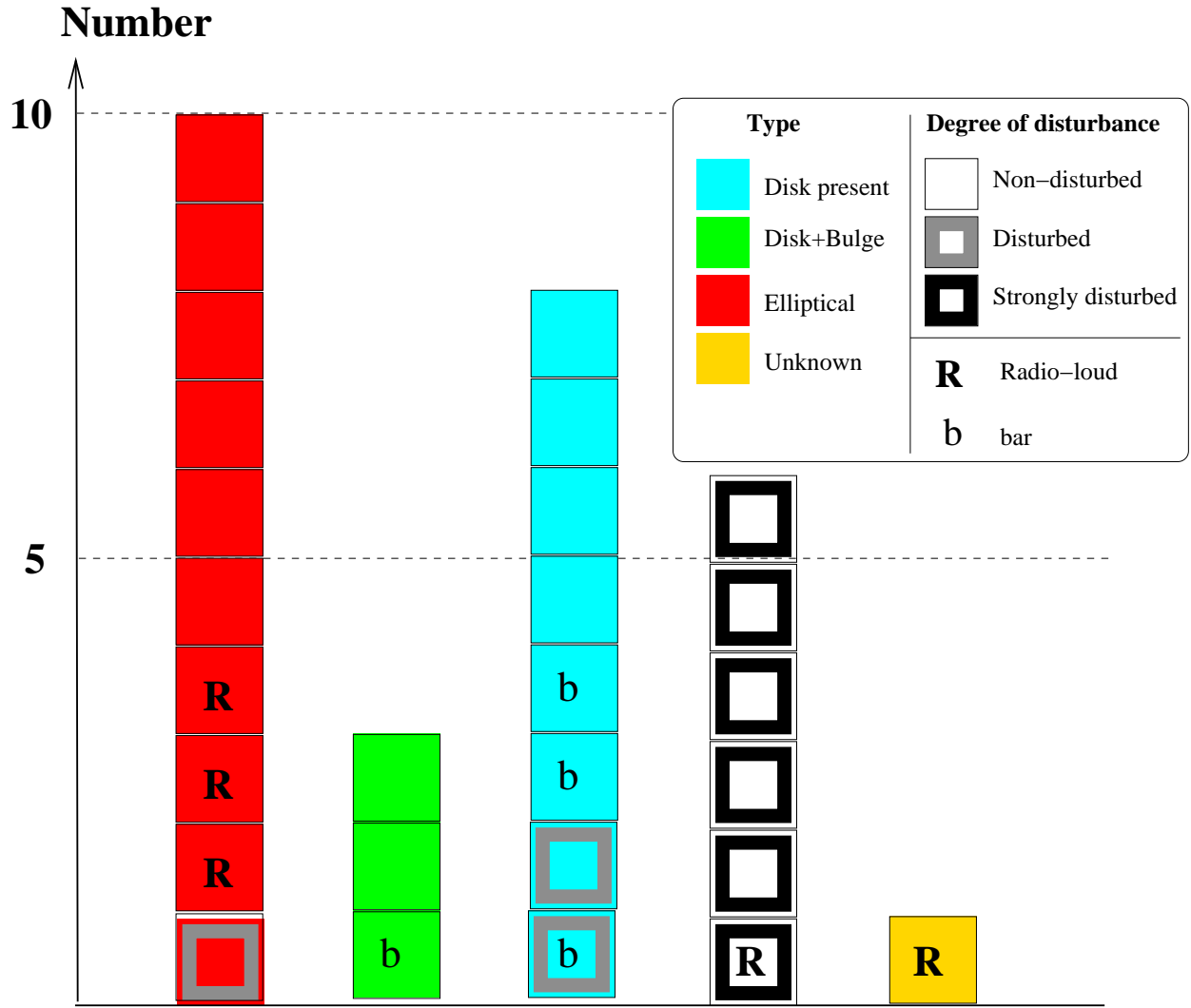}
\caption{\label{fig:hosttype} Distribution of ``host type", and ``degree of disturbance" within each ``host type" for 28 out of the 32 objects ($28 = 32 - 2$ non detections $-$ 2 ``bad PSF'') listed in Table \ref{tab:hostclass}.}
\end{figure}

\begin{figure}
\epsscale{0.7}
\plotone{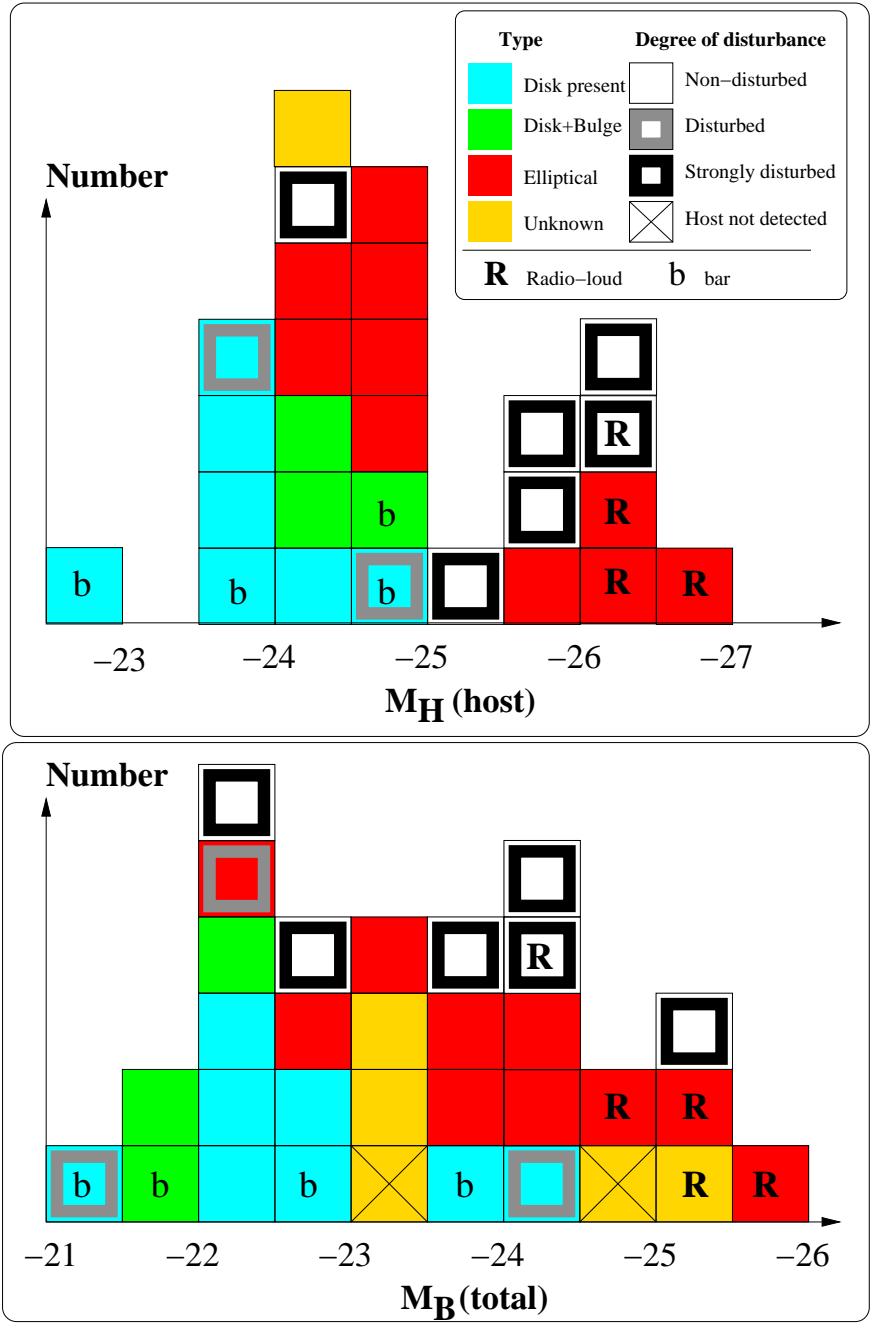}
\caption{\label{fig:MHhMB_class} (top)\ Distribution of host types as a function of the absolute $H$-band 
magnitude of the host.  Only those 27 objects with hosts characterized (32 images $-$ 2 ``too 
faint to measure host" $-$ 2 ``bad PSF'' $-$ PG0050+124) are included in the histogram. If no $H$ band images were acquired, $H$ magnitudes for the QSO and host galaxy were derived from $J$ or $K^\prime$, assuming $J\!-\!H=0.94$ and $H\!-\!K=0.97$ for the QSO nucleus (average values derived from the photometry obtained in our sample). For the host galaxy, $J\!-\!H = 0.72-0.4\times z$ and $H\!-\!K = 0.22 + 2.0 \times z$ were assumed, where $z$ is the redshift.\ \ (bottom)\  Distribution of host types as a function 
of the absolute {\it total} $B$-band magnitude for all 32 QSOs.    As described in the 
text, host types (disk present, bulge+disk or elliptical) are not assigned to strongly disturbed hosts.}
\end{figure}

\clearpage
\begin{figure}
\epsscale{1.0}
\plotone{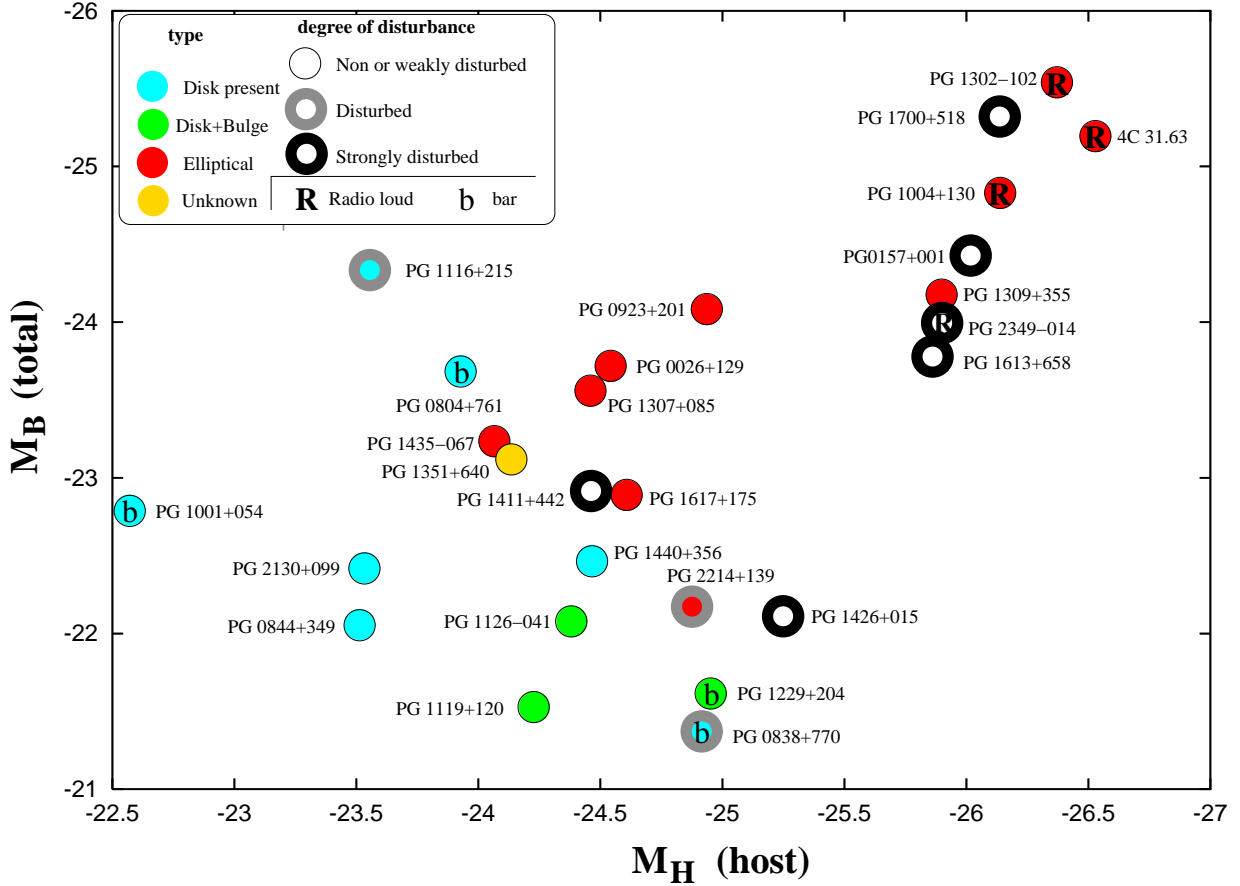}
\caption{\label{fig:MBvsMH_class}  Absolute {\it total} $B$-band magnitude of the QSO versus the absolute $H$-band 
magnitude of the QSO host galaxy.  Only those 27 objects with hosts detected and characterized (32 images $-$ 2 ``too 
faint to measure host" $-$ 2 ``bad PSF'' $-$ 1 ``no reliable 2-D fit'') are shown (same subsample as top panel in Fig. \ref{fig:MHhMB_class}). If no $H$ band images were acquired, $H$ magnitudes for the QSO and host galaxy were derived from $J$ or $K^\prime$, assuming $J\!-\!H=0.94$ and $H\!-\!K=0.97$ for the QSO nucleus (average values derived from the photometry obtained in our sample). For the host galaxy, $J\!-\!H = 0.72-0.4\times z$ and $H\!-\!K = 0.22 + 2.0 \times z$ were assumed, where $z$ is the redshift. Radio-loud objects are labeled in bold red.}
\end{figure}
\clearpage

\begin{figure}
\epsscale{0.5}
\plotone{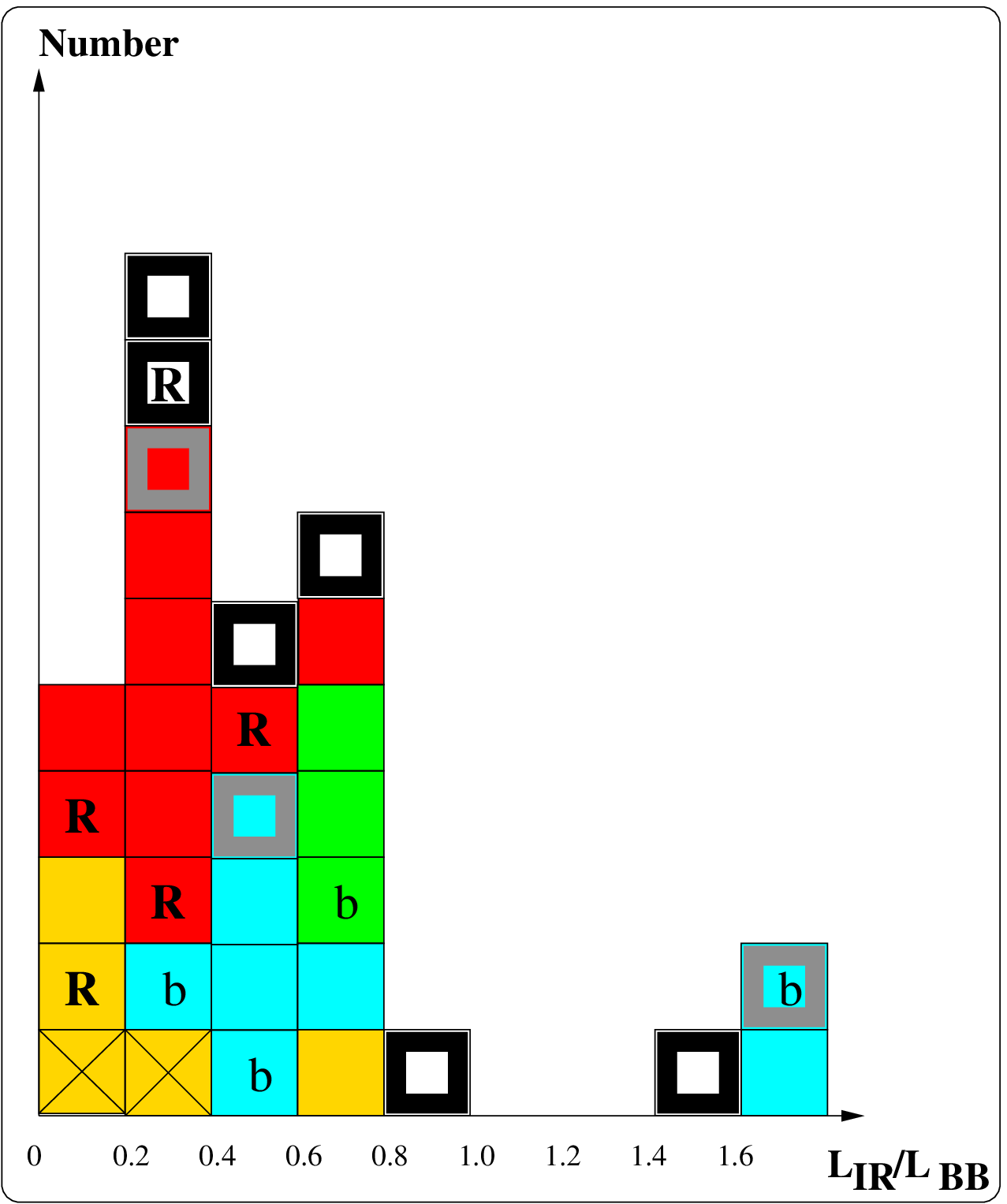}
\caption{\label{fig:LirLbb_class} Distribution of host types as a function of  ``infrared" ($L_{\rm IR}$) to ``optical/UV" ($L_{\rm BB}$) 
luminosity ratio of the QSO, where $L_{\rm IR}$ and $L_{\rm BB}$ are meant to represent the total far-infrared luminosity 
and the luminosity of the ``blue-bump" in the spectral energy distributions of  QSOs (see text for the exact frequency 
limits used to estimate $L_{\rm IR}$ and $L_{\rm BB}$).  As described in the text, host types (disk present, bulge+disk or 
elliptical) are not assigned to strongly disturbed hosts. All 32 objects are shown in this figure, and PG2251+113 ($L_{\rm IR}/L_{\rm BB} < 0.25$) is shown here in the lowest $L_{\rm IR}/L_{\rm BB}$ bin.}
\end{figure}

\begin{figure}
\epsscale{1}
\plotone{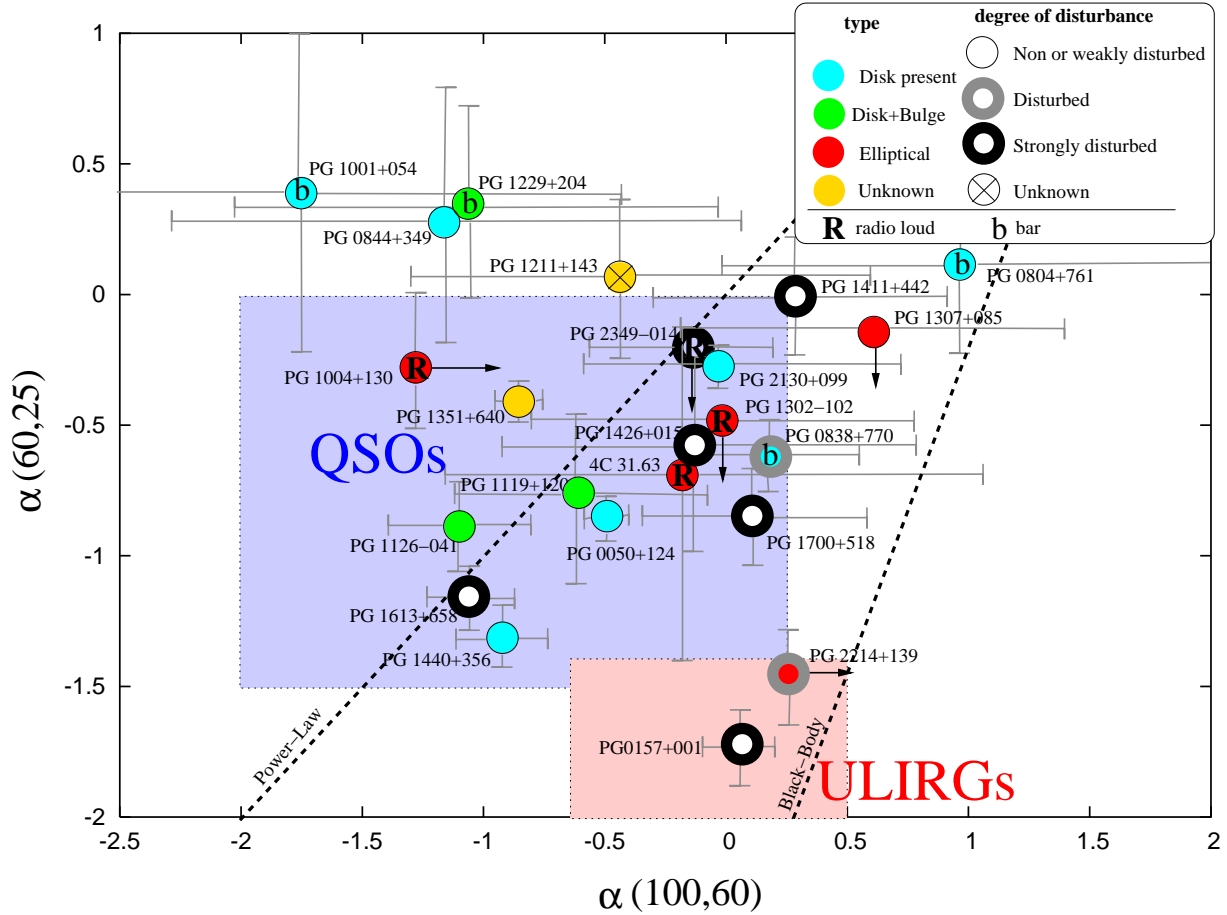}
\caption{\label{fig:IRcolcol} IR color-color diagram (adapted from \citet{lip94} and \citet{can01}). Spectral 
indexes $\alpha(100,60) \equiv -{\rm log}(f_{100 \mu m}/f_{60 \mu m})/{\rm log}(100 \mu m/60 \mu m)$ and 
$\alpha(60,25) \equiv -{\rm log}(f_{60 \mu m}/f_{25 \mu m})/{\rm log}(60 \mu m/25 \mu m)$ 
are derived from Table \ref{tab:rad_prop}. The light blue and light red areas show the approximate empirical 
locations of QSOs (close to a power law SED) and ULIRGs (dominated by thermal black body emission 
peaking in the mid-IR). Due to incomplete IR photometry, only 23 out of 32 objects are shown in this figure. 
Objects not detected (upper limit) in only one of the 3 IR wavelengths are shown, with a black arrow 
indicating possible positions on this figure. Lines corresponding to power-law emission and single 
temperature black body emission are also shown. Radio-loud objects are labeled in bold red.}
\end{figure}

\clearpage
\begin{center}
\begin{deluxetable}{ l c c c c c  }
\tablewidth{0pt}
\tabletypesize{\footnotesize}
\tablecaption{\label{tab:sourcelist}Source List}
\tablehead{
\colhead{Name} & \colhead{J2000 Coordinates} & \colhead{Redshift\tablenotemark{a}} & \colhead{$B$} & \colhead{$M_{\rm B}$\tablenotemark{b}} & \colhead{Radio\tablenotemark{c}}  \\ 
            &  \colhead{(h m s)  (\arcdeg~~\arcmin~~\arcsec)} & & \colhead{(mag)} & \colhead{(mag)} & \colhead{(Q,L)}
}
\startdata
PG0026+129 & 00 29 13.81 +13 16 04.5 & 0.142 & $\phn15.3\phn\pm0.1$\tablenotemark{d}\phd & $-23.68$ &Q\\
PG0050+124 & 00 53 35.08 +12 41 34.4 & 0.061 & $14.50\pm0.20$\tablenotemark{e} & $-22.56$ &Q\\
PG0157+001 & 01 59 50.24 +00 23 40.9 & 0.163 & $14.90\pm0.20$\tablenotemark{e} & $-24.41$ &Q\\
PG0804+761 & 08 10 58.55 +76 02 41.8 & 0.100 & $\phn14.5\phn \pm 0.1$\tablenotemark{d}\phd & $-23.66$ &Q\\
PG0838+770 & 08 44 45.63 +76 53 09.4 & 0.131 & $\phn17.4\phn \pm 0.1$\tablenotemark{d}\phd & $-21.39$ &Q\\ 
PG0844+349 & 08 47 42.51 +34 45 04.6 & 0.064 & $\phn15.1\phn \pm 0.1$\tablenotemark{d}\phd & $-22.07$ &Q\\
PG0923+201 & 09 25 54.87 +19 54 06.7 & 0.190 & $\phn15.6\phn\pm0.1$\tablenotemark{d}\phd & $-24.08$ & Q\\
PG0953+414 & 09 56 52.46 +41 15 22.0 & 0.234 & $\phn15.4\phn\pm0.1$\tablenotemark{d}\phd & $-24.79$ &Q\\
PG1001+054 & 10 04 20.15 +05 13 00.3 & 0.160 &  $\phn16.5\phn\pm0.1$\tablenotemark{d}\phd & $-22.77$ &Q\\
PG1004+130 & 10 07 26.11 +12 48 56.2 & 0.240 & $15.41\pm0.05$\tablenotemark{f} & $-24.84$ &L\\
PG1116+215 & 11 19 08.70 +21 19 18.0 & 0.177 & $15.17\pm0.20$\tablenotemark{g} & $-24.34$ &Q\\
PG1119+120 & 11 21 47.17 +11 44 18.1 & 0.050 & $15.09\pm0.12$\tablenotemark{h} & $-21.53$ &Q\\
PG1126$-$041 & 11 29 16.73 $-$04 24 07.8 & 0.060 & $14.92\pm0.12$\tablenotemark{h} &$ -22.10$ & Q\\
PG1211+143 & 12 14 17.67 +14 03 12.4 & 0.081 & $\phn14.4\phn\pm0.1$\tablenotemark{d}\phd & $-23.30$ &Q\\
PG1229+204 & 12 32 03.67 +20 09 29.1 & 0.063 & $\phn15.5\phn\pm0.1$\tablenotemark{d}\phd & $-21.63$ & Q\\
PG1302$-$102 & 13 05 33.01 $-$10 33 19.4 & 0.278 & $15.09\pm0.20$\tablenotemark{g} & $-25.53$ &L\\
PG1307+085 & 13 09 47.03 +08 19 49.3 & 0.155 & $\phn15.6\phn\pm0.1$\tablenotemark{d}\phd & $-23.59$ &Q\\
PG1309+355 & 13 12 17.74 +35 15 21.3 & 0.184 & $\phn15.5\phn\pm0.1$\tablenotemark{d}\phd & $-24.10$ &Q\\
PG1351+640 & 13 53 15.85 +63 45 44.9 & 0.088 & $\phn14.7\phn\pm0.1$\tablenotemark{d}\phd & $-23.16$ &Q\\
PG1411+442 & 14 13 48.47 +44 00 13.6 & 0.090 & $\phn15.0\phn\pm0.1$\tablenotemark{d}\phd & $-22.91$ &Q\\
PG1426+015 & 14 29 06.67 +01 17 06.3 & 0.086 & $\phn15.7\phn\pm0.1$\tablenotemark{d}\phd & $-22.13$ &Q\\
PG1435$-$067 & 14 38 16.25 $-$06 58 21.5 & 0.126 & $15.54\pm0.20$\tablenotemark{g} & $-23.16$ &Q\\
PG1440+356 & 14 42 07.54 +35 26 22.9 & 0.079 & $15.15\pm0.20$\tablenotemark{h} & $-22.49$ &Q\\
PG1613+658 & 16 13 57.27 +65 43 09.7 & 0.129 & $\phn14.9\phn\pm0.1$\tablenotemark{d}\phd & $-23.85$ &Q\\
PG1617+175 & 16 20 11.29 +17 24 27.6 & 0.112 & $\phn15.5\phn\pm0.1$\tablenotemark{d}\phd & $-22.92$ &Q\\
PG1626+554 & 16 27 56.03 +55 22 31.2 & 0.133 & $\phn15.7\phn\pm0.1$\tablenotemark{d}\phd & $-23.13$ &Q\\
PG1700+518 & 17 01 24.87 +51 49 21.0 & 0.292 & $\phn15.4\phn\pm0.1$\tablenotemark{d}\phd & $-25.35$ & Q\\
PG2130+099 & 21 32 27.92 +10 08 18.7 & 0.063 & $\phn14.7\phn\pm0.1$\tablenotemark{d}\phd & $-22.43$ &Q\\
4C 31.63 & 22 03 15.98 +31 45 38.3 & 0.295 & $15.58\pm0.05$\tablenotemark{f} & $-25.19$ &L\\
PG2214+139 & 22 17 12.35 +14 14 21.2 & 0.066 & $15.05\pm0.03$\tablenotemark{i} & $-22.19$ &Q\\
PG2251+113 & 22 54 10.50 +11 36 39.0 & 0.326 & $\phn16.0\phn\pm0.1$\tablenotemark{d}\phd & $-25.02$ & L\\
PG2349$-$014 & 23 51 56.13 $-$01 09 13.4 & 0.174 & $15.45\pm0.20$\tablenotemark{j} & $-24.02$ &L\\
\enddata
\tablenotetext{a}{Redshift given by NED}
\tablenotetext{b}{$H_0 = 75$, $\Omega_{\rm M}= 0.3$, $\Omega_\Lambda = 0.7$.}
\tablenotetext{c}{Q - radio quiet, L - radio loud:  following \cite{kel94} where sources with a 5GHz 
radio luminosity greater than $10^{25}$ W Hz$^{-1}$ ($H_0 = 50$ km s$^{-1}$ Mpc$^{-1}$, $q_0 = 0.5$) were referred to as 
radio loud and the weaker ones as radio quiet.}
\tablenotetext{d}{$B$ magnitude from \citet{giv99}.}
\tablenotetext{e}{$B$ magnitude from \citet{sur00}.}
\tablenotetext{f}{$B$ magnitude from \citet{neu79}.}
\tablenotetext{g}{$B$ magnitude from \citet{sch83}.}
\tablenotetext{h}{$B$ magnitude from \citet{sur01}.}
\tablenotetext{i}{$B$ magnitude from \citet{mca83}.}
\tablenotetext{j}{$B$ magnitude = 15.45 (Simbad).}
\end{deluxetable}
\end{center}

\begin{deluxetable}{ l c c c c  c }
\tablewidth{0pt}
\tabletypesize{\footnotesize}
\tablecaption{\label{tab:observations_log}Summary of Observations}
\tablehead{
 \colhead{Name} &  \colhead{Obs. Date} &  \colhead{Exposure\tablenotemark{a}} &  \colhead{Filter} &  \colhead{FWHM} &  \colhead{Telescope} \\ 
            & \colhead{(UT)}     &  \colhead{(s)}              &           &  \colhead{(\arcsec)} &
}
\startdata
PG0026+129 & 2001 Jun 27 & 2670 & $H\phantom{'}$ & 0.13 & Gemini\\
PG0050+124 & 2000 Nov 17 & (1200) & $H\phantom{'}$ & 0.32 & Gemini\\
           & 2000 Nov 17 & (1620) & $K^\prime$ & 0.23 & Gemini\\
           & 2000 Nov 19 & (1560) & $H\phantom{'}$ & 0.16 & Gemini\\
           & 2000 Nov 19 & (1200) & $K^\prime$ & 0.15 & Gemini\\
           & 2000 Nov 20 & (2300) & $H\phantom{'}$ & 0.23 & Gemini\\
           & 2000 Nov 20 & (1130) & $K^\prime$ & 0.21 & Gemini\\
PG0157+001 & 2001 Sep 14 & 4050 & $H\phantom{'}$ & 0.19 & Gemini\\
PG0804+761 & 2000 Dec 21 & 2910 & $H\phantom{'}$ & 0.15 & Gemini\\
           & 2000 Dec 21 & 2620 & $K^\prime$ & 0.15 & Gemini\\
PG0838+770 & 2003 Feb 12 & 1460 & $H\phantom{'}$ & 0.32 & Subaru\\ 
PG0844+349 & 2001 Oct 10 & \phn600 & $H\phantom{'}$ & 0.38 & Gemini\\
PG0923+201 & 2001 Feb 21 & 3930 & $H\phantom{'}$ & 0.20 & Gemini\\
           &  2001 Feb 23 & (5700) & $J\phantom{'}$ & 0.17 & Gemini\\
PG0953+414 & 2000 Dec 20 & 4590 & $K^\prime$ & 0.13 & Gemini\\
           &  2001 Feb 23 & (1380) & $J\phantom{'}$ & 0.23 & Gemini\\
PG1001+054 & 2000 Dec 26 & 4140 & $H\phantom{'}$ & 0.20 & Gemini\\
PG1004+130 & 2003 Feb 12 & 2400 & $H\phantom{'}$ & 0.16 & Subaru\\
PG1116+215 & 2002 Apr 21 & (3030) & $H\phantom{'}$ & 0.18 & Gemini\\
PG1119+120 & 2001 Feb 23 & (3960) & $H\phantom{'}$ & 0.15 & Gemini\\
PG1126$-$041 & 2001 Feb 21 & 4950 & $H\phantom{'}$ & 0.15 & Gemini\\
PG1211+143 & 2002 Apr 21 & (2970) & $H\phantom{'}$ & 0.17 & Gemini\\
PG1229+204 & 2002 Apr 23 & \phn(630) & $H\phantom{'}$ & 0.24 & Gemini\\
PG1302$-$102 & 2003 Feb 12 & 1620 & $H\phantom{'}$ & 0.15 & Subaru\\
PG1307+085 & 2003 Feb 12 & 2420 & $H\phantom{'}$ & 0.15 & Subaru\\
PG1309+355 & 2002 Apr 23 & (2490) & $H\phantom{'}$ & 0.22 & Gemini\\
PG1351+640 & 2002 Apr 21 & (2100) & $K^\prime$ & 0.23 & Gemini\\
PG1411+442 & 2001 Feb 23 & (1800) & $H\phantom{'}$ & 0.13 & Gemini\\
           &  2001 Feb 23 & (1620) & $J\phantom{'}$ & 0.15 & Gemini\\
           &  2001 Feb 25 & 3870 & $H\phantom{'}$ & 0.17 & Gemini\\
           & 2001  Apr 25 & 2730 & $K^\prime$ & 0.13 & Gemini\\
PG1426+015 & 2001 Apr 24 & (4080) & $K^\prime$ & 0.18 & Gemini\\
PG1435$-$067 & 2001 Feb 23 & (2100) & $J\phantom{'}$ & 0.21 & Gemini\\
PG1440+356 & 2001 Jun 28 & \phn(960) & $K^\prime$ & 0.16 & Gemini\\
PG1613+658 & 2001 Feb 25 & 4500 & $K^\prime$ & 0.30 & Gemini\\ 
           &  2001 Apr 25 & 3420 & $J\phantom{'}$ & 0.18 & Gemini\\          
           & 2001 Apr 27 & 1380 & $H\phantom{'}$ & 0.18 & Gemini\\
           &  2001 Apr 27 & 1110 & $K^\prime$ & 0.28 & Gemini\\          
PG1617+175 &  2002 Apr 21 & (4140) & $K^\prime$ & 0.23 & Gemini\\
PG1626+554 & 2001 Apr 27 & 1170 & $K^\prime$ & 0.18 & Gemini\\
           &  2001 Jun 27 & 4530 & $H\phantom{'}$ & 0.16 & Gemini\\
PG1700+518 &  2001 Sep 12 & \phn990 & $H\phantom{'}$ & 0.13 & Gemini\\            
           &  2001 Sep 14 & 3270 & $H\phantom{'}$ & 0.14 & Gemini\\
           & 2001 Oct 10 & 1350 & $K^\prime$ & 0.15 & Gemini\\
PG2130+099 & 2001 Jun 28 & (3180) & $K^\prime$ & 0.18 & Gemini\\
           &  2001 Oct 10 & 1080 & $H\phantom{'}$ & 0.15 & Gemini\\
4C 31.63 & 2001 Jun 27 & 3540 & $H\phantom{'}$ & 0.13 & Gemini\\
PG2214+139 & 2001 Sep 12 & 2070 & $H\phantom{'}$ & 0.17 & Gemini\\
PG2251+113 & 2001 Sep 14 & 2910 & $H\phantom{'}$ & 0.14 & Gemini\\
PG2349$-$014 & 2001 Sep 12 & 3540 & $H\phantom{'}$ & 0.13 & Gemini\\
\enddata
\tablenotetext{a}{Values in parentheses indicate poor photometric conditions, hence the sensitivity may not scale as $\sqrt t$}.

\end{deluxetable}

\clearpage

\begin{deluxetable}{ l c c c c c c }
\tablewidth{0pt}
\tabletypesize{\footnotesize}
\tablecaption{\label{tab:photom} Near-Infrared Photometry}
\tablehead{
 \colhead{Name} &  \colhead{Band} &  \colhead{2MASS\tablenotemark{a}} &  \colhead{N79,N87\tablenotemark{b,c}} & \colhead{Others} & \colhead{This Work\tablenotemark{c}} & \colhead{Adopted Value\tablenotemark{d}} \\
            &            &     \colhead{(mJy)}     &   \colhead{(mJy)}  &   \colhead{(mJy)}  &   \colhead{(mJy)}  &   \colhead{(mJy)} \\
}
\startdata
PG0026+129 & $H\phantom{'}$ & \phn$6.24\pm0.2$ &  \phn$6.02 \pm 1.0$  & \nodata & \phn$6.38$ &  \phn$6.38$\\
PG0050+124 & $J\phantom{'}$ & $16.69\pm0.5$& \nodata & \nodata & \nodata & \nodata \\
PG0050+124 & $H\phantom{'}$ & $25.88\pm1.0$ & \nodata & \nodata & \nodata & \nodata \\
PG0050+124 & $K^\prime$ & $47.42\pm1.0$ & \nodata & \nodata & \nodata & \nodata \\
PG0157+001 & $H\phantom{'}$ &  \phn$6.82 \pm 0.2$ &  \phn$5.37 \pm 0.3$ & \nodata &  \phn$6.16$ &  \phn$6.16$\\
PG0804+761 & $H\phantom{'}$ & $14.95 \pm 0.2$ & $12.88 \pm 0.6$& \nodata & $15.95$ & $15.95$\\
PG0804+761 & $K^\prime$ & $26.57 \pm 0.2$ & $25.11 \pm 2.0$ & \nodata & $29.50$ & $29.50$\\
PG0838+770 & $H\phantom{'}$ &  \phn$3.11 \pm 0.2$ &  \phn$3.09 \pm 0.2$ & \nodata &  \phn$2.94$ &  \phn$2.94$\\
PG0844+349 & $H\phantom{'}$ &  \phn$8.73 \pm 0.1$ &  \phn$9.54 \pm 0.5$ & \nodata &  \phn$9.31$ &  \phn$9.31$\\
PG0923+201 & $J\phantom{'}$ &   \phn\phn$3.12 \pm 0.06$ & \nodata & \nodata &  \phn$>3.07$\phd\phs &  \phn$3.12$ \\
PG0923+201 & $H\phantom{'}$ &  \phn$4.54 \pm 0.2$ & \nodata & \nodata &  \phn$4.73$ &  \phn$4.73$\\
PG0953+414 & $J\phantom{'}$    &  \phn$3.40 \pm 0.1$ &  \phn$3.39 \pm 0.3$ & \nodata &  \phn$> 2.73$\phd\phs &  \phn$3.40$\\ 
PG0953+414 & $K^\prime$   &  \phn$6.52 \pm 0.2$ &  \phn$7.76 \pm 0.5$ & \nodata &  \phn$6.72$ &  \phn$6.72$\\
PG1001+054 & $H\phantom{'}$    &  \phn$2.18 \pm 0.2$ &  \phn$7.76 \pm 0.5$ &  \phn$2.19 \pm 0.1$\tablenotemark{e}&  \phn$2.50$ &  \phn$2.50$\\
PG1004+130 & $H\phantom{'}$    &  \phn$3.74 \pm 0.3$ &  \phn$4.07 \pm 0.4$ &  \phn$5.1\phn \pm 0.2$\tablenotemark{f} &  \phn$3.62$ &  \phn$3.62$\\ 
PG1116+215 & $H\phantom{'}$    &  \phn$8.84 \pm 0.2$ & \nodata & \nodata & $>10.54$\phd\phs &  \phn$8.84$\\
PG1119+120 & $H\phantom{'}$    &  \phn$9.35 \pm 0.1$ &  \phn$9.77 \pm 0.3$ & \nodata &  \phn$>8.74$\phd\phs &  \phn$9.35$\\
PG1126$-$041 & $H\phantom{'}$    & $15.78 \pm 0.6$ & \nodata & \nodata & $15.49$ & $15.49$\\
PG1211+143 & $H\phantom{'}$    & $12.23 \pm 0.5$ & $14.8 \pm 1.0$ & \nodata &  \phn$>8.46$\phd\phs & $12.23$\\
PG1229+204 & $H\phantom{'}$    &  \phn$8.67 \pm 0.1$ &  \phn$8.51 \pm 0.6$ & \nodata &  \phn$>6.23$\phd\phs &  \phn$8.67$\\ 
PG1302$-$102 & $H\phantom{'}$    &  \phn$3.83 \pm 0.2$ & \nodata &  \phn$3.74 \pm 0.2$\tablenotemark{e} &  \phn$3.99$ &  \phn$3.99$\\
PG1307+085 & $H\phantom{'}$    &  \phn$4.44 \pm 0.2$ &  \phn$4.32 \pm 0.4$ & \nodata &  \phn$3.65$ &  \phn$3.65$ \\
PG1309+355 & $H\phantom{'}$    &  \phn$4.06 \pm 0.1$ & \nodata & \nodata &  \phn$>3.36$\phd\phs &  \phn$4.06$\\
PG1351+640 & $K^\prime$   & $12.12 \pm 0.1$ & $10.96 \pm 0.8$ & \nodata &  \phn$>5.65$\phd\phs & $12.12$\\
PG1411+442 & $J\phantom{'}$    &  \phn$8.03 \pm 0.1$ &  \phn$5.62 \pm 0.4$ & \nodata &  \phn$>7.19$\phd\phs &  \phn$8.03$\\
PG1411+442 & $H\phantom{'}$    & $10.57 \pm 0.1$ &  \phn$8.32 \pm 0.8$ & \nodata & $12.23$ & $12.23$\\
PG1411+442 & $K^\prime$   & $16.90 \pm 0.2$ & $17.38 \pm 1.3$ & \nodata & $16.8$\phn &  $16.8$\phn\\
PG1426+015 & $K^\prime$   & $24.59 \pm 0.4$ & $15.85 \pm 1.1$ & \nodata & $> 16.27$\phd\phs & $24.59$\\
PG1435$-$067 & $J\phantom{'}$    &  \phn$2.76 \pm 0.1$ & \nodata & \nodata &  \phn$>2.18$\phd\phs &  \phn$2.76$\\
PG1440+356 & $K^\prime$   & $24.73 \pm 0.2$ & $25.12 \pm 1.8$ & \nodata & $>17.56$\phd\phs & $24.73$\\
PG1613+658 & $J\phantom{'}$ & $11.16 \pm 0.2$ &  \phn$4.57 \pm 0.3$ & \nodata &  \phn$7.87$ &  \phn$7.87$\\
PG1613+658 & $H\phantom{'}$ & $14.29 \pm 0.2$ &  \phn$6.31 \pm 0.6$ & \nodata & $11.23$ & $11.23$\\
PG1613+658 & $K^\prime$ & $25.16 \pm 0.7$ & $10.72 \pm 0.8$ & \nodata & $17.85$ & $17.85$\\
PG1617+175 & $K^\prime$ & $24.59 \pm 0.4$ & $12.30 \pm 0.9$ & \nodata &  \phn$>7.21$\phd\phs & $24.59$\\
PG1626+554 & $H\phantom{'}$ &  \phn$3.70 \pm 0.3$ & \nodata & \nodata &  \phn$4.13$ &  \phn$4.13$\\
PG1626+554 & $K^\prime$ &  \phn$6.15 \pm 0.1$ & \nodata & \nodata &  \phn$5.37$ &  \phn$5.37$\\
PG1700+518 & $H\phantom{'}$ &  \phn$6.44 \pm 0.2$ &  \phn$6.53 \pm 0.4$ & \nodata &  \phn$6.19$ &  \phn$6.19$\\
PG1700+518 & $K^\prime$ & $10.69 \pm 0.1$ & $12.74 \pm 0.9$ & \nodata &  \phn$9.38$ &  \phn$9.38$\\
PG2130+099 & $H\phantom{'}$ & $21.89 \pm 0.6$ & $ 15.37 \pm 1.1$ & \nodata & $18.74$ & $18.74$\\
PG2130+099 & $K^\prime$ & $36.64 \pm 0.5$ & $ 27.33 \pm 2.0$ & \nodata & $>19.88$\phd\phs & $36.64$ \\
4C 31.63 & $H\phantom{'}$ &  \phn$4.65 \pm 0.1$ &  \phn$6.17 \pm 0.5$ & \nodata & \phn$4.5$\phn &  \phn$4.5$\phn\\
PG2214+139 & $H\phantom{'}$ & $13.82 \pm 0.2$ & \nodata & $13.20 \pm 0.8$\tablenotemark{g} & $16.44$ & $16.44$\\
PG2251+113 & $H\phantom{'}$ &  \phn$4.38 \pm 0.2$ &  \phn$4.57 \pm 0.3$ & \nodata &  \phn$4.48$ &  \phn$4.48$\\
PG2349$-$014 & $H\phantom{'}$ &  \phn$5.33 \pm 0.2$ &  \phn$5.75 \pm 0.3$ &  \phn$5.35 \pm 0.9$\tablenotemark{h} &  \phn$5.58$ &  \phn$5.58$\\
\hline
\hline
\enddata
\tablenotetext{a}{Flux enclosed within a 8\arcsec\ diameter disk.}
\tablenotetext{b}{\citet{neu79}, \citet{neu87}. Flux enclosed within a 5\farcs5 diameter disk}
\tablenotetext{c}{Flux enclosed within a 5\farcs5 diameter disk. Typical $1\sigma$ uncertainty $\pm 5\%$. 
Lower limits were derived in non-photometric conditions, and are accurate to $\pm 5\%$ ($1\sigma$ level).}
\tablenotetext{d}{If available, the photometry done in this work is adopted. If not, 2MASS values are adopted.}
\tablenotetext{e}{\citet{hyl82}}
\tablenotetext{f}{\citet{sit82}}
\tablenotetext{g}{\citet{bal81}}
\tablenotetext{h}{\citet{rud82}}
\end{deluxetable}

\clearpage

\begin{deluxetable}{lccc}
\tablewidth{0pt}
\tabletypesize{\footnotesize}
\tablecaption{\label{tab:2Dfitresult1} Results of Host 2-D Fits}
\tablehead{
	 \colhead{Name/}  &  \colhead{$\chi_{\rm r^{1/4}}^2$} &  \colhead{$\chi_{\rm exp}^2$} &  \colhead{$\chi_{\rm r^{1/4}+exp}^2$\tablenotemark{a}} \\
  \colhead{Filter}  &  &  &
}
\startdata
PG0026+129 $H$  & \phn1.05 & \phn1.08 & 1.04 [0.96/0.04] \\
PG0157+001 $H$  & \phn3.79 & \phn4.26 & 3.12 [0.64/0.36] \\
PG0804+761 $H$  & \phn1.54 & \phn1.90 & 1.05 [0.82/0.18] \\
PG0804+761 $K^\prime$ & \phn1.01 & \phn1.10 & 0.95 [0.68/0.32] \\
PG0838+770 $H$  & \phn4.43 & \phn5.05 & 3.58 [0.78/0.22] \\
PG0844+349 $H$  & \phn2.33 & \phn3.07 & 2.31 [0.69/0.31] \\
PG0923+201 $J$  & \phn1.12 & \phn1.24 & 1.15 [0.89/0.11] \\
PG0923+201 $H$  & \phn1.07 & \phn1.24 & 1.07 [0.94/0.06] \\
PG0953+414 $J$  & \phn0.94 & \phn0.90 & 0.88 [0.32/0.68] \\
PG0953+414 $K^\prime$ & \phn0.92 & \phn0.93 & 0.91 [0.98/0.02] \\
PG1001+054 $H$  & \phn0.97 & \phn1.03 & 0.93 [0.48/0.52] \\
PG1004+130 $H$  & \phn1.13 & \phn1.79 & 1.06 [0.65/0.35] \\
PG1116+215 $H$  & \phn1.20 & \phn1.13 & 1.13 [0.00/1.00] \\
PG1119+120 $H$  & 11.38 & 10.86 & 6.57 [0.70/0.30] \\
PG1126$-$041 $H$  & 11.61 & 17.00 & 9.66 [0.64/0.36] \\
PG1211+143 $H$  & \phn1.28 & \phn1.28 & 1.30 [0.46/0.54] \\
PG1229+204 $H$  & \phn1.97 & \phn2.46 & 1.41 [0.65/0.35] \\
PG1302$-$102 $H$  & \phn2.20 & \phn2.46 & 2.00 [0.74/0.26] \\
PG1307+085 $H$  & \phn1.42 & \phn2.15 & 1.19 [0.74/0.26] \\
PG1309+355 $H$  & \phn0.98 & \phn1.28 & 0.97 [0.92/0.08] \\
PG1351+640 $K^\prime$ & \phn0.88 & \phn0.88 & 0.87 [0.48/0.52] \\
PG1411+442 $J$  & \phn2.08 & \phn2.05 & 2.00 [0.86/0.14] \\
PG1411+442 $H$  & \phn9.20 & \phn8.77 & 8.57 [0.61/0.39] \\
PG1411+442 $K^\prime$ & \phn1.48 & \phn2.26 & 1.47 [0.96/0.04] \\
PG1426+015 $K^\prime$ & \phn1.53 & \phn1.52 & 1.37 [0.91/0.09] \\
PG1435$-$067 $J$  & \phn0.98 & \phn1.04 & 0.96 [0.92/0.08] \\
PG1440+356 $K^\prime$ & \phn2.14 & \phn1.93 & 1.65 [0.84/0.16] \\
PG1613+658 $H$  & \phn4.30 & \phn6.67 & 3.95 [0.90/0.10] \\
PG1613+658 $J$  & \phn5.14 & \phn5.61 & 5.15 [0.99/0.01] \\
PG1613+658 $K^\prime$ & \phn6.15 & 13.17 & 5.97 [0.98/0.02] \\
PG1617+175 $K^\prime$ & \phn1.05 & \phn1.62 & 1.06 [0.98/0.02] \\
PG1626+554 $H$  & \phn0.89 & \phn0.86 & 0.84 [0.52/0.48] \\
PG1626+554 $K^\prime$ & \phn0.93 & \phn0.92 & 0.92 [0.79/0.21] \\
PG1700+518 $H$  & \phn5.48 & \phn5.52 & 5.38 [0.80/0.20] \\
PG1700+518 $K^\prime$ & \phn2.53 & \phn2.34 & 2.34 [0.00/1.00] \\
PG2130+099 $H$  & \phn4.57 & \phn3.13 & 3.28 [0.00/1.00] \\
PG2130+099 $K^\prime$ & \phn5.25 & \phn3.98 & 3.87 [0.42/0.58] \\
4C 31.63 \phn\phn\phn\phn$H$ & \phn1.07 & \phn1.49 & 1.06 [0.84/0.16] \\
PG2214+139 $H$  & \phn1.79 & \phn3.26 & 1.49 [0.76/0.24] \\
PG2251+113 $H$  & \phn1.04 & \phn1.02 & 0.95 [0.53/0.47] \\
PG2349$-$014 $H$  & \phn4.31 & \phn5.70 & 4.13 [0.82/0.18] \\
\enddata
\tablenotetext{a}{The numbers in brackets give the fraction of the host's luminosity originating from the bulge component (first number) and the disk component (second number) for the bulge+disk 2-D fit.}
\end{deluxetable}

\clearpage

\begin{deluxetable}{l c c c c c c c l}
\tablewidth{0pt}
\tabletypesize{\scriptsize}
\tablecaption{\label{tab:hostclass}Host Morphologies\tablenotemark{a}}
\tablehead{
 \colhead{Name} &  \multicolumn{3}{c}{$\chi^2$-Type\tablenotemark{b}} &  \multicolumn{3}{c}{Visual Features} &  \colhead{Adopted Type\tablenotemark{d}} &  \colhead{Notes\tablenotemark{e}} \\
           &   \colhead{$J$} &   \colhead{$H$} &   \colhead{$K^\prime$} &  \colhead{Bar} &  \colhead{Arms} &  \colhead{Dist\tablenotemark{c}} & 
}
\startdata
PG0026+129 & \nodata & ($r^{1/4}$) & \nodata & no  & no  & no & E & \\
PG0050+124 & \nodata & \nodata & \nodata    & no  & yes & no & D$_{\rm p}$  & \\
PG0157+001 & \nodata & $r^{1/4}$+exp.& \nodata & no  & yes & str & ? & host asymmetry\\
PG0804+761 & \nodata & $r^{1/4}$ &$r^{1/4}$ & yes & no  & no & D$_{\rm p}$  & well-defined bar\\
PG0838+770 & \nodata & $r^{1/4}$ & \nodata & yes & yes & yes & D$_{\rm p}$ &twisted bar+ring, well-defined bar\\
PG0844+349 & \nodata & $r^{1/4}$\tablenotemark{f} &\nodata & no & yes & no & D$_{\rm p}$ &\\
PG0923+201 & $r^{1/4}$ & $r^{1/4}$ & \nodata & no  & no  & no & E   &\\
PG0953+414 & (exp) & \nodata & ?  & no  & no  & -- & -- & host too faint to be classified\\
PG1001+054 & \nodata & ($r^{1/4}$) & \nodata & yes & no  & no & D$_{\rm p}$   & well-defined bar\\
PG1004+130 & \nodata & $r^{1/4}$ & \nodata & no  & no  & no & E   &\\
PG1116+215 & \nodata & exp  & \nodata       & no  & no & yes & D$_{\rm p}$   & CC, host asymmetry\\
PG1119+120 & \nodata & $r^{1/4}$+exp. & \nodata & no & no & no & B+D & CC\\
PG1126$-$041 & \nodata & $r^{1/4}$+exp. & \nodata & no  & yes & no & B+D & CC\\
PG1211+143 & \nodata & ? & \nodata & no & no & -- & -- & host too faint to be classified \\
PG1229+204 & \nodata & $r^{1/4}$+exp. & \nodata & yes & yes & no & B+D & well-defined bar\\
PG1302$-$102 & \nodata & $r^{1/4}$  & \nodata & no  & no  & no & E   & CC\\
PG1307+085 & \nodata & $r^{1/4}$ & \nodata & no  & no  & no & E  & \\
PG1309+355 & \nodata & $r^{1/4}$ & \nodata & no  & no  & no & E   & \\
PG1351+640 & \nodata & \nodata & ?    & no  & no  & no & ?   & \\
PG1411+442 & exp & $r^{1/4}$+exp &$r^{1/4}$   & no  & ?\tablenotemark{g} & str & ?    & CC\\
PG1426+015 & \nodata & \nodata & $r^{1/4}$ & no  & ?\tablenotemark{g} & str & ?    & CC\\
PG1435$-$067 & $r^{1/4}$ & \nodata & \nodata & no  & no  & no & E   & \\
PG1440+356 & \nodata & \nodata & exp & no & yes  & no & D$_{\rm p}$   &\\
PG1613+658 & $r^{1/4}$ & $r^{1/4}$ & $r^{1/4}$ & no  & ?\tablenotemark{g} & str & ?    & CC\\
PG1617+175 & \nodata & \nodata & $r^{1/4}$\tablenotemark{f} & no & no & no & E &\\
PG1626+554 & \nodata & (exp)\tablenotemark{f} & ?\tablenotemark{f} & no & no & no & -- & bad PSF; unreliable 2-D fit\\
PG1700+518 & \nodata & $r^{1/4}$ & exp     & no  & ?\tablenotemark{g} & str & ?    & CC \\
PG2130+099 & \nodata & exp & $r^{1/4}$+exp & no & yes  & no  & D$_{\rm p}$ &\\
4C 31.63& \nodata & $r^{1/4}$ & \nodata & no  & no  & no & E   & \\
PG2214+139 & \nodata & $r^{1/4}$ & \nodata & no  & no & yes & E   & concentric shells\\
PG2251+113 & \nodata & ?\tablenotemark{f} & \nodata & no  & no & no & -- & bad PSF; unreliable 2-D fit\\
PG2349$-$014 & \nodata & $r^{1/4}$ & \nodata & no & ?\tablenotemark{g}  & str & ?  & CC\\
\enddata
\tablenotetext{a}{The radial profile in the central 1\arcsec\ to 2\arcsec\ radius of each host galaxy (200 pc to 2 kpc, depending on redshift) is usually not reliable, as PSF variations dominate this inner region. Classifications presented in this table therefore apply only to the region outside this central radius.}
\tablenotetext{b}{Best 2-D fit(s) correspond to the fit with minimum $\chi^2$ in each band as listed in Table \ref{tab:2Dfitresult}.  Parentheses indicate 2-D fits with low confidence level (difference between $\chi^2_{\rm r^{1/4}}$ and $\chi^2_{\rm exp}$ is between 0.02 and 0.05).}
\tablenotetext{c}{Apparent degree of disturbance from visual inspection of the images: Disturbed (``yes''), Non disturbed (``no'') or Strongly disturbed (``str'').}
\tablenotetext{d}{D$_{\rm p}$: disk present, E: Elliptical, B+D: Bulge+disk. Type ``disk present'' indicates the 
presence of a disk component, {\it but does not exclude a small bulge in the inner part of the host}.  The final adopted host morphology is based upon minimum $\chi^2$ 2-D fitting of images {\it modified} by visual inspection of the image for signs of features indicative of the presence of a disk component. }
\tablenotetext{e}{CC: Close (within 10kpc projected separation) compact companion (see text)}
\tablenotetext{f}{Bad PSF;  2-D fit is uncertain.}
\tablenotetext{g}{Unclear whether features seen are arms or tidal debris.}
\tablenotetext{h}{No reliable 2-D fit available: does not appear in table \ref{tab:2Dfitresult}.}
\end{deluxetable}

\clearpage

\begin{deluxetable}{lccccccccccccc}
\tablewidth{0pt}
\tabletypesize{\tiny}
\tablecaption{\label{tab:2Dfitresult}  Host 2-D Fit Parameters for the Adopted Type.}
\tablehead{
	 \colhead{Name/}  &  \multicolumn{2}{c}{Nuclear magn} &  \multicolumn{2}{c}{Host magn} & \colhead{$I_e$\tablenotemark{a}} &  \colhead{$r_e$\tablenotemark{a}} & \colhead{$e_{e}$\tablenotemark{a}} &  \colhead{$PA_{e}$\tablenotemark{a}} &  \colhead{$I_0$\tablenotemark{a}} &  \colhead{$r_0$\tablenotemark{a}} &  \colhead{$e_0$\tablenotemark{a}} &  \colhead{$PA_0$\tablenotemark{a}} \\
  \colhead{Filter}    &  \colhead{app.}  &  \colhead{abs.}  &   \colhead{app.}  &  \colhead{abs.}    &   \colhead{m/\arcsec$^2$} &  \colhead{(\arcsec)} &        &  \colhead{(rad)}     &  \colhead{m/\arcsec$^2$} &  \colhead{(\arcsec)} &        &  \colhead{(rad)}     &  
}
\startdata
PG0026+129 $H$  & 13.05 & $-$25.93 & 14.46 & $-$24.52 & 22.16 & 10.62 & 0.21 & 0.01 & \nodata & \nodata & \nodata & \nodata  \\
PG0157+001 $H$  & 13.41 & $-$25.90 & 13.30 & $-$26.01 & 18.79 & 1.94 & 0.43 & 0.60 & 18.55 & 3.75 & 0.43 & 1.73  \\
PG0804+761 $H$  & 12.22 & $-$25.94 & 14.24 & $-$23.92 & \nodata & \nodata & \nodata & \nodata & 17.08 & 1.74 & 0.28 & 0.82  \\
PG0804+761 $K^\prime$ & 10.67 & $-$27.49 & 14.44 & $-$23.72 & \nodata & \nodata & \nodata & \nodata & 18.02 & 3.39 & 0.61 & 0.82  \\
PG0838+770 $H$  & 14.31 & $-$24.48 & 13.89 & $-$24.90 & \nodata & \nodata & \nodata & \nodata & 17.26 & 2.97 & 0.59 & 2.85  \\
PG0844+349 $H$  & 13.35 & $-$23.82 & 13.65 & $-$23.52 & \nodata & \nodata & \nodata & \nodata & 16.04 & 1.38 & 0.25 & 0.53  \\
PG0923+201 $J$  & 14.52 & $-$25.16 & 15.46 & $-$24.22 & 22.25 & \phn5.73 & 0.14 & 1.18 & \nodata & \nodata & \nodata & \nodata  \\
PG0923+201 $H$  & 13.53 & $-$26.15 & 14.73 & $-$24.95 & 21.20 & \phn4.53 & 0.07 & 0.79 & \nodata & \nodata & \nodata & \nodata  \\
PG0953+414 $J$  & 14.39 & $-$25.80 & 15.25 & $-$27.77 & 22.25 & \phn3.58 & 0.42 & 3.07 & 20.95 & 7.91 & 0.40 & 1.62  \\
PG0953+414 $K^\prime$ & 12.42 & $-$24.94 & 13.86 & $-$26.33 & 18.29 & \phn0.95 & 0.90 & 1.55 & 22.02 & 8.75 & 0.90 & 1.98  \\
PG1001+054 $H$  & 14.21 & $-$25.06 & 16.68 & $-$22.59 & \nodata & \nodata & \nodata & \nodata & 19.59 & 1.61 & 0.12 & 0.56  \\
PG1004+130 $H$  & 13.84 & $-$26.41 & 14.12 & $-$26.13 & 22.00 & 10.04 & 0.10 & 2.94 & \nodata & \nodata & \nodata & \nodata  \\
PG1116+215 $H$  & 12.61 & $-$26.90 & 15.96 & $-$23.55 & \nodata & \nodata & \nodata & \nodata & 17.50 & 0.99 & 0.33 & 0.04  \\
PG1119+120 $H$  & 13.30 & $-$23.32 & 12.39 & $-$24.23 & 17.22 & \phn0.91 & 0.43 & 2.37 & 16.07 & 1.49 & 0.37 & 1.38  \\
PG1126$-$041 $H$  & 12.49 & $-$24.53 & 12.63 & $-$24.39 & 17.94 & \phn2.65 & 0.79 & 1.21 & 16.05 & 1.65 & 0.52 & 1.00  \\
PG1211+143 $H$  & 12.37 & $-$25.33 & 14.44 & $-$23.26 & 21.98 & \phn4.99 & 0.00 & 2.41 & 20.19 & 7.49 & 0.51 & 0.85  \\
PG1229+204 $H$  & 13.56 & $-$23.57 & 12.18 & $-$24.95 & 17.59 & \phn1.57 & 0.24 & 1.51 & 16.94 & 3.38 & 0.59 & 2.67  \\
PG1302$-$102 $H$  & 13.74 & $-$26.88 & 14.25 & $-$26.37 & 19.89 & \phn3.14 & 0.46 & 0.81 & \nodata & \nodata & \nodata & \nodata  \\
PG1307+085 $H$  & 13.97 & $-$25.22 & 14.71 & $-$24.48 & 21.38 & \phn4.70 & 0.00 & 0.64 & \nodata & \nodata & \nodata & \nodata  \\
PG1309+355 $H$  & 14.03 & $-$25.57 & 13.71 & $-$25.89 & 19.53 & \phn3.15 & 0.18 & 1.46 & \nodata & \nodata & \nodata & \nodata  \\
PG1351+640 $K^\prime$ & 12.02 & $-$25.84 & 13.77 & $-$24.09 & 20.83 & \phn3.82 & 0.00 & 0.64 & 15.48 & 0.63 & 0.00 & 3.04  \\
PG1411+442 $J$  & 13.48 & $-$24.43 & 14.39 & $-$23.52 & 21.01 & \phn6.32 & 0.52 & 1.28 & 17.20 & 0.82 & 0.55 & 2.36  \\
PG1411+442 $H$  & 12.45 & $-$25.46 & 13.45 & $-$24.46 & 18.82 & \phn1.30 & 0.15 & 1.09 & 17.83 & 2.85 & 0.56 & 1.35  \\
PG1411+442 $K^\prime$ & 11.52 & $-$26.39 & 13.54 & $-$24.37 & 19.84 & \phn5.90 & 0.57 & 1.28 & 15.99 & 0.62 & 0.83 & 2.50  \\
PG1426+015 $K^\prime$ & 11.37 & $-$26.46 & 12.17 & $-$25.66 & 18.61 & \phn5.33 & 0.42 & 2.45 & 14.88 & 0.65 & 0.60 & 0.36  \\
PG1435$-$067 $J$  & 14.72 & $-$23.98 & 15.25 & $-$23.45 & 21.13 & \phn2.97 & 0.00 & 2.55 & \nodata & \nodata & \nodata & \nodata  \\
PG1440+356 $K^\prime$ & 11.51 & $-$26.12 & 12.80 & $-$24.84 & \nodata & \nodata & \nodata & \nodata & 14.63 & 1.02 & 0.18 & 2.35  \\
PG1613+658 $J$  & 13.56 & $-$25.19 & 13.76 & $-$24.99 & 20.98 & \phn7.90 & 0.22 & 0.82 & 15.99 & 0.30 & 0.90 & 2.14  \\
PG1613+658 $H$  & 12.90 & $-$25.85 & 12.88 & $-$25.87 & 19.98 & \phn6.94 & 0.25 & 0.95 & 15.91 & 0.68 & 0.46 & 2.91  \\
PG1613+658 $K^\prime$ & 11.88 & $-$26.99 & 12.11 & $-$26.64 & 18.34 & \phn4.00 & 0.14 & 0.88 & 17.08 & 0.86 & 0.61 & 2.77  \\
PG1617+175 $K^\prime$ & 11.12 & $-$27.30 & 13.40 & $-$25.02 & 21.73 & 14.34 & 0.00 & 0.08 & \nodata & \nodata & \nodata & \nodata  \\
PG1626+554 $H$  & 13.71 & $-$25.12 & 14.33 & $-$25.87 & 20.34 & \phn2.08 & 0.13 & 1.69 & 18.53 & 1.99 & 0.07 & 0.32  \\
PG1626+554 $K^\prime$ & 12.96 & $-$24.50 & 13.11 & $-$25.72 & 17.76 & \phn0.30 & 0.01 & 1.13 & 15.61 & 0.59 & 0.03 & 1.30  \\
PG1700+518 $H$  & 13.28 & $-$27.47 & 14.62 & $-$26.13 & 20.85 & \phn4.29 & 0.44 & 1.98 & 15.80 & 0.60 & 0.74 & 1.59  \\
PG1700+518 $K^\prime$ & 12.04 & $-$28.71 & 15.56 & $-$25.19 & inf & \phn2.11 & 0.54 & 1.24 & 15.87 & 1.04 & 0.81 & 1.70  \\
PG2130+099 $H$  & 12.05 & $-$25.08 & 13.60 & $-$23.53 & \nodata & \nodata & \nodata & \nodata & 15.54 & 1.25 & 0.40 & 2.27  \\
PG2130+099 $K^\prime$ & 10.76 & $-$26.37 & 12.72 & $-$24.41 & \nodata & \nodata & \nodata & \nodata & 14.42 & 1.08 & 0.35 & 2.24  \\
4C 31.63 \phn\phn\phn\phn$H$ & 13.68 & $-$27.09 & 14.25 & $-$26.52 & 20.97 & \phn6.47 & 0.37 & 0.64 & \nodata & \nodata & \nodata & \nodata \\
PG2214+139 $H$  & 12.41 & $-$24.83 & 12.37 & $-$24.87 & 18.07 & \phn2.72 & 0.07 & 3.02 & \nodata & \nodata & \nodata & \nodata\\
PG2251+113 $H$  & 13.51 & $-$27.51 & 14.27 & $-$26.75 & 19.77 & \phn1.43 & 0.34 & 0.21 & 20.79 & 10.00 & 0.16 & 0.56  \\
PG2349$-$014 $H$  & 13.54 & $-$25.93 & 13.57 & $-$25.90 & 20.41 & \phn6.19 & 0.38 & 0.22 & 19.32 & \phn3.60 & 0.51 & 1.67  \\
\enddata
\tablenotetext{a}{Definitions of the parameters are given in equation \ref{equ:hostfit}. The bulge component is characterized by $I_e$ (effective surface brightness), $r_e$ (effective radius), $e_e$ (elongation) and $PA_e$ (position angle). The disk component is characterized by $I_0$ (peak surface brightness), $r_0$ (effective radius), $e_0$ (elongation) and $PA_0$ (position angle).}
\tablenotetext{b}{The host type adopted in Table \ref{tab:hostclass} does not necessarily match the lowest $\chi^2$ value shown here, as morphological structures in the images is also taken into account. As discussed in the text, classification from $\chi^2$ values alone is unreliable. In columns 2 (Nuclear Mad.) to 11 ($PA_0$), the results of the fit corresponding to the final host classification in Table \ref{tab:hostclass} are given. For hosts that are unclassified (PG0953+414, PG1211+143, PG1411+442, PG1426+015, PG1613+658, PG1626+554, PG1700+518 and PG2349$-$014), the results of the bulge+disk fits are shown.}
\end{deluxetable}

\clearpage

\begin{deluxetable}{ l c c c c c c c c c c }
\tablewidth{0pt}
\tabletypesize{\footnotesize}
\tablecaption{\label{tab:hostclassstatnir} Host Classification Statistics:  $H$-band luminosities\tablenotemark{a}}
\tablehead{
           & \colhead{Number}\tablenotemark{b} & \multicolumn{3}{c}{Nucleus $M_{\rm H}$} & \multicolumn{3}{c}{Host $M_{\rm H}$} & \multicolumn{3}{c}{$L_{\rm nucl.}/L_{\rm host}$}\\
           &                &  \colhead{med.} &  \colhead{ave.} &  \colhead{$\sigma_{ave}$} &  \colhead{med.} &  \colhead{ave.} &  \colhead{$\sigma_{ave}$} &  \colhead{med.} &  \colhead{ave.} &  \colhead{$\sigma_{ave}$} 
}
\startdata
Elliptical & 10 & $-$26.04 & $-$25.93 & 0.24 & $-$24.91 & $-$25.24 & 0.27 & 1.83 & 2.21 & 0.40\\

Disk present & \phn7 & $-$25.08 & $-$25.20 & 0.35 & $-$23.55 & $-$23.78 & 0.26 & 4.17 & 6.58 & 2.61\\

Bulge+Disk & \phn3 & $-$23.57 & $-$23.81 & 0.30 & $-$24.39 & $-$24.52 & 0.18 & 0.43 & 0.62 & 0.22\\

Unknown & \phn7 & $-$25.85 & $-$25.99 & 0.24 & $-$25.87 & $-$25.39 & 0.28 & 1.23 & 2.16 & 0.54\\
\hline
Non-disturbed & 18 & $-$25.40 & $-$25.38 & 0.25 & $-$24.47 & $-$24.62 & 0.24 & 1.93 & 2.86 & 0.56\\

Disturbed & \phn3 & $-$24.83 & $-$25.40 & 0.62 & $-$24.87 & $-$24.44 & 0.36 & 0.96 & 7.84 & 5.73\\

Strongly disturbed & \phn6 & $-$25.88 & $-$26.02 & 0.27 & $-$25.88 & $-$25.61 & 0.26 & 1.13 & 1.68 & 0.39\\

\hline
\quad\quad\quad\quad\quad\quad Total & 27 & $-$25.57 & $-$25.52 & 0.20 & $-$24.58 & $-$24.82 & 0.19 & 1.69 & 3.15 & 0.81\\
\enddata
\tablenotetext{a}{If no $H$ band images were acquired, $H$ magnitudes for the QSO and host galaxy were derived from $J$ or $K^\prime$, assuming $J\!-\!H=0.94$ and $H\!-\!K=0.97$ for the QSO nucleus (average values derived from the photometry obtained in our sample). For the host galaxy, $J\!-\!H = 0.72-0.4\times z$ and $H\!-\!K = 0.22 + 2.0 \times z$ were assumed, where $z$ is the redshift.}
\tablenotetext{a}{Only objects for which a meaningful fit was obtained are included in this table: PG0050+124, PG0953+414, PG1211+143, PG1626+554 and PG2251+113 were excluded. This subsample of 27 objects is identical to the one shown in the top panel of Fig. \ref{fig:MHhMB_class} and in
Fig. \ref{fig:MBvsMH_class}.}
\end{deluxetable}

\clearpage

\begin{deluxetable}{ l c c c c c c }
\tablewidth{0pt}
\tabletypesize{\scriptsize}
\tablecaption{\label{tab:rad_prop} Mid- and Far-Infrared Source Properties\tablenotemark{a}}
\tablehead{
\colhead{Name} & \colhead{$f_{12 \mu m}$} & \colhead{$f_{25 \mu m}$} & \colhead{$f_{60 \mu m}$} & \colhead{$f_{100 \mu m}$} & \colhead{log $L_{\rm IR}$} & \colhead{$L_{\rm IR}/L_{\rm BB}$\tablenotemark{b}}\\ 
  & \colhead{(mJy)} &  \colhead{(mJy)} &   \colhead{(mJy)}  &   \colhead{(mJy)}  &  \colhead{($L_{\odot}$)} & \\ 
}
\startdata
PG0026+129 & \phn\phn\phn\phn$\phantom{<}22 \pm  4\phantom{0}\phantom{0}$ & $\phn\phn<40\phantom{\pm000}$ & $\phn\phn<27\phantom{\pm000}$ & $\phn\phn<80\phantom{\pm000}$ & $\phantom{<0}10.93\pm0.04$ & $\phantom{<0}0.06\pm0.01$ \\
PG0050+124 & \phn\phn\phn$\phantom{<}549 \pm 11\phantom{0}$ & $\phn\phn\phantom{<}1097 \pm 20\phantom{0}$ & $\phn\phn\phantom{<}2293 \pm 17\phantom{0}$ & $\phn\phn\phantom{<}2959 \pm 51\phantom{0}$ & $\phantom{<0}11.91\pm0.00$ & $\phantom{<0}1.71\pm0.15$ \\
PG0157+001 & \phn\phn\phn$\phantom{<}136 \pm 28\phantom{0}$ & $\phn\phn\phn\phantom{<}525 \pm 55\phantom{0}$ & $\phn\phn\phantom{<}2376 \pm 56\phantom{0}$ & $\phn\phn\phantom{<}2308 \pm 127$ & $\phantom{<0}12.59\pm0.01$ & $\phantom{<0}1.46\pm0.13$ \\
PG0804+761 & \phn\phn\phn$\phantom{<}171 \pm 22\phantom{0}$ & $\phn\phn\phn\phantom{<}210 \pm 25\phantom{0}$ & $\phn\phn\phn\phantom{<}190 \pm 34\phantom{0}$ & $\phn\phn\phn\phantom{<}116 \pm 41\phantom{0}$ & $\phantom{<0}11.62\pm0.02$ & $\phantom{<0}0.31\pm0.02$ \\
PG0838+770 & \phn\phn\phn\phn$\phantom{<}34 \pm  5\phantom{0}\phantom{0}$ & $\phn\phn\phn\phantom{<}102 \pm  7\phantom{0}\phantom{0}$ & $\phn\phn\phn\phantom{<}174 \pm  9\phantom{0}\phantom{0}$ & $\phn\phn\phn\phantom{<}159 \pm 21\phantom{0}$ & $\phantom{<0}11.47\pm0.01$ & $\phantom{<0}1.79\pm0.09$ \\
PG0844+349 & \phn\phn\phn$\phantom{<}126 \pm 29\phantom{0}$ & $\phn\phn\phn\phantom{<}209 \pm 35\phantom{0}$ & $\phn\phn\phn\phantom{<}163 \pm 41\phantom{0}$ & $\phn\phn\phn\phantom{<}294 \pm 97\phantom{0}$ & $\phantom{<0}11.16\pm0.03$ & $\phantom{<0}0.47\pm0.04$ \\
PG0923+201 & \phd\phn\phn$<98\phantom{\pm000}$ & $\phn\phn\phn\phantom{<}150 \pm 45\phantom{0}$ & $\phn\phn\phn\phantom{<}340 \pm 57\phantom{0}$ & $\phn\phn\phantom{<}1020 \pm 184$ & $\phantom{<0}12.10\pm0.03$ & $\phantom{<0}0.64\pm0.05$ \\
PG0953+414 & \phd\phn\phn$<89\phantom{\pm000}$ & $\phn<107\phantom{\pm000}$ & $\phn\phn\phn\phantom{<}170 \pm 26\phantom{0}$ & $\phn<315\phantom{\pm000}$ & $\phantom{<0}11.80\pm0.06$ & $\phantom{<0}0.17\pm0.03$ \\
PG1001+054 & \phn\phn\phn\phn$\phantom{<}34 \pm  6\phantom{0}\phantom{0}$ & $\phn\phn\phn\phn\phantom{<}45 \pm 11\phantom{0}$ & $\phn\phn\phn\phn\phantom{<}32 \pm  9\phantom{0}\phantom{0}$ & $\phn\phn\phn\phn\phantom{<}78 \pm 27\phantom{0}$ & $\phantom{<0}11.39\pm0.03$ & $\phantom{<0}0.42\pm0.03$ \\
PG1004+130 & \phd\phn\phn$<91\phantom{\pm000}$ & $\phn<150\phantom{\pm000}$ & $\phn\phn\phn\phantom{<}191 \pm 42\phantom{0}$ & $\phn<284\phantom{\pm000}$ & $\phantom{<0}11.88\pm0.06$ & $\phantom{<0}0.19\pm0.03$ \\
PG1116+215 & \phn\phn\phn$\phantom{<}152 \pm 46\phantom{0}$ & $\phn\phn\phn\phantom{<}187 \pm 56\phantom{0}$ & $\phn<219\phantom{\pm000}$ & $\phn<285\phantom{\pm000}$ & $\phantom{<0}12.06\pm0.04$ & $\phantom{<0}0.46\pm0.06$ \\
PG1119+120 & \phn\phn\phn$\phantom{<}120 \pm 40\phantom{0}$ & $\phn\phn\phn\phantom{<}280 \pm 53\phantom{0}$ & $\phn\phn\phn\phantom{<}546 \pm 51\phantom{0}$ & $\phn\phn\phn\phantom{<}746 \pm 126$ & $\phantom{<0}11.12\pm0.02$ & $\phantom{<0}0.70\pm0.05$ \\
PG1126$-$041 & \phn\phn\phn$\phantom{<}104 \pm 19\phantom{0}$ & $\phn\phn\phn\phantom{<}309 \pm 35\phantom{0}$ & $\phn\phn\phn\phantom{<}669 \pm 26\phantom{0}$ & $\phn\phn\phantom{<}1172 \pm 134$ & $\phantom{<0}11.33\pm0.01$ & $\phantom{<0}0.68\pm0.04$ \\
PG1211+143 & \phn\phn\phn$\phantom{<}167 \pm 30\phantom{0}$ & $\phn\phn\phn\phantom{<}347 \pm 44\phantom{0}$ & $\phn\phn\phn\phantom{<}327 \pm 50\phantom{0}$ & $\phn\phn\phn\phantom{<}408 \pm 140$ & $\phantom{<0}11.55\pm0.02$ & $\phantom{<0}0.38\pm0.02$ \\
PG1229+204 & \phn\phn\phn\phn$\phantom{<}76 \pm 25\phantom{0}$ & $\phn\phn\phn\phantom{<}246 \pm 30\phantom{0}$ & $\phn\phn\phn\phantom{<}183 \pm 36\phantom{0}$ & $\phn\phn\phn\phantom{<}317 \pm 95\phantom{0}$ & $\phantom{<0}11.09\pm0.03$ & $\phantom{<0}0.60\pm0.04$ \\
PG1302$-$102 & \phn\phn\phn\phn$\phantom{<}32 \pm  6\phantom{0}\phantom{0}$ & $\phn<180\phantom{\pm000}$ & $\phn\phn\phn\phantom{<}343 \pm 68\phantom{0}$ & $\phn\phn\phn\phantom{<}343 \pm 68\phantom{0}$ & $\phantom{<0}12.28\pm0.03$ & $\phantom{<0}0.25\pm0.03$ \\
PG1307+085 & \phd\phn\phn$<54\phantom{\pm000}$ & $\phn<153\phantom{\pm000}$ & $\phn\phn\phn\phantom{<}212 \pm 42\phantom{0}$ & $\phn\phn\phn\phantom{<}155 \pm 31\phantom{0}$ & $\phantom{<0}11.46\pm0.05$ & $\phantom{<0}0.23\pm0.03$ \\
PG1309+355 & \phn\phn\phn\phn$\phantom{<}60 \pm 18\phantom{0}$ & $\phn\phn\phn\phantom{<}102 \pm 31\phantom{0}$ & $\phn<140\phantom{\pm000}$ & $\phn<192\phantom{\pm000}$ & $\phantom{<0}11.76\pm0.04$ & $\phantom{<0}0.29\pm0.03$ \\
PG1351+640 & \phn\phn\phn$\phantom{<}172 \pm  5\phantom{0}\phantom{0}$ & $\phn\phn\phn\phantom{<}532 \pm  6\phantom{0}\phantom{0}$ & $\phn\phn\phn\phantom{<}757 \pm  8\phantom{0}\phantom{0}$ & $\phn\phn\phantom{<}1167 \pm 26\phantom{0}$ & $\phantom{<0}11.81\pm0.00$ & $\phantom{<0}0.77\pm0.03$ \\
PG1411+442 & \phn\phn\phn$\phantom{<}115 \pm 10\phantom{0}$ & $\phn\phn\phn\phantom{<}160 \pm 12\phantom{0}$ & $\phn\phn\phn\phantom{<}162 \pm 17\phantom{0}$ & $\phn\phn\phn\phantom{<}140 \pm 28\phantom{0}$ & $\phantom{<0}11.38\pm0.01$ & $\phantom{<0}0.36\pm0.02$ \\
PG1426+015 & \phn\phn\phn$\phantom{<}127 \pm 26\phantom{0}$ & $\phn\phn\phn\phantom{<}195 \pm 40\phantom{0}$ & $\phn\phn\phn\phantom{<}323 \pm 43\phantom{0}$ & $\phn\phn\phn\phantom{<}346 \pm 100$ & $\phantom{<0}11.48\pm0.02$ & $\phantom{<0}0.92\pm0.06$ \\
PG1435$-$067 & \phn\phn\phn\phn$\phantom{<}50 \pm 15\phantom{0}$ & $\phn<126\phantom{\pm000}$ & $\phn\phn\phn\phantom{<}141 \pm 52\phantom{0}$ & $\phn<315\phantom{\pm000}$ & $\phantom{<0}11.34\pm0.04$ & $\phantom{<0}0.26\pm0.04$ \\
PG1440+356 & \phn\phn\phn$\phantom{<}101 \pm 12\phantom{0}$ & $\phn\phn\phn\phantom{<}208 \pm 15\phantom{0}$ & $\phn\phn\phn\phantom{<}651 \pm 21\phantom{0}$ & $\phn\phn\phantom{<}1042 \pm 62\phantom{0}$ & $\phantom{<0}11.52\pm0.01$ & $\phantom{<0}0.73\pm0.07$ \\
PG1613+658 & \phn\phn\phn\phn$\phantom{<}87 \pm 12\phantom{0}$ & $\phn\phn\phn\phantom{<}231 \pm 14\phantom{0}$ & $\phn\phn\phn\phantom{<}635 \pm 19\phantom{0}$ & $\phn\phn\phantom{<}1090 \pm 59\phantom{0}$ & $\phantom{<0}11.96\pm0.01$ & $\phantom{<0}0.57\pm0.03$ \\
PG1617+175 & \phn\phn\phn\phn$\phantom{<}65 \pm 13\phantom{0}$ & $\phn\phn\phn\phn\phantom{<}71 \pm 17\phantom{0}$ & $\phn\phn\phn\phantom{<}102 \pm 20\phantom{0}$ & $\phn<252\phantom{\pm000}$ & $\phantom{<0}11.32\pm0.03$ & $\phantom{<0}0.31\pm0.02$ \\
PG1626+554 & \phn\phn\phn\phn$\phantom{<}33 \pm 10\phantom{0}$ & $\phn\phn\phn\phn\phantom{<}39 \pm 12\phantom{0}$ & $\phn\phn<70\phantom{\pm000}$ & $\phn\phn\phn\phn\phantom{<}70 \pm 21\phantom{0}$ & $\phantom{<0}11.16\pm0.04$ & $\phantom{<0}0.18\pm0.02$ \\
PG1700+518 & \phn\phn\phn$\phantom{<}111 \pm 18\phantom{0}$ & $\phn\phn\phn\phantom{<}222 \pm 20\phantom{0}$ & $\phn\phn\phn\phantom{<}466 \pm 34\phantom{0}$ & $\phn\phn\phn\phantom{<}441 \pm 69\phantom{0}$ & $\phantom{<0}12.68\pm0.01$ & $\phantom{<0}0.76\pm0.04$ \\
PG2130+099 & \phn\phn\phn$\phantom{<}186 \pm  9\phantom{0}\phantom{0}$ & $\phn\phn\phn\phantom{<}380 \pm 10\phantom{0}$ & $\phn\phn\phn\phantom{<}479 \pm 12\phantom{0}$ & $\phn\phn\phn\phantom{<}485 \pm 145$ & $\phantom{<0}11.39\pm0.01$ & $\phantom{<0}0.58\pm0.03$ \\
4C 31.63 & \phn\phn\phn\phn$\phantom{<}54 \pm 11\phantom{0}$ & $\phn\phn\phn\phantom{<}106 \pm 40\phantom{0}$ & $\phn\phn\phn\phantom{<}193 \pm 30\phantom{0}$ & $\phn\phn\phn\phantom{<}212 \pm 82\phantom{0}$ & $\phantom{<0}12.36\pm0.03$ & $\phantom{<0}0.42\pm0.03$ \\
PG2214+139 & \phn\phn\phn\phn$\phantom{<}61 \pm  7\phantom{0}\phantom{0}$ & $\phn\phn\phn\phn\phantom{<}95 \pm 12\phantom{0}$ & $\phn\phn\phn\phantom{<}337 \pm 11\phantom{0}$ & $\phn<282\phantom{\pm000}$ & $\phantom{<0}11.00\pm0.01$ & $\phantom{<0}0.29\pm0.01$ \\
PG2251+113 & \phd\phn\phn$<36\phantom{\pm000}$ & $\phn\phn<66\phantom{\pm000}$ & $\phn\phn<67\phantom{\pm000}$ & $\phn<214\phantom{\pm000}$ & $<12.04\phantom{\pm0.00}$ & $<0.25\phantom{\pm0.00}$ \\
PG2349$-$014 & \phd\phn$<119\phantom{\pm000}$ & $\phn<180\phantom{\pm000}$ & $\phn\phn\phn\phantom{<}271 \pm 51\phantom{0}$ & $\phn\phn\phn\phantom{<}290 \pm 50\phantom{0}$ & $\phantom{<0}11.75\pm0.05$ & $\phantom{<0}0.30\pm0.04$ \\
\enddata
\tablenotetext{a}{IRAS and ISO fluxes, in mJy, from \citet{san89}, \citet{haa00}, \citet{pol00}, and \citet{haa03} . When several independent measurements exist, we list the variance weighted mean value. Uncertainties are 1 $\sigma$; all upper limits are 3 $\sigma$.}
\tablenotetext{b}{See text for definitions of $L_{\rm IR}$, the ``Infrared Bump'' luminosity, and $L_{\rm BB}$, the ``Blue Bump'' luminosity.}
\end{deluxetable}

\clearpage

\begin{deluxetable}{ l c c c c c c}
\tablewidth{0pt}
\tabletypesize{\footnotesize}
\tablecaption{\label{tab:hostclassstat} Host Classification Statistics:  FIR luminosities}
\tablehead{
           & \colhead{Number} & \colhead{Median\tablenotemark{a}} & \colhead{Median\tablenotemark{a}} & \colhead{Median\tablenotemark{a}} & \colhead{Mean [$\sigma$] \tablenotemark{b}}\\
           &                &  \colhead{$M_{\rm B}$ (mag)} &  \colhead{log $L_{\rm IR}$ ($L_{\odot}$)} &  \colhead{$L_{\rm IR}/L_{\rm BB}$}&  \colhead{$L_{\rm IR}/L_{\rm BB}$}
}
\startdata
Elliptical & 10 & $-$23.88 & 11.61 & 0.27 & 0.29 [0.14]\\

Disk present & \phn8 & $-$22.52 & 11.49 & 0.53 & 0.81 [0.56]\\

Bulge+Disk & \phn3 & $-$21.63 & 11.12 & 0.68 & 0.66 [0.04]\\

Unknown \tablenotemark{c}& \phn11 & $-$23.85 & 11.77\tablenotemark{d} & 0.38 & 0.54 [0.39]\\
\hline
Non-disturbed & 21 & $-$23.16 & 11.42\tablenotemark{d} & 0.42 & 0.47 [0.35]\\

Disturbed & \phn3 & $-$22.19 & 11.47 & 0.46 & 0.85 [0.67]\\

Strongly disturbed & \phn6 & $-$23.93 & 11.85 & 0.67 & 0.73 [0.39]\\

Unknown & \phn2 & $-$24.04 & 11.67 & 0.27 & 0.27 [0.10]\\
\hline

\quad\quad\quad\quad\quad\quad Total & 32 & $-$23.23 & 11.52\tablenotemark{d} & 0.42 & 0.54 [0.42]\\
\enddata
\tablenotetext{a}{The median of an even number of values is the average of the two central values.}
\tablenotetext{b}{For the QSO with $L_{\rm IR}/L_{\rm BB}$ upper limit (PG2251+113), the 1 $\sigma$ $L_{\rm IR}/L_{\rm BB}$ upper limit is used to derive both the mean and its standard deviation.}
\tablenotetext{c}{Includes 6 hosts classified in Table 5 as ``strongly disturbed''.}
\tablenotetext{d}{The median was determined without including PG 2251+113, for which there is only an upper limit on $L_{\rm IR}$.}
\end{deluxetable}

\end{document}